\documentclass[twocolumn,showpacs,preprintnumbers,superscriptaddress,amsmath,amssymb,aps,pre,10pt]{revtex4-1}

\usepackage{graphicx}
\usepackage{dcolumn}
\usepackage{bm}
\usepackage{color}
\usepackage{multirow}
\usepackage{listings}

\begin{document}


\title{Comparing the performance of FA, DFA and DMA using different synthetic long-range correlated time series}

\author{Ying-Hui Shao}
 \affiliation{School of Business, East China University of Science and Technology, Shanghai 200237, China} %
 \affiliation{Research Center for Econophysics, East China University of Science and Technology, Shanghai 200237, China} %

\author{Gao-Feng Gu}
 \affiliation{School of Business, East China University of Science and Technology, Shanghai 200237, China} %
 \affiliation{Research Center for Econophysics, East China University of Science and Technology, Shanghai 200237, China} %

\author{Zhi-Qiang Jiang}
 \affiliation{School of Business, East China University of Science and Technology, Shanghai 200237, China} %
 \affiliation{Research Center for Econophysics, East China University of Science and Technology, Shanghai 200237, China} %

\author{Wei-Xing Zhou}
 \email{wxzhou@ecust.edu.cn}
 \affiliation{School of Business, East China University of Science and Technology, Shanghai 200237, China} %
 \affiliation{Research Center for Econophysics, East China University of Science and Technology, Shanghai 200237, China} %
 \affiliation{School of Science, East China University of Science and Technology, Shanghai 200237, China} %

\author{Didier Sornette}
\email{dsornette@ethz.ch}
\affiliation{Department of Management, Technology and Economics, ETH Zurich, Zurich, Switzerland}
\affiliation{Swiss Finance Institute, c/o University of Geneva, Geneva, Switzerland}

\date{\today}

\begin{abstract}
Notwithstanding the significant efforts to develop estimators of long-range correlations (LRC) and to compare their performance, no clear consensus exists on what is the best method and under which conditions. In addition, synthetic tests suggest that the performance of LRC estimators varies when using different generators of LRC time series. Here, we compare the performances of four estimators [Fluctuation Analysis (FA), Detrended Fluctuation Analysis (DFA), Backward Detrending Moving Average (BDMA), and centred Detrending Moving Average (CDMA)]. We use three different generators [Fractional Gaussian Noises, and two ways of generating Fractional Brownian Motions]. We find that CDMA has the best performance and DFA is only slightly worse in some situations, while FA performs the worst. In addition, CDMA and DFA are less sensitive to the scaling range than FA. Hence, CDMA and DFA remain ``The Methods of Choice'' in determining the Hurst index of time series.
\end{abstract}


\maketitle



A complex system, be it ecological, biological, technological, social, economic or financial, is usually embedded in a complex network, which is composed of a large number of interacting heterogeneous constituents linked via interwoven nonlinear heterogenous ties \cite{Albert-Barabasi-2002-RMP}. The observed signals of the physical quantities characterizing a complex system often exhibit long-range correlations \cite{Sornette-2004}. It is of crucial importance and significance to quantify such long-range correlations to have a deep understanding of the dynamics of the underlying complex systems. More than ten techniques have been invented to detect long-range correlations in time series \cite{Taqqu-Teverovsky-Willinger-1995-Fractals,Kantelhardt-2009-ECSS}, such as the rescaled range (R/S) analysis \cite{Hurst-1951-TASCE}, the wavelet transform module maxima (WTMM) approach
\cite{Holschneider-1988-JSP,Muzy-Bacry-Arneodo-1991-PRL,Bacry-Muzy-Arneodo-1993-JSP,Muzy-Bacry-Arneodo-1993-PRE,Muzy-Bacry-Arneodo-1994-IJBC}, the fluctuation analysis (FA) \cite{Peng-Buldyrev-Goldberger-Havlin-Sciortino-Simons-Stanley-1992-Nature}, the detrended fluctuation analysis (DFA) \cite{Peng-Buldyrev-Havlin-Simons-Stanley-Goldberger-1994-PRE}, the detrending moving average analysis (DMA) \cite{Alessio-Carbone-Castelli-Frappietro-2002-EPJB}, and so on.

Our work focuses on three methods (FA, DFA and DMA) that are very popular especially in the econophysics community. Consider a time series $\{x(t): t=1,2,\cdots,N\}$ with zero mean and its profile $y(t)$ constructed as the cumulative sum of $x(t)$. The three methods proceed to obtain fluctuation functions $F(s)$ specific to a timescale $s$. For long-range correlated time series, we have
\begin{equation}
  F(s) \sim s^{\alpha},
\end{equation}
where $\alpha$ is a scaling exponent. In FA, the fluctuation function is computed as follows \cite{Peng-Buldyrev-Goldberger-Havlin-Sciortino-Simons-Stanley-1992-Nature}
\begin{equation}
  F(s) = \sqrt{\langle[y(t+s)-y(t)]^2\rangle},
  \label{Eq:F:s:FA}
\end{equation}
which is actually a special case of the structure function in turbulence \cite{Kolmogorov-1962-JFM}.
In contrast, both DFA and DMA adopt detrending techniques. The time series $y(t)$ is covered by $N_s$ disjoint boxes of size $s$. When the whole time series $y(t)$ cannot be completely covered by $N_s$ boxes, we can utilize $2N_s$ boxes to cover the time series by starting from both ends of the time series. In each box, a trend function $g(t)$ of the sub-series is determined. The residuals are calculated by
\begin{equation}
  \epsilon(t) = y(t)-g(t),
  \label{Eq:epsilon}
\end{equation}
where the trend $g(t)$ is a polynomial function in the DFA algorithm \cite{Peng-Buldyrev-Havlin-Simons-Stanley-Goldberger-1994-PRE} and a moving average function over $s$ data points in
the DMA method  \cite{Alessio-Carbone-Castelli-Frappietro-2002-EPJB}. The fluctuation function $F(s)$ is
then obtained as the r.m.s. of the residual time series:
\begin{equation}
  F(s) = \sqrt{\frac{1}{N}\sum_{t=1}^{N} \left[\epsilon(t)\right]^2}.
  \label{Eq:F:s:DFA:DMA}
\end{equation}
Note that all these methods have a multifractal version \cite{Ghashghaie-Breymann-Peinke-Talkner-Dodge-1996-Nature,CastroESilva-Moreira-1997-PA,Weber-Talkner-2001-JGR,Kantelhardt-Zschiegner-KoscielnyBunde-Havlin-Bunde-Stanley-2002-PA,Gu-Zhou-2010-PRE} and can be generalized to handle high-dimensional fractals and multifractals \cite{Gu-Zhou-2006-PRE,Carbone-2007-PRE,Gu-Zhou-2010-PRE}. When $y(t)$ is a fractional Brownian motion (FBM), the scaling exponent $\alpha$ is identical to the Hurst index $H$ \cite{Talkner-Weber-2000-PRE,Heneghan-McDarby-2000-PRE,Kantelhardt-KoscielnyBunde-Rego-Havlin-Bunde-2001-PA,Arianos-Carbone-2007-PA}.

\begin{figure*}[t]
\centering
\includegraphics[width=3.3cm]{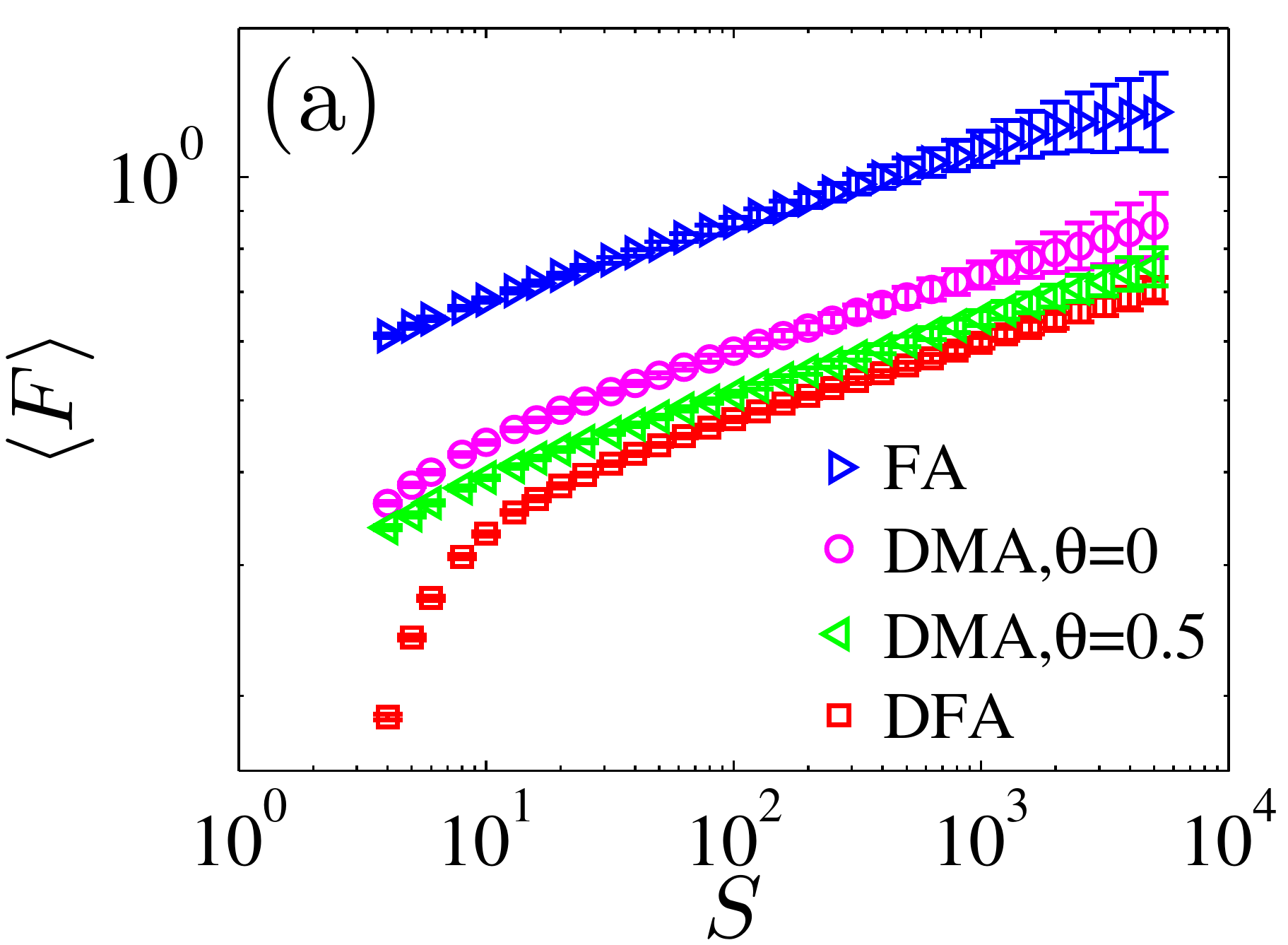}
\includegraphics[width=3.3cm]{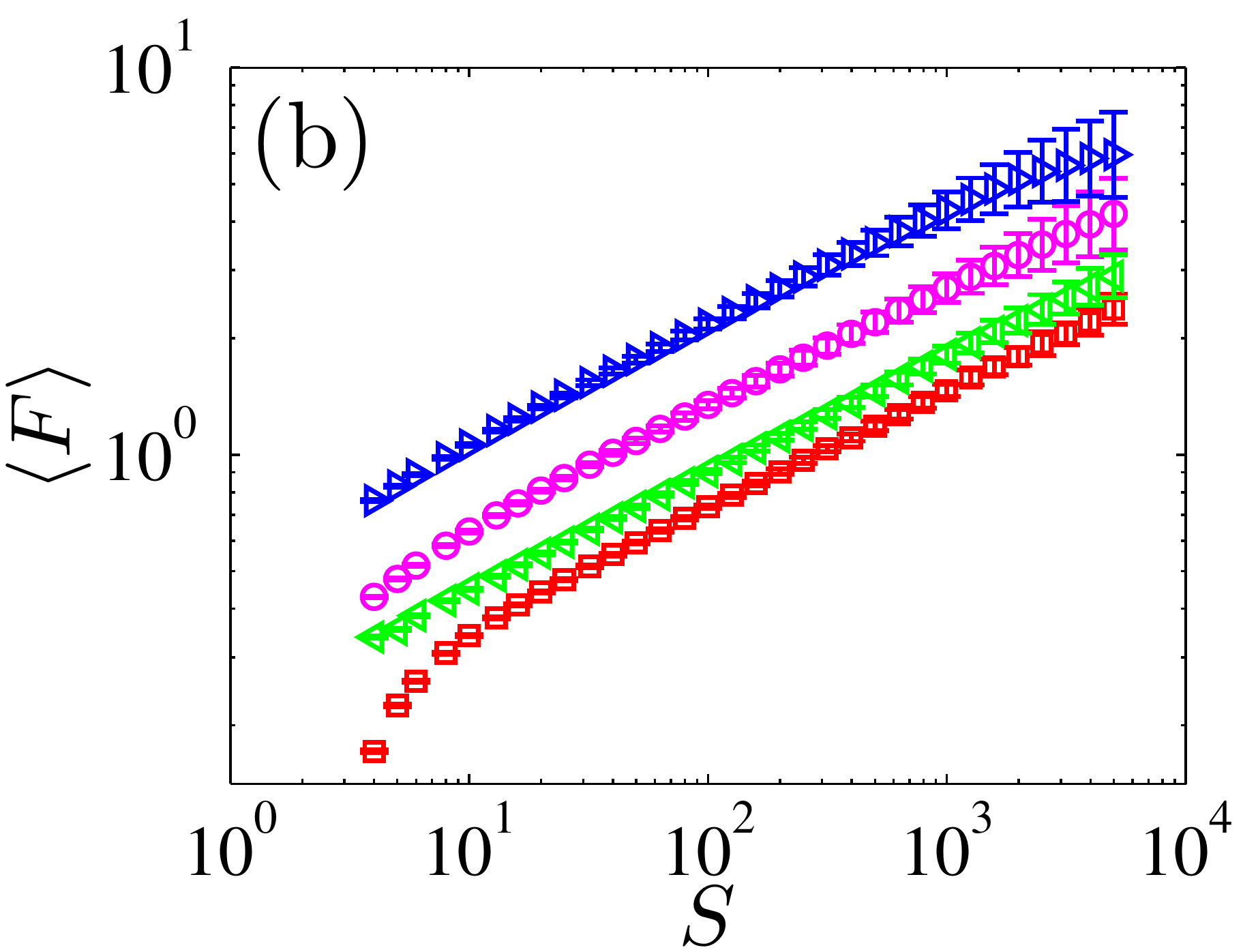}
\includegraphics[width=3.3cm]{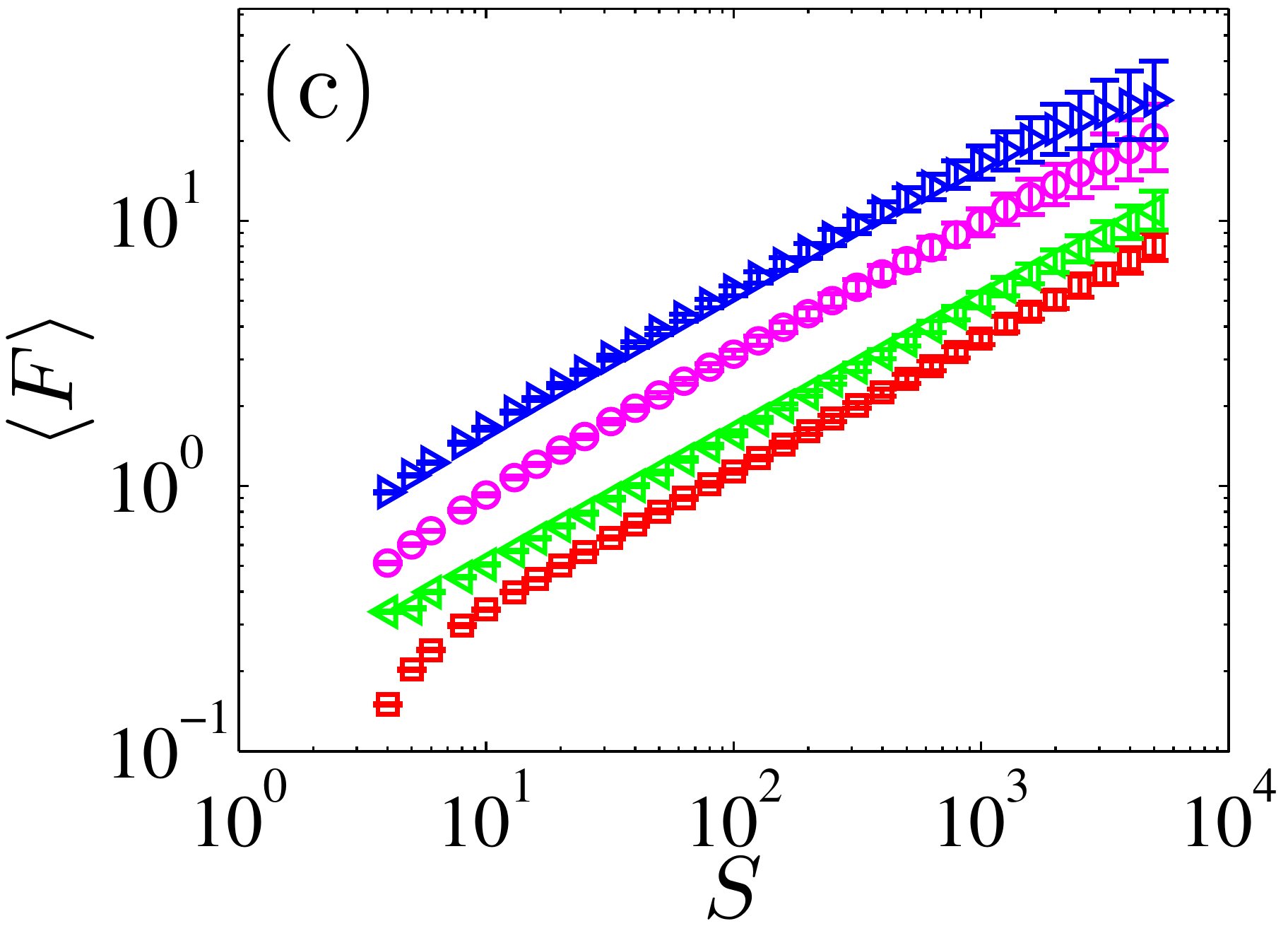}
\includegraphics[width=3.3cm]{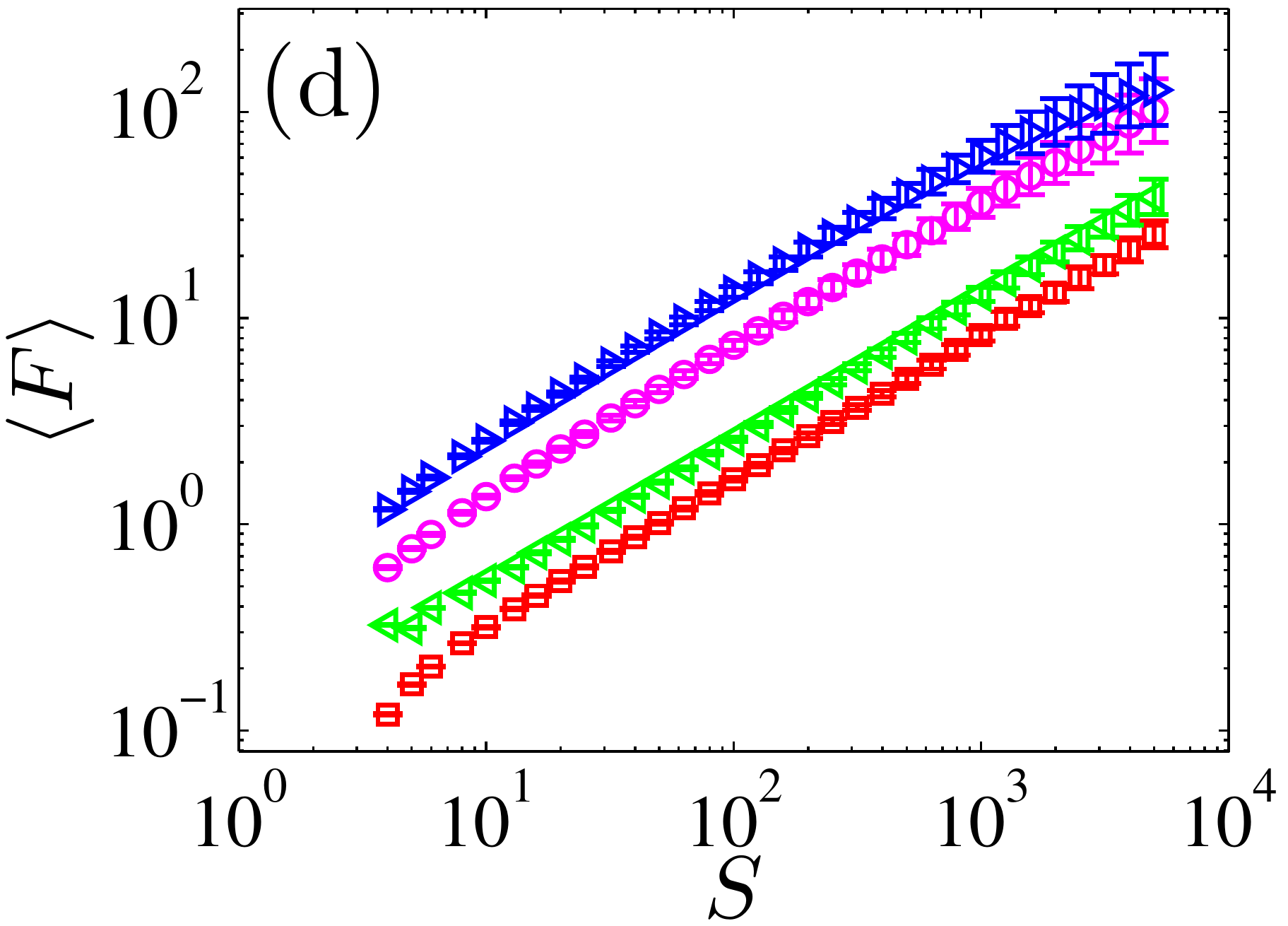}
\includegraphics[width=3.3cm]{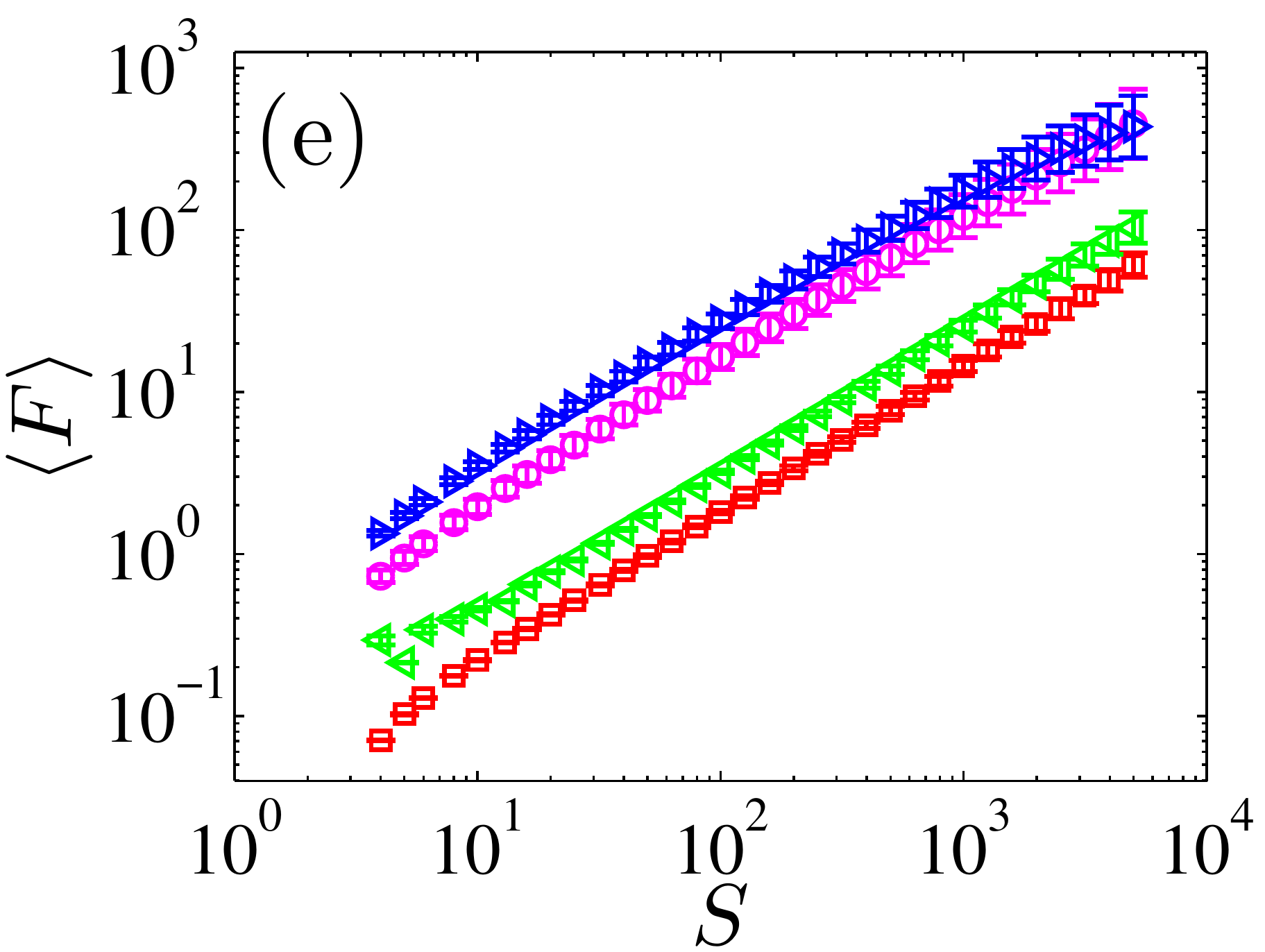}
\includegraphics[width=3.3cm]{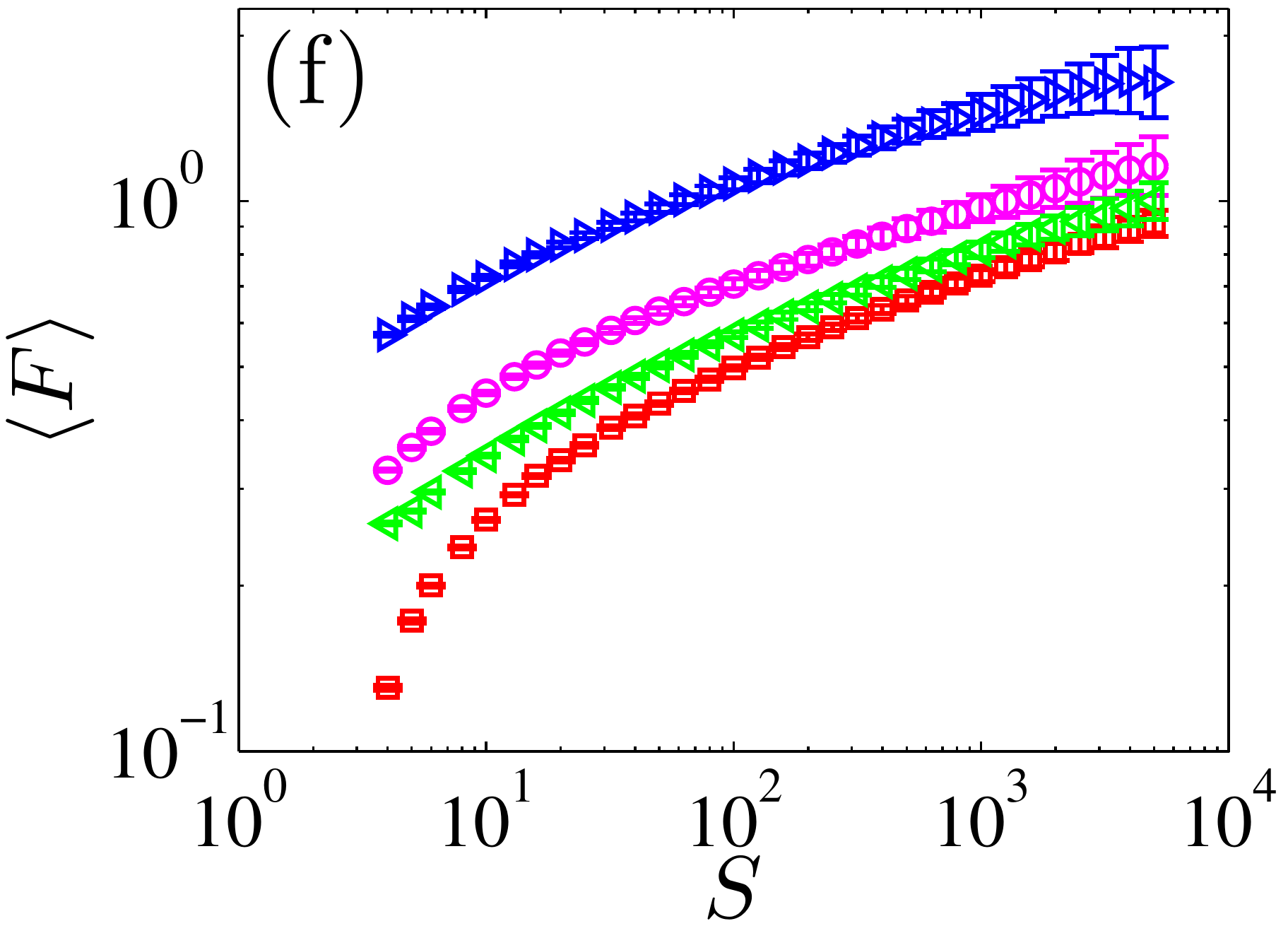}
\includegraphics[width=3.3cm]{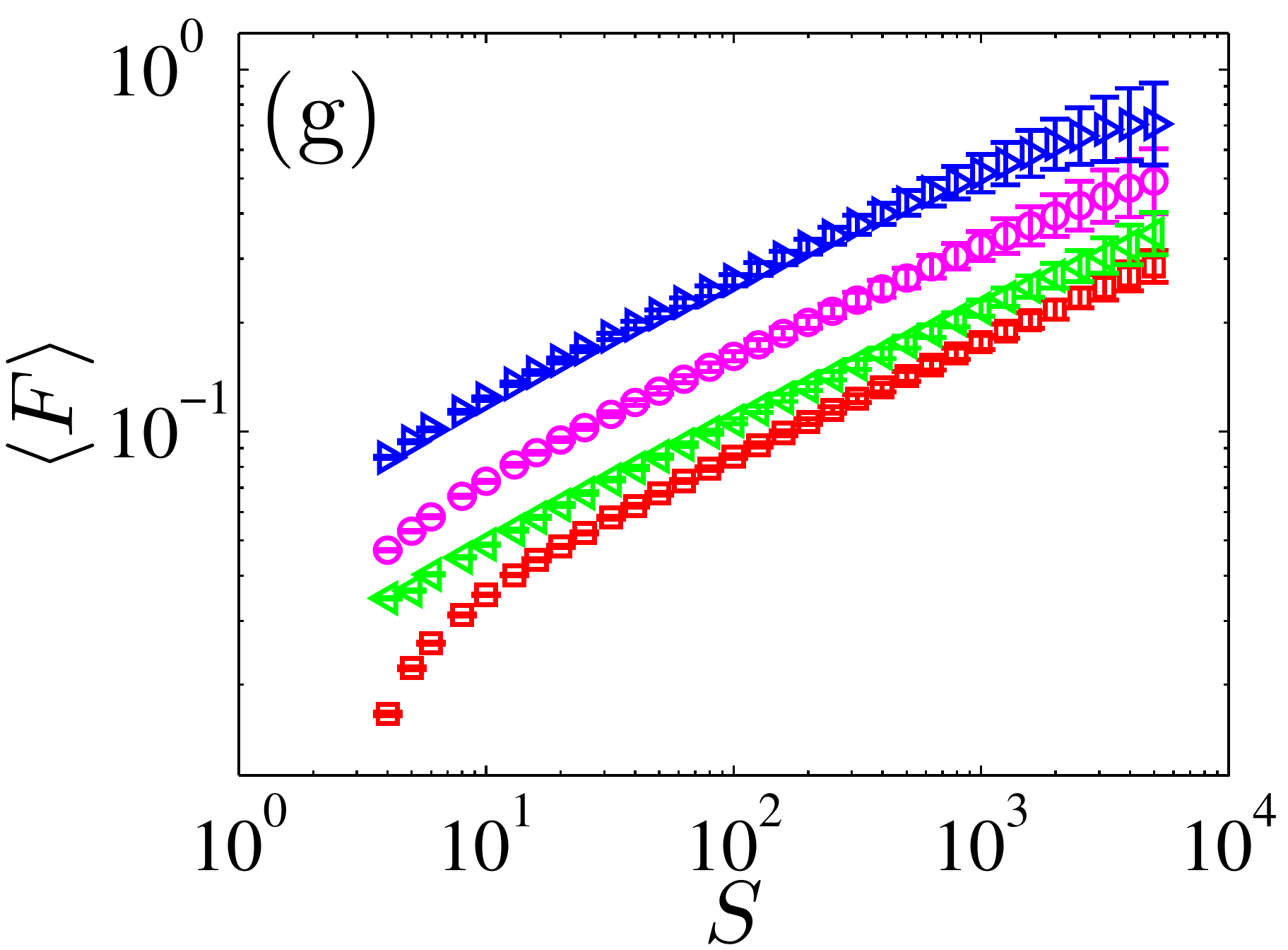}
\includegraphics[width=3.3cm]{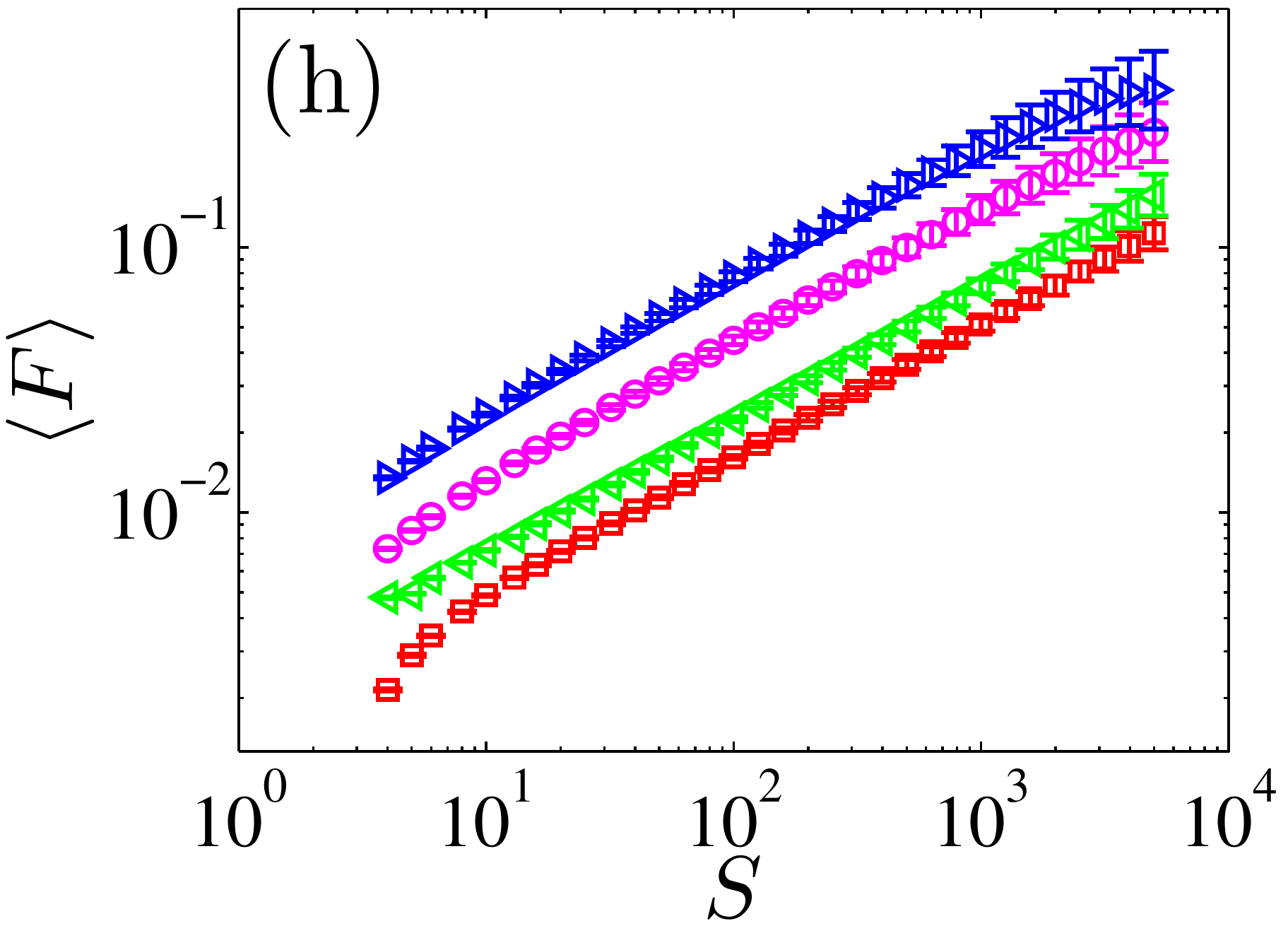}
\includegraphics[width=3.3cm]{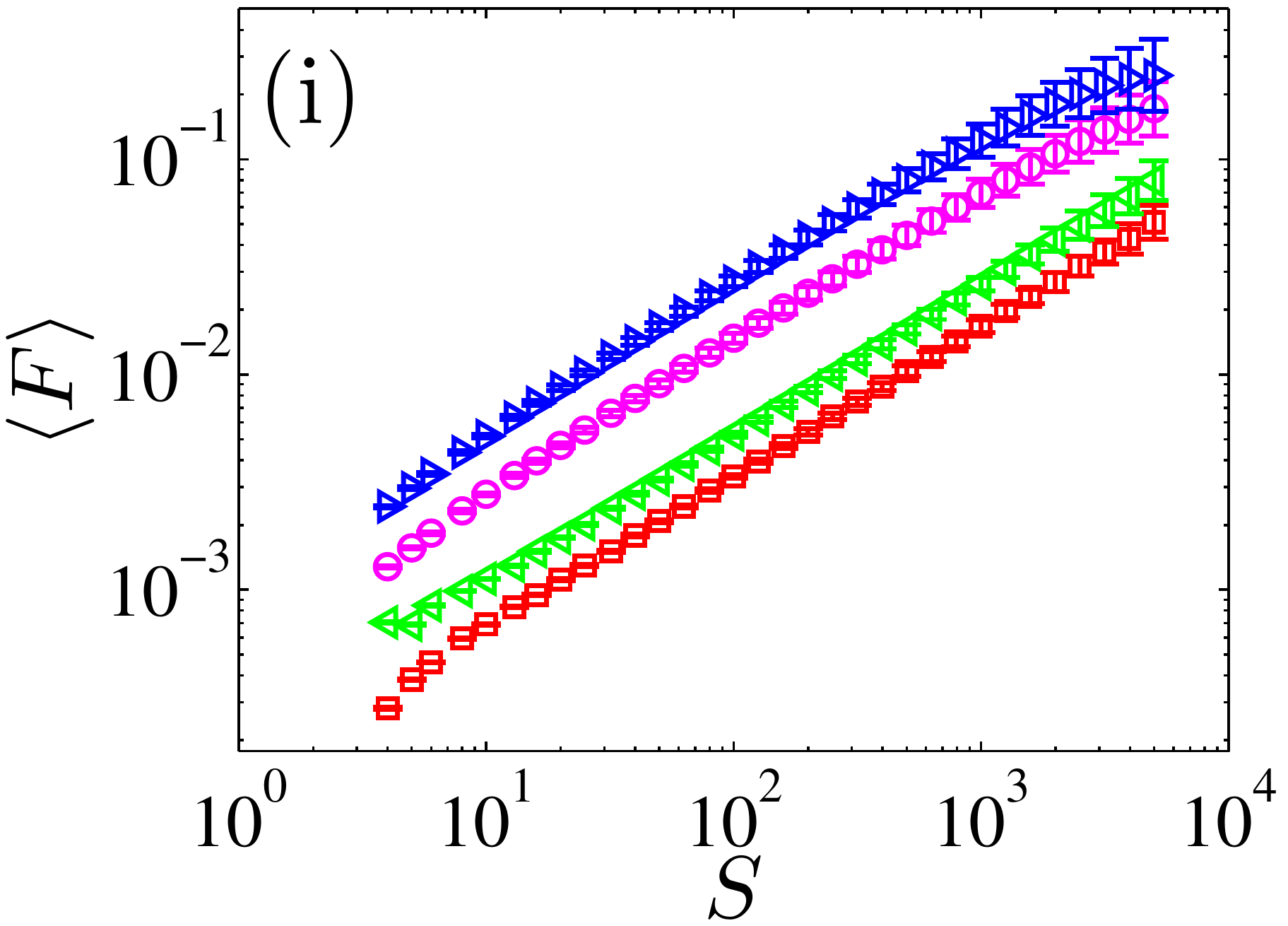}
\includegraphics[width=3.3cm]{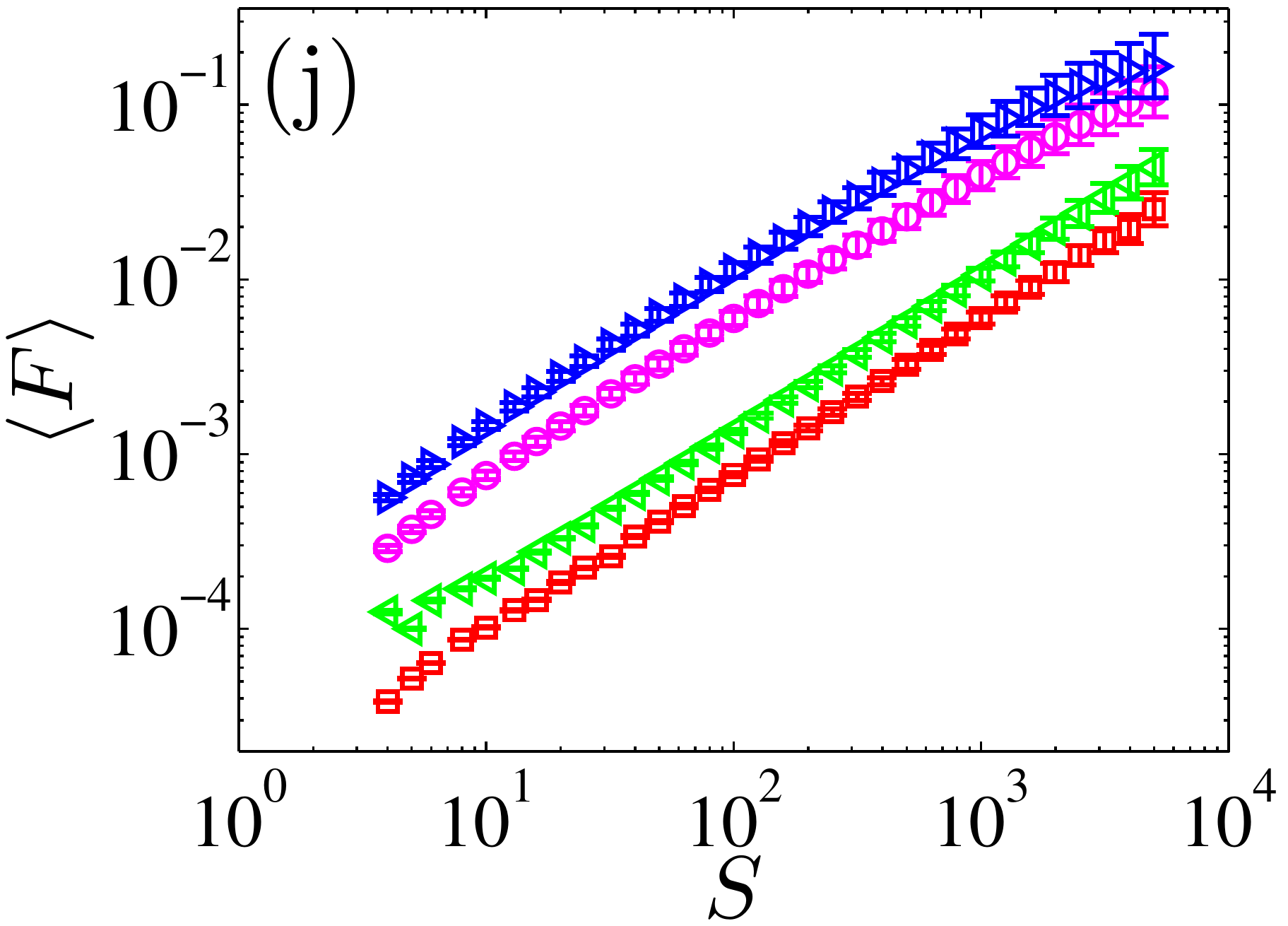}
\includegraphics[width=3.3cm]{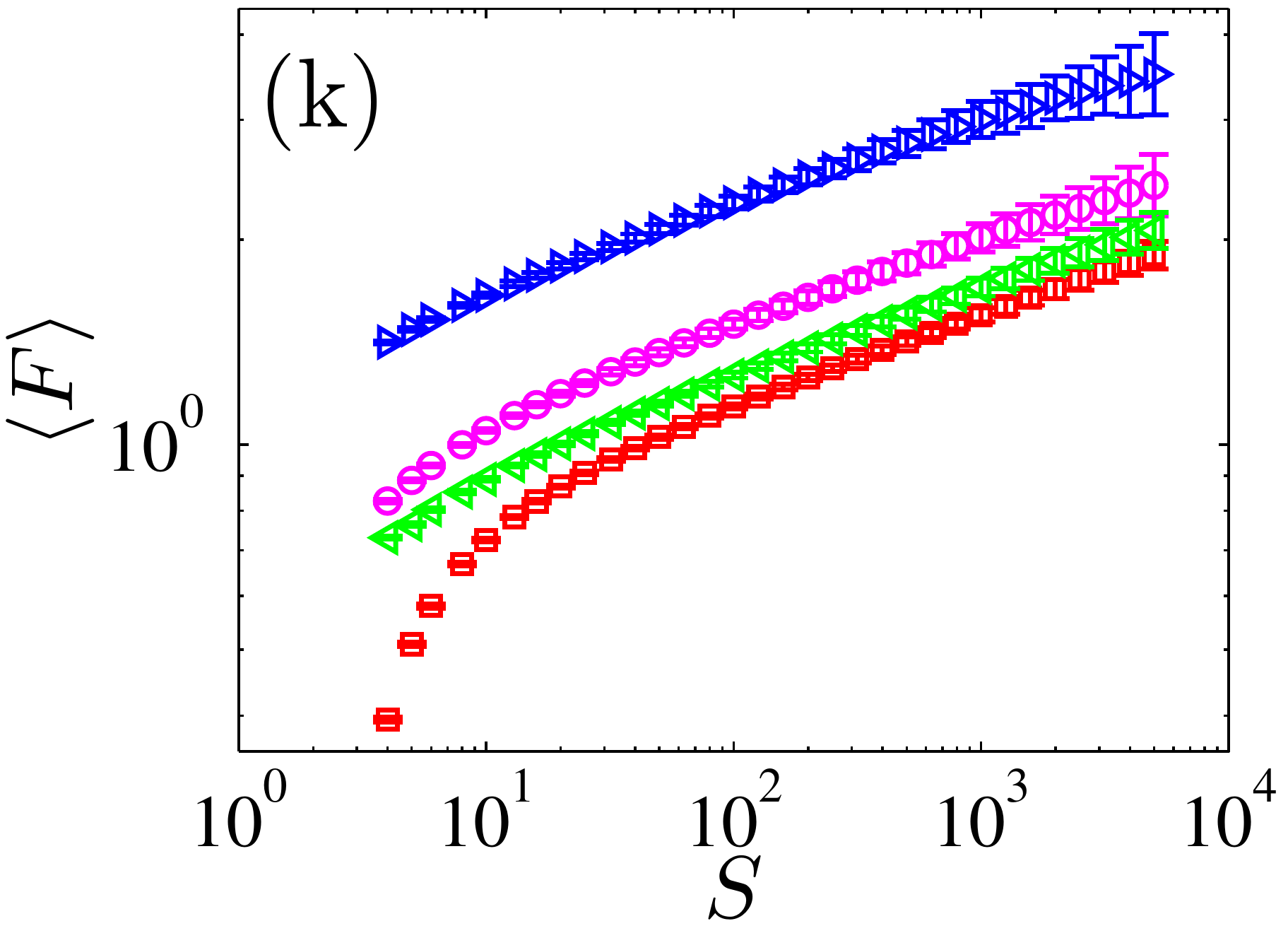}
\includegraphics[width=3.3cm]{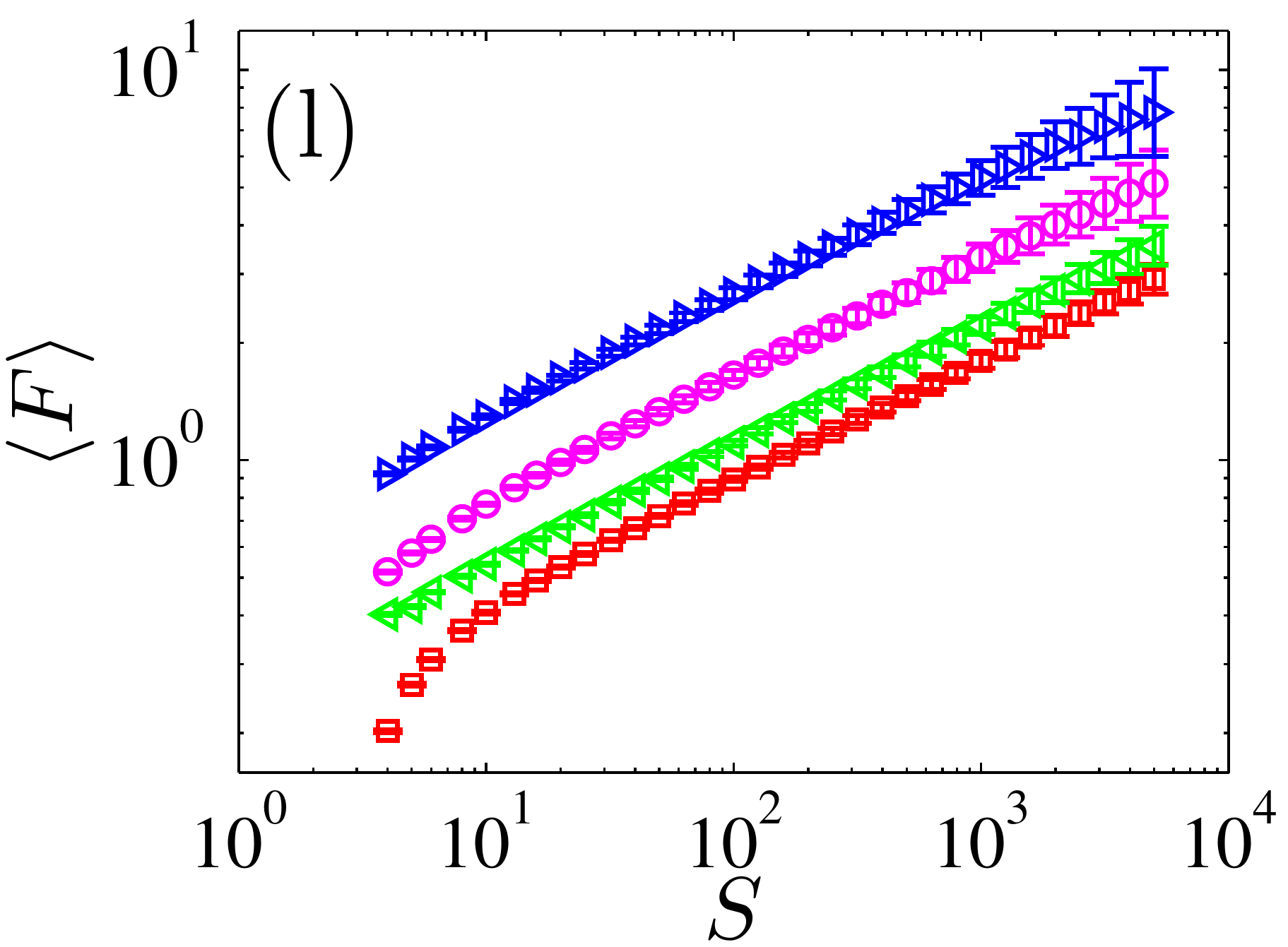}
\includegraphics[width=3.3cm]{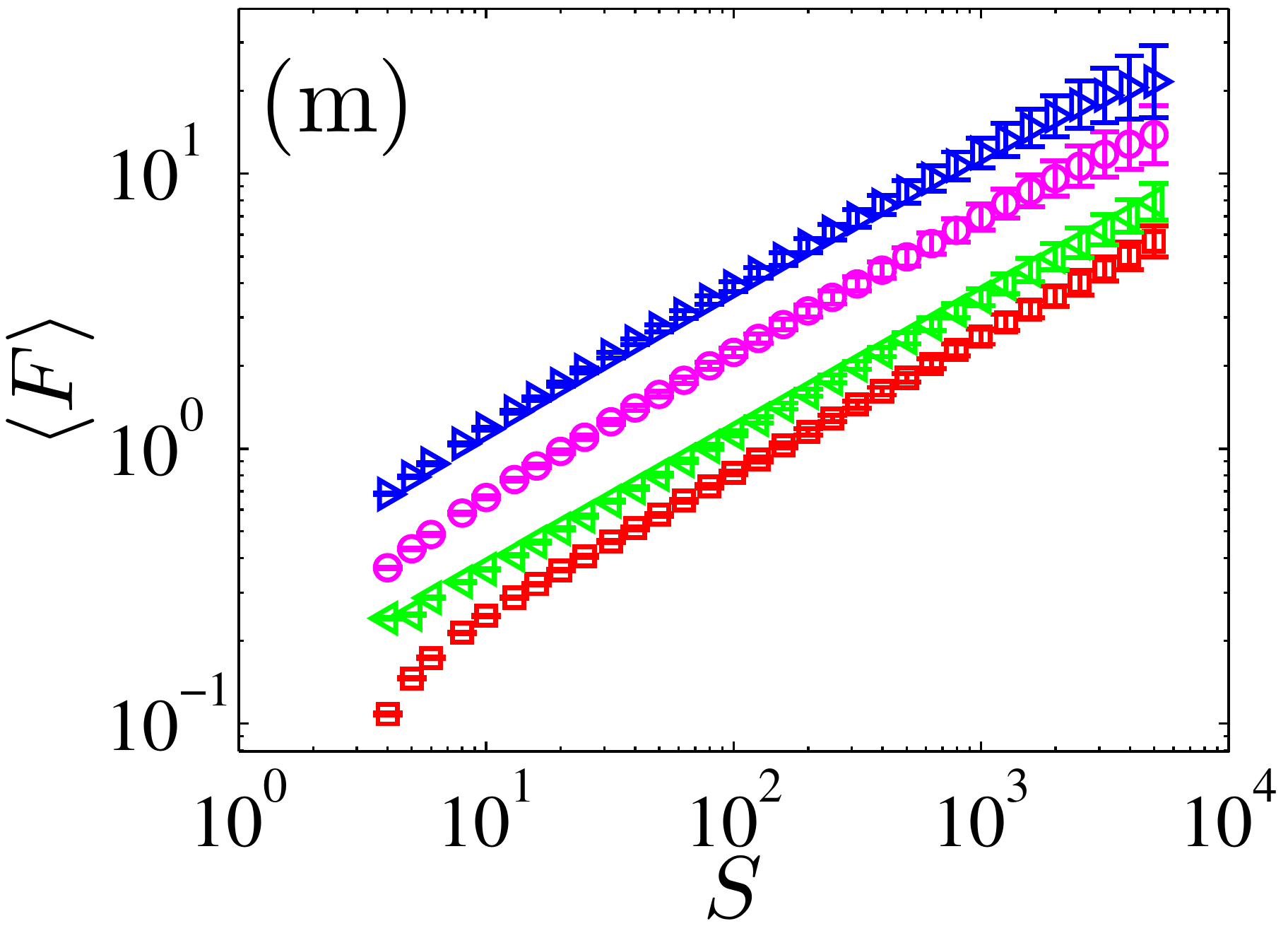}
\includegraphics[width=3.3cm]{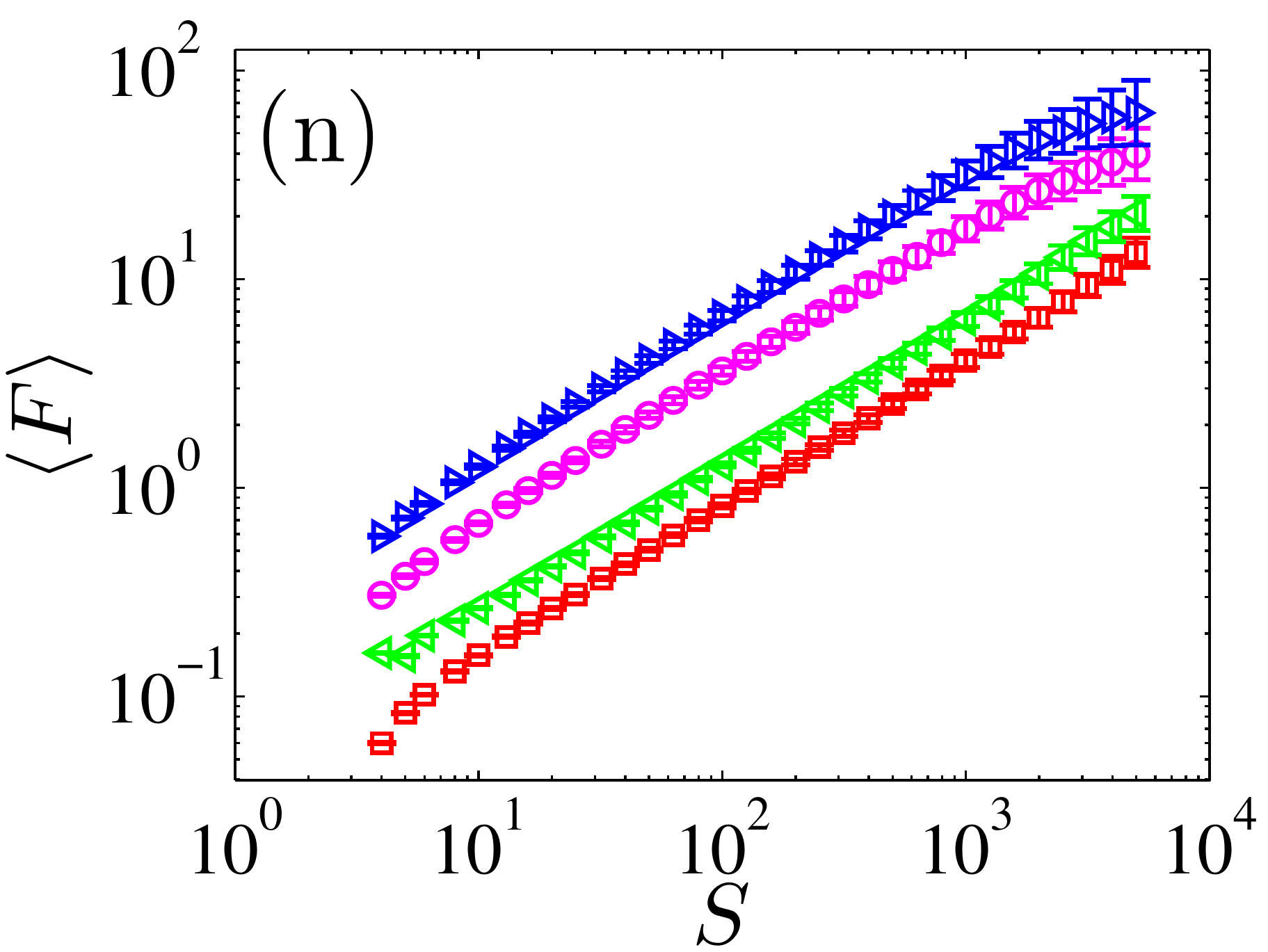}
\includegraphics[width=3.3cm]{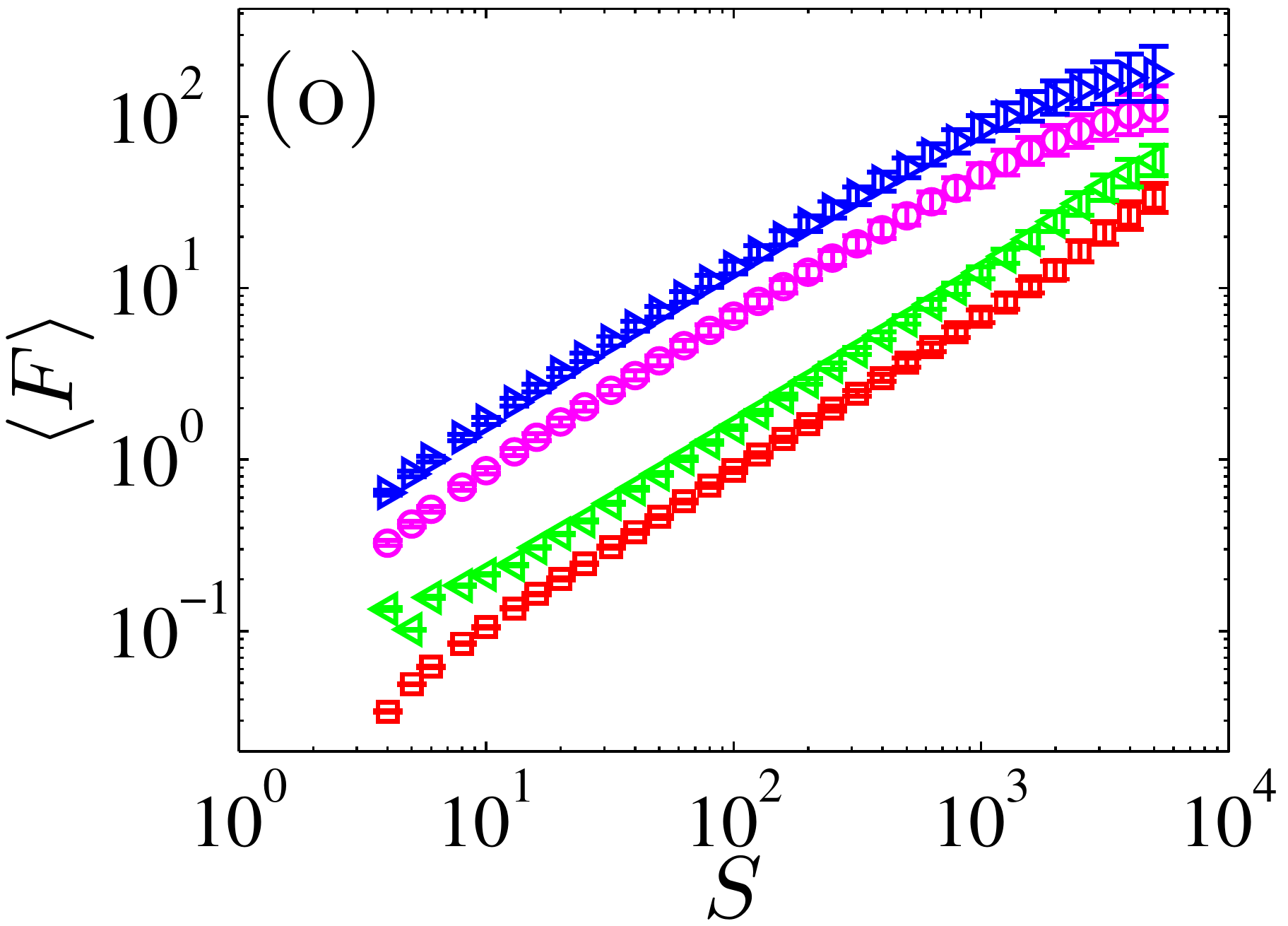}
\caption{{\textbf{Scaling plots of $\langle{F}\rangle$ against $s$.}} Each plot contains four curves obtained from four different analysis methods (FA, BDMA, CDMA and DFA) and each curve represents a fluctuation function averaged over 100 repeated simulated time series with the error bars showing the standard deviations. The three rows correspond to three generators (FGN-DH, FBM-RMD and WFBM from top to bottom). Each column corresponds to a fixed Hurst index ($H_{\mathrm{in}}=0.1,0.3,0.5,0.7$, and $0.9$ from left to right). The curves have been shifted vertically for better visibility.}
\label{Fig:FluctuationFunction}
\end{figure*}

Several groups have attempted to assess the performance and relative merits of these techniques.
Xu et al. \cite{Xu-Ivanov-Hu-Chen-Carbone-Stanley-2005-PRE} compare the performances of DFA and DMA on long-range power-law correlated time series synthesized using the modified Fourier filtering method \cite{Makse-Havlin-Schwartz-Stanley-1996-PRE}, and find that DFA is superior to different DMA variants. Bashan et al.  \cite{Bashan-Bartsch-Kantelhardt-Havlin-2008-PA} observe that the centred DMA performs as well as DFA for long time series with weak trends and slightly outperforms DFA for short data with weak trends.
They conclude that DFA ``remains the method of choice'' when the trend is not {\it{a priori}} known.
Serinaldi \cite{Serinaldi-2010-PA} uses the Davies-Harte algorithm to generate fractional Gaussian noises (FGNs) and FBMs by summing the FGNs \cite{Davis-Harte-1987-Bm}, and find that DFA and DMA have comparable performances. Jiang and Zhou \cite{Jiang-Zhou-2011-PRE} report that DFA and the centred DMA perform similarly and both of them outperform the backward and forward DMA methods, when the FBMs are generated using the Fourier-based Wood-Chan algorithm \cite{Wood-Chan-1994-JCGS}. Huang et al.  \cite{Huang-Schmitt-Hermand-Gagne-Lu-Liu-2011-PRE} find comparative performances of FA and DFA for FBMs with $H = 1/3$, which are generated with the Wood-Chan algorithm \cite{Wood-Chan-1994-JCGS}. In contrast, Bryce and Sprague \cite{Bryce-Sprague-2012-SR} argue that FA outperforms DFA, for FGNs with $H=0.3$ that are generated using the Davies-Harte algorithm \cite{Davis-Harte-1987-Bm}.

We notice that these studies concentrate on DFA versus DMA or DFA versus FA and report
what appears to be contradictory results when considered together.
A careful reading unveils that these studies cannot be directly
compared because they have adopted different synthesis algorithms (or generators)
for the long-range correlated time series to be tested. Indeed, comparing the performances of long-range correlation detection methods is not an easy task for the following reasons.
Firstly, there are many algorithms to generate FGNs and FBMs \cite{Zhou-Sornette-2002-IJMPC}, and one should
be careful not to draw too rapid conclusions on the relative performance of long-range correlation detection methods
that may be sensitive to the micro-structure of the generated time series that depend
on the specific synthesis algorithm. Secondly, real time series may contain {\it{a priori}} unknown nontrivial trends \cite{Montanari-Taqqu-Teverovsky-1999-MCM,Hu-Ivanov-Chen-Carpena-Stanley-2001-PRE,Chen-Ivanov-Hu-Stanley-2002-PRE,Chen-Hu-Carpena-Bernaola-Galvan-Stanley-Ivanov-2005-PRE}, which
complicates significantly the detection of long-range correlations, because trends
and long-range correlations often lead to similar signals.
Thirdly, there is no consensus on an objective determination approach of the scaling range, which plays a crucial role in the estimation of the scaling exponents. Often, studies use quite short scaling ranges (a decade or less), which
is an hindrance for determining the genuine presence of long-range correlations \cite{Malcai-Lidar-Biham-Avnir-1997-PRE,Mandelbrot-1998-Science,Avnir-Biham-Lidar-Malcai-1998-Science}.

In this work, we focus on comparing FA, DFA and two versions of DMA, where a linear detrending is adopted in DFA and the backward and centred versions of DMA (denoted BDMA and CDMA respectively) are investigated since the forward DMA performs the worst according to the literature. The comparison between FA, DFA and
two versions of DMA is conducted on time series generated using three different algorithms,
thus generating a $3 \times 4$ matrix of comparisons: (1) FGNs using the Davies-Harte algorithm (FGN-DH) \cite{Davis-Harte-1987-Bm} so that we can compare with the analysis by Bryce and Sprague \cite{Bryce-Sprague-2012-SR}, (2) FBMs using a wavelet-based generator (WFBM) \cite{Abry-Sellan-1996-ACHA}, which input Hurst indexes are very close to the estimated DFA exponents even when $H<0.5$ \cite{Ni-Jiang-Zhou-2009-PLA}, and (3) FBMs using the random midpoint displacement algorithm (FBM-RMD) \cite{Mandelbrot-1983}, because the numerical results of the generated time series are in excellent agreement with the analytical results for DMA \cite{Arianos-Carbone-2007-PA}. Besides, we do not consider trends or other hidden nonlinear structures.


\vspace{5mm}

\noindent{\Large\textbf{Results}}
\vspace{3mm}

\begin{figure*}
\centering
\includegraphics[width=3.3cm]{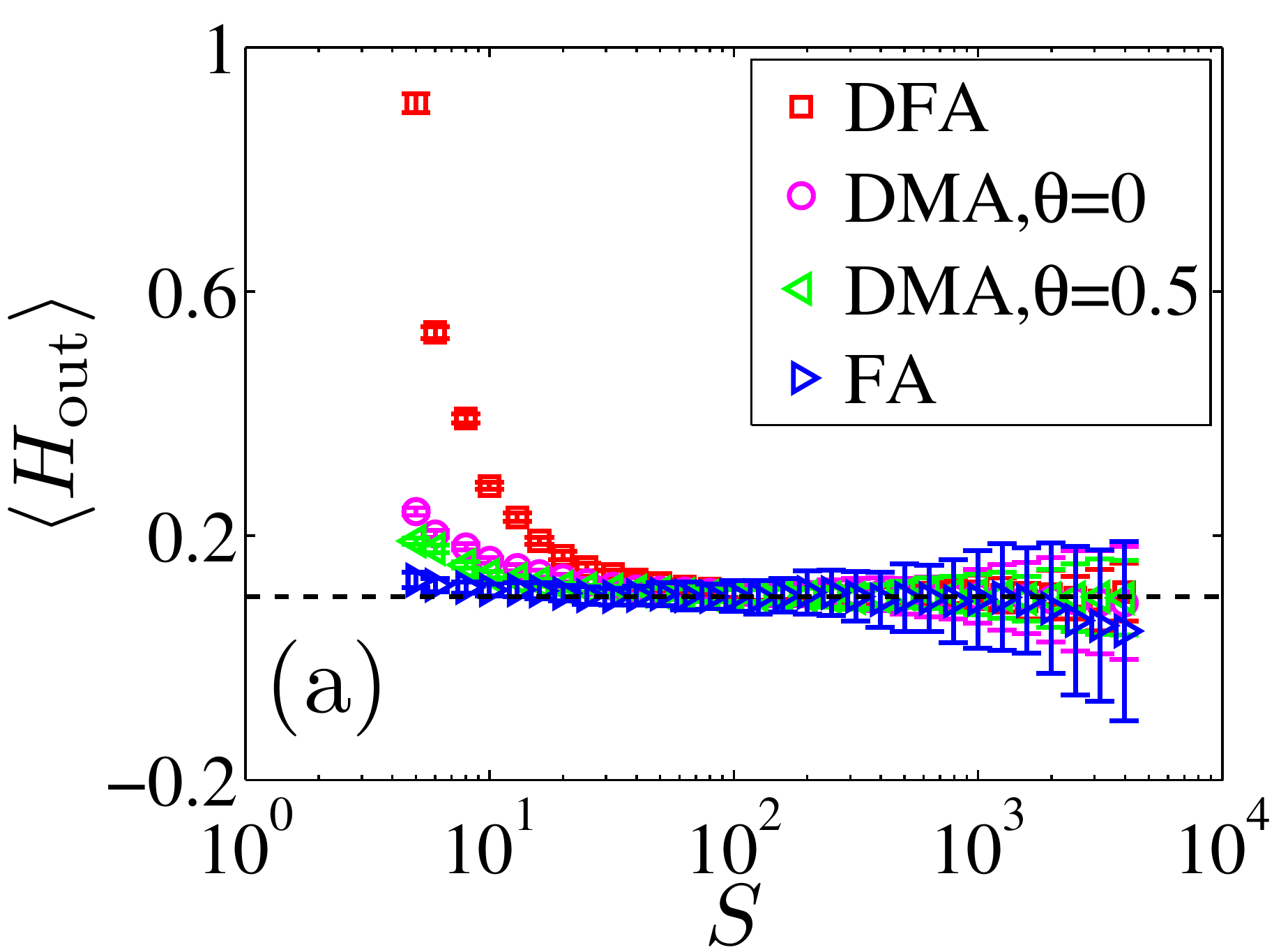}
\includegraphics[width=3.3cm]{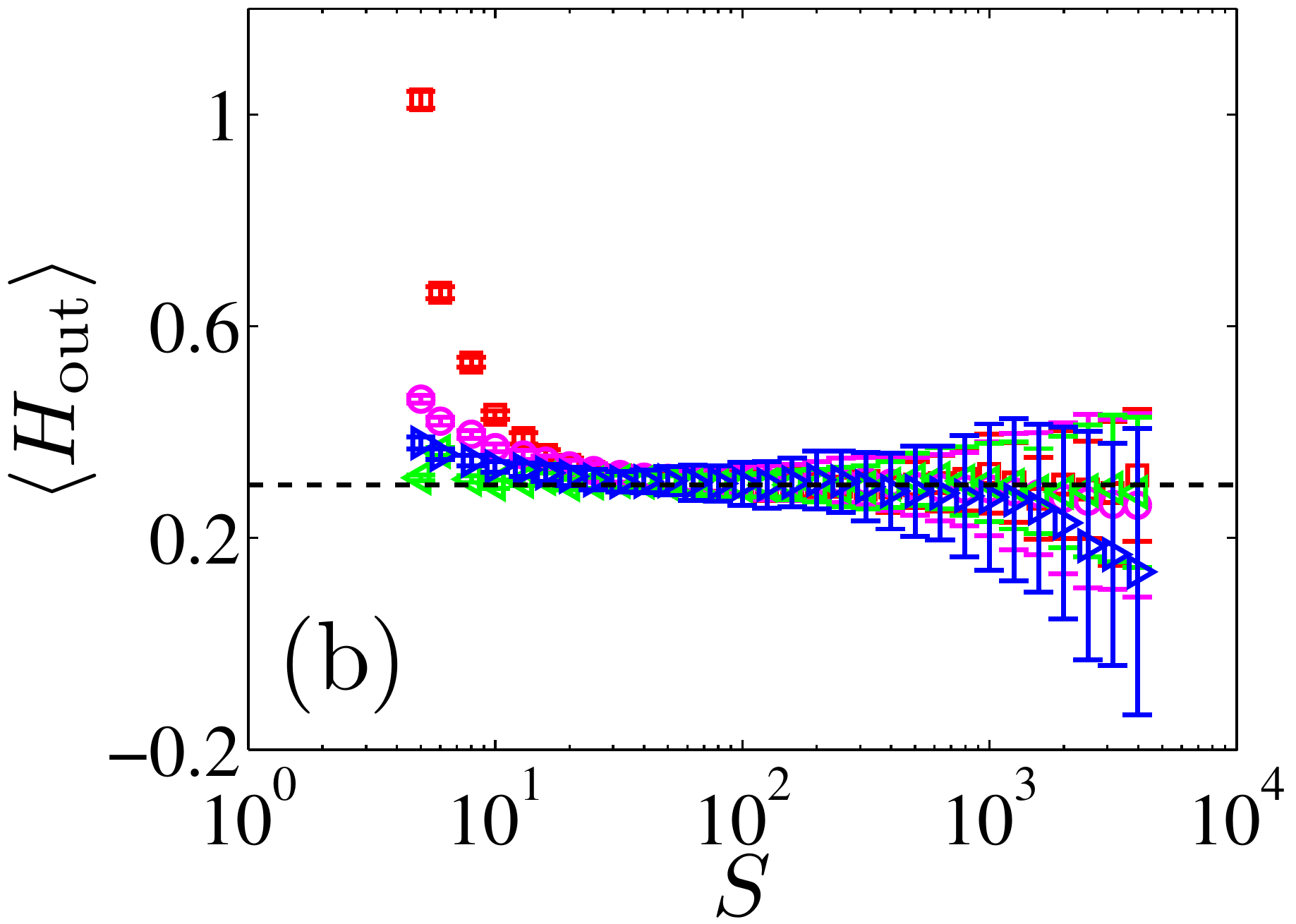}
\includegraphics[width=3.3cm]{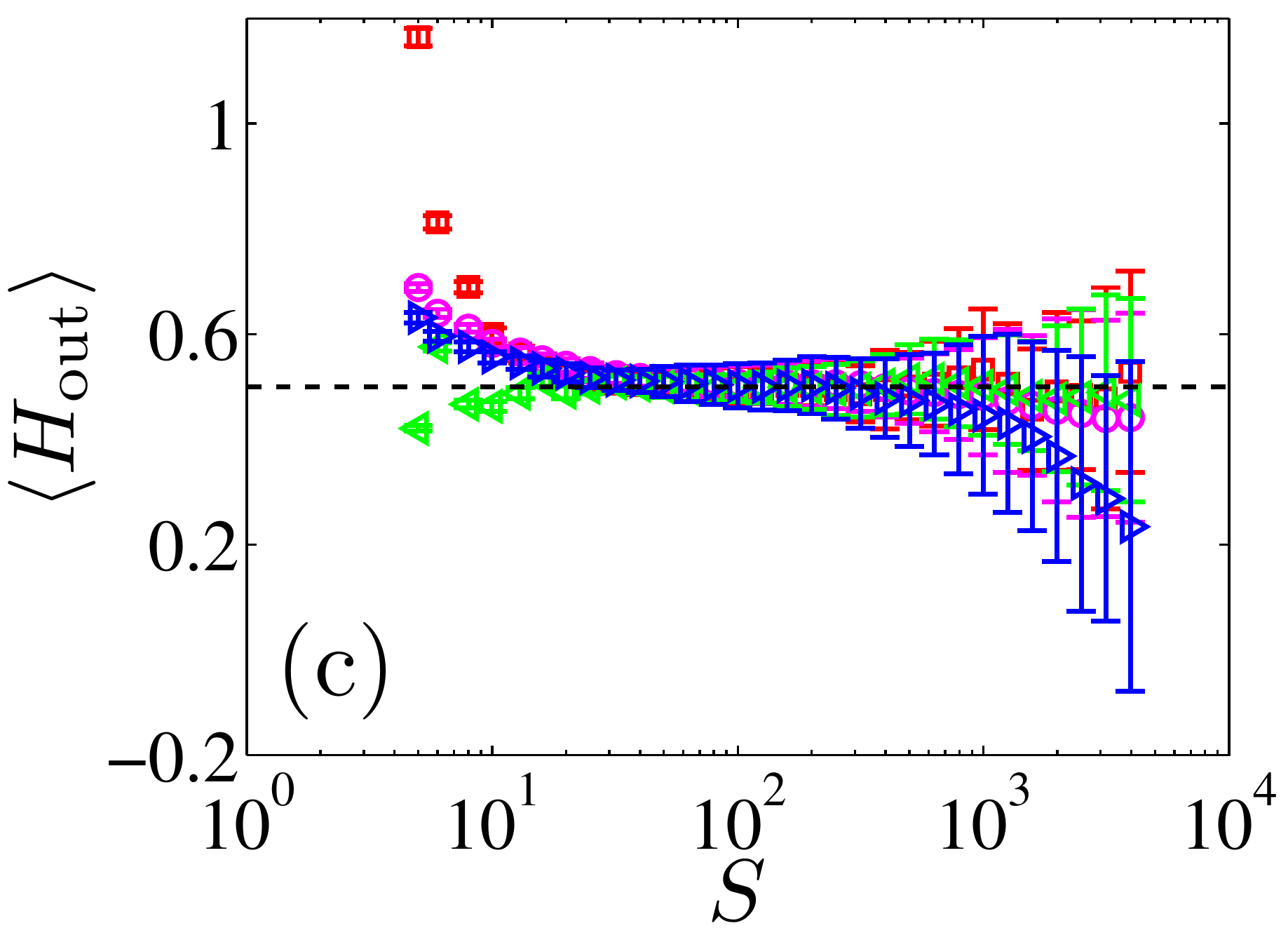}
\includegraphics[width=3.3cm]{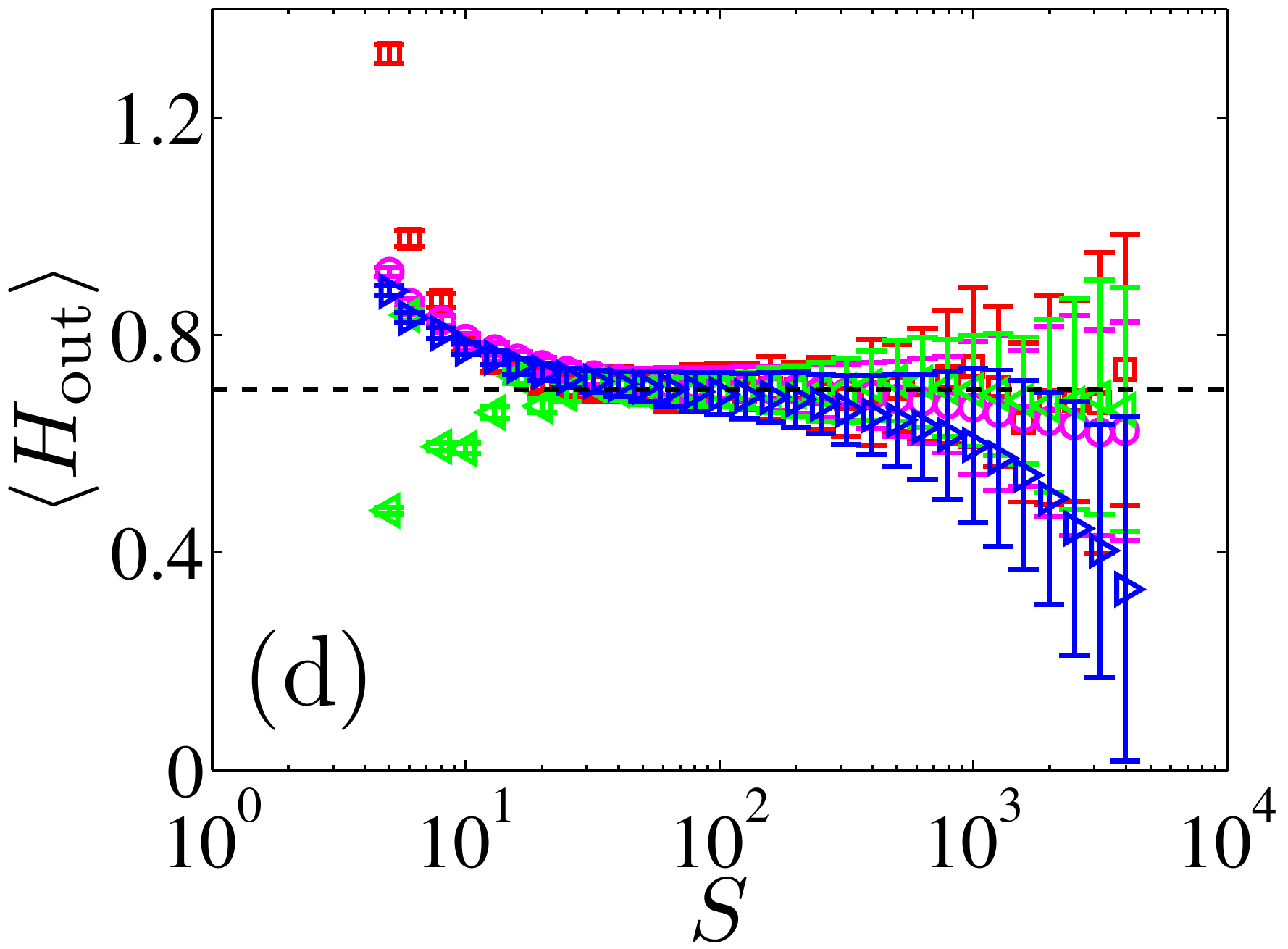}
\includegraphics[width=3.3cm]{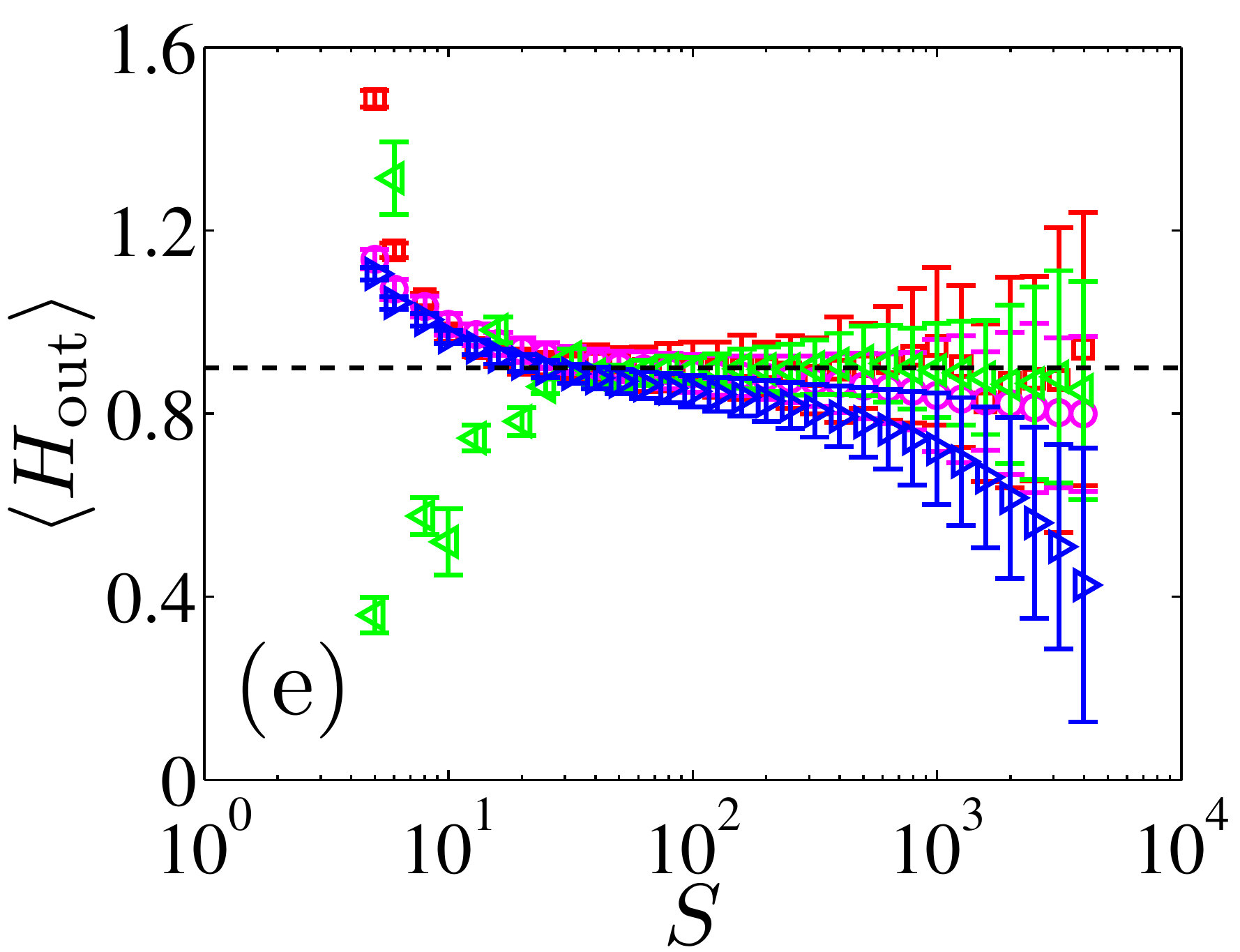}
\includegraphics[width=3.3cm]{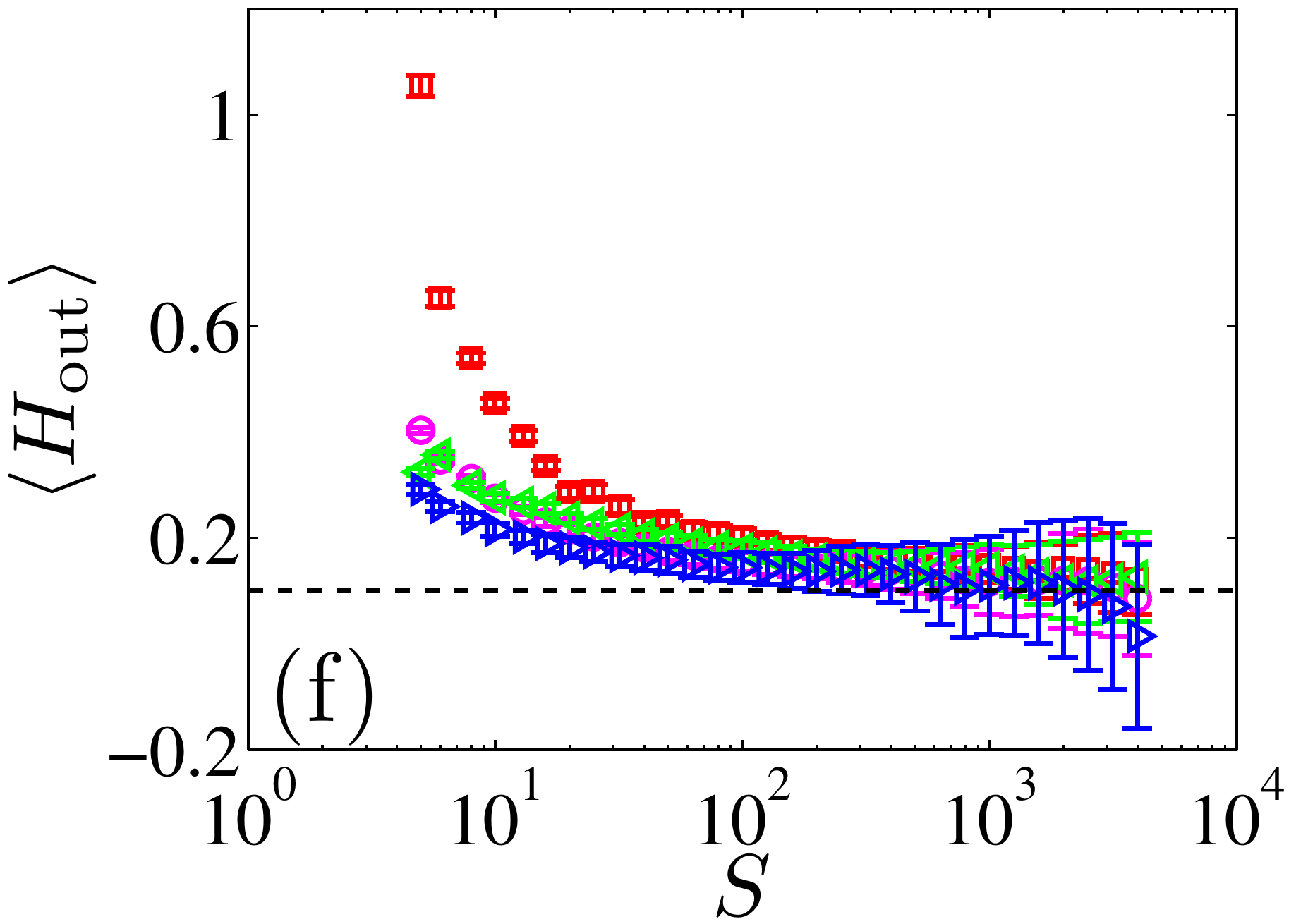}
\includegraphics[width=3.3cm]{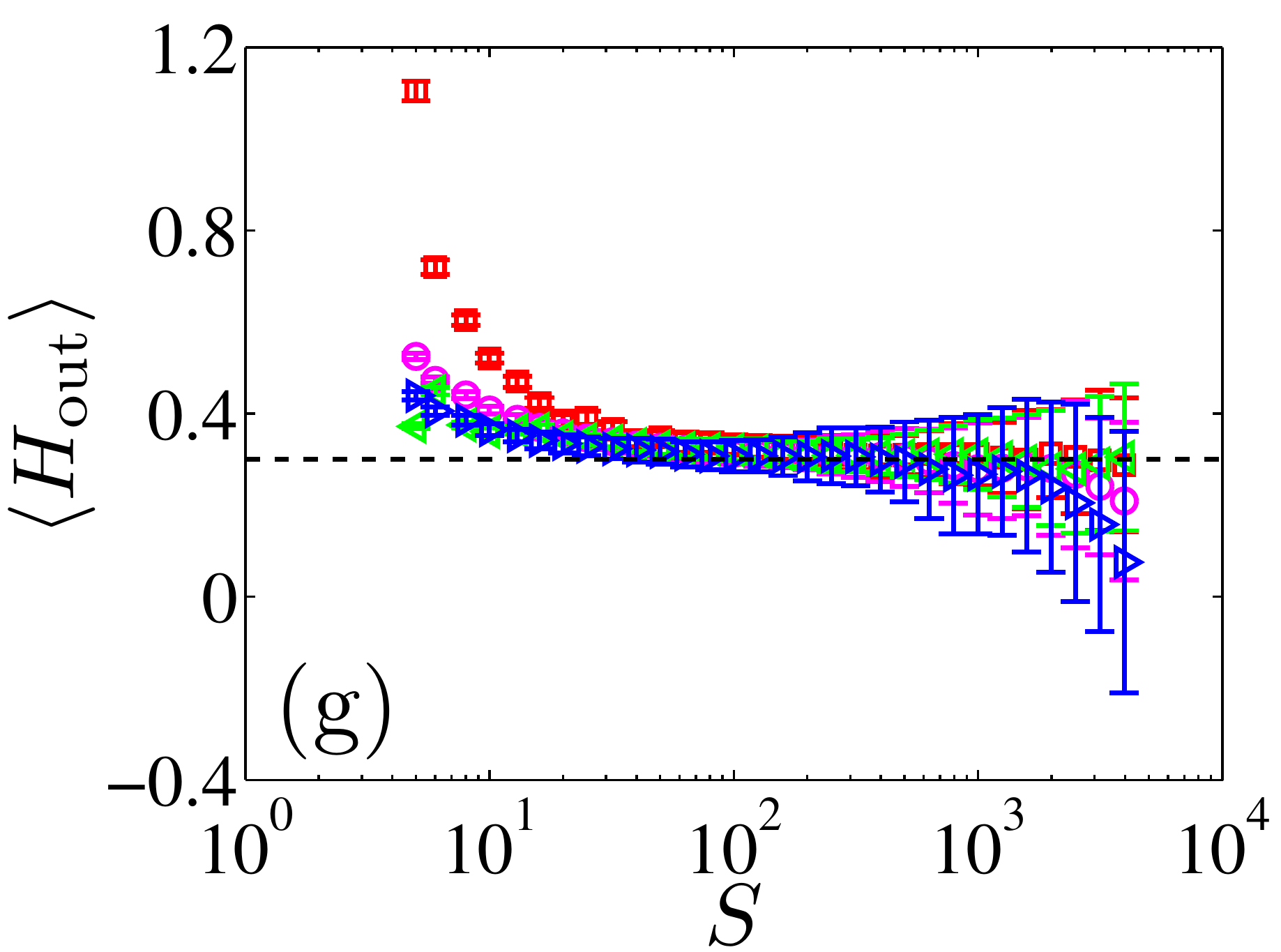}
\includegraphics[width=3.3cm]{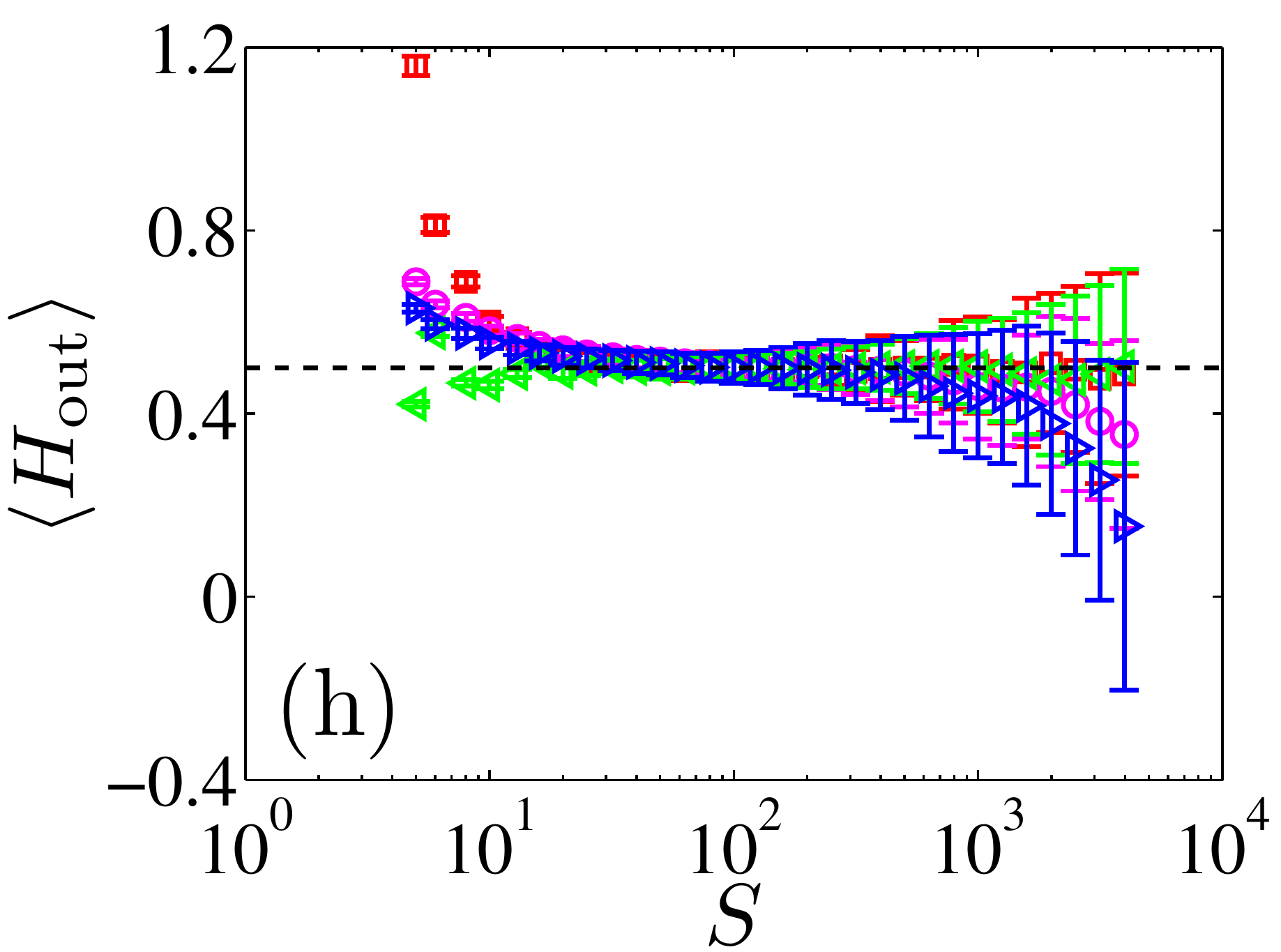}
\includegraphics[width=3.3cm]{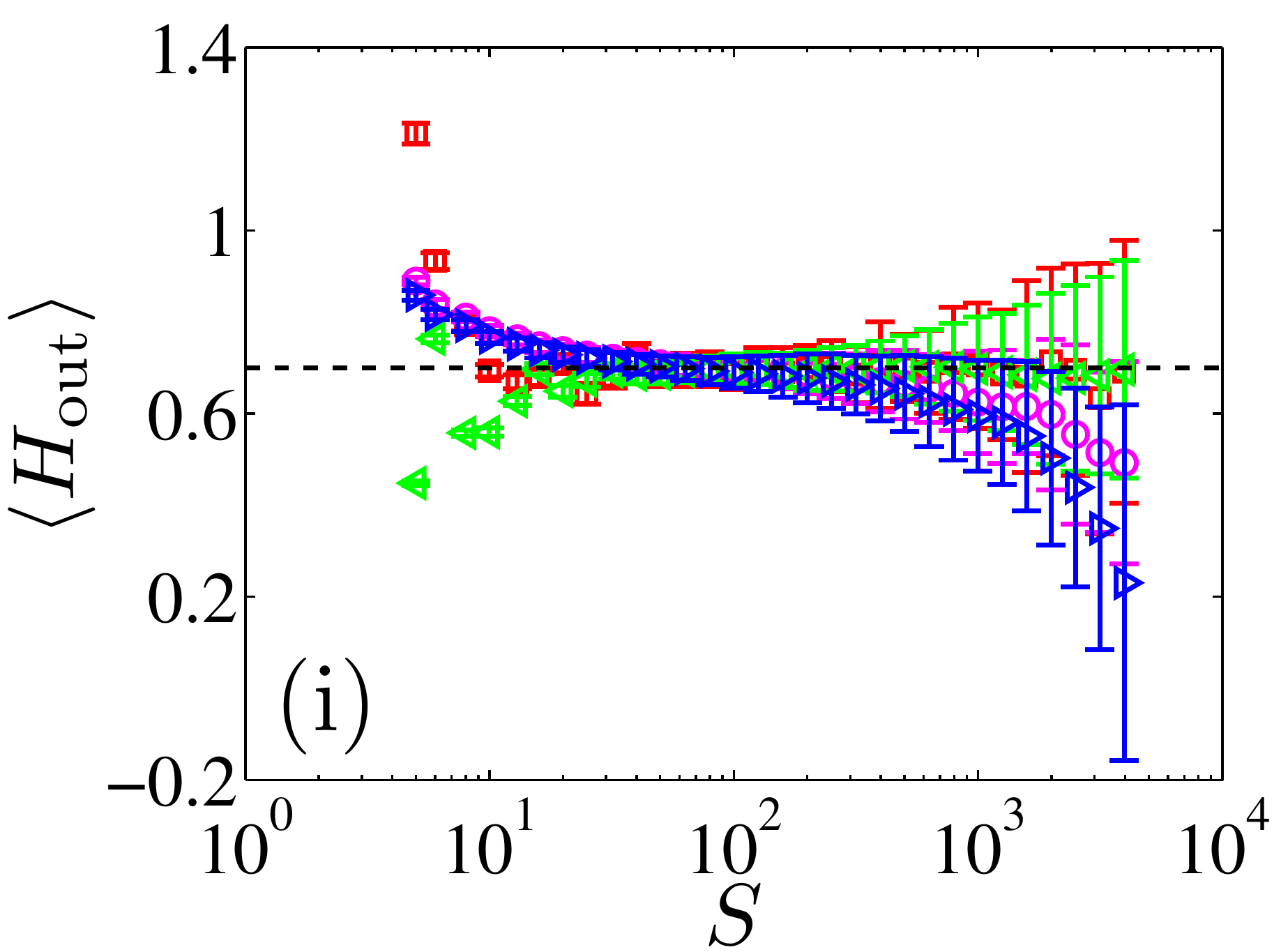}
\includegraphics[width=3.3cm]{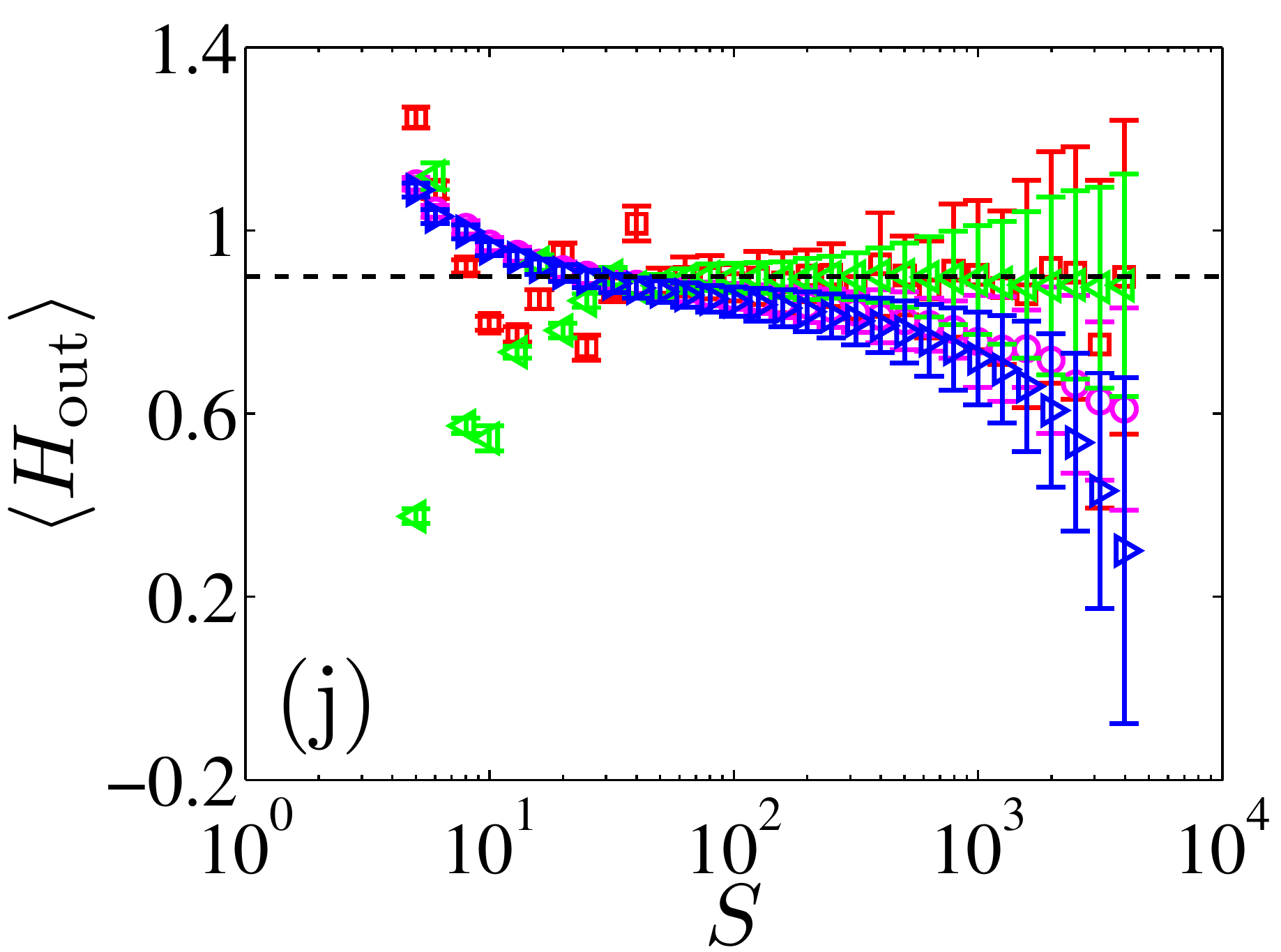}
\includegraphics[width=3.3cm]{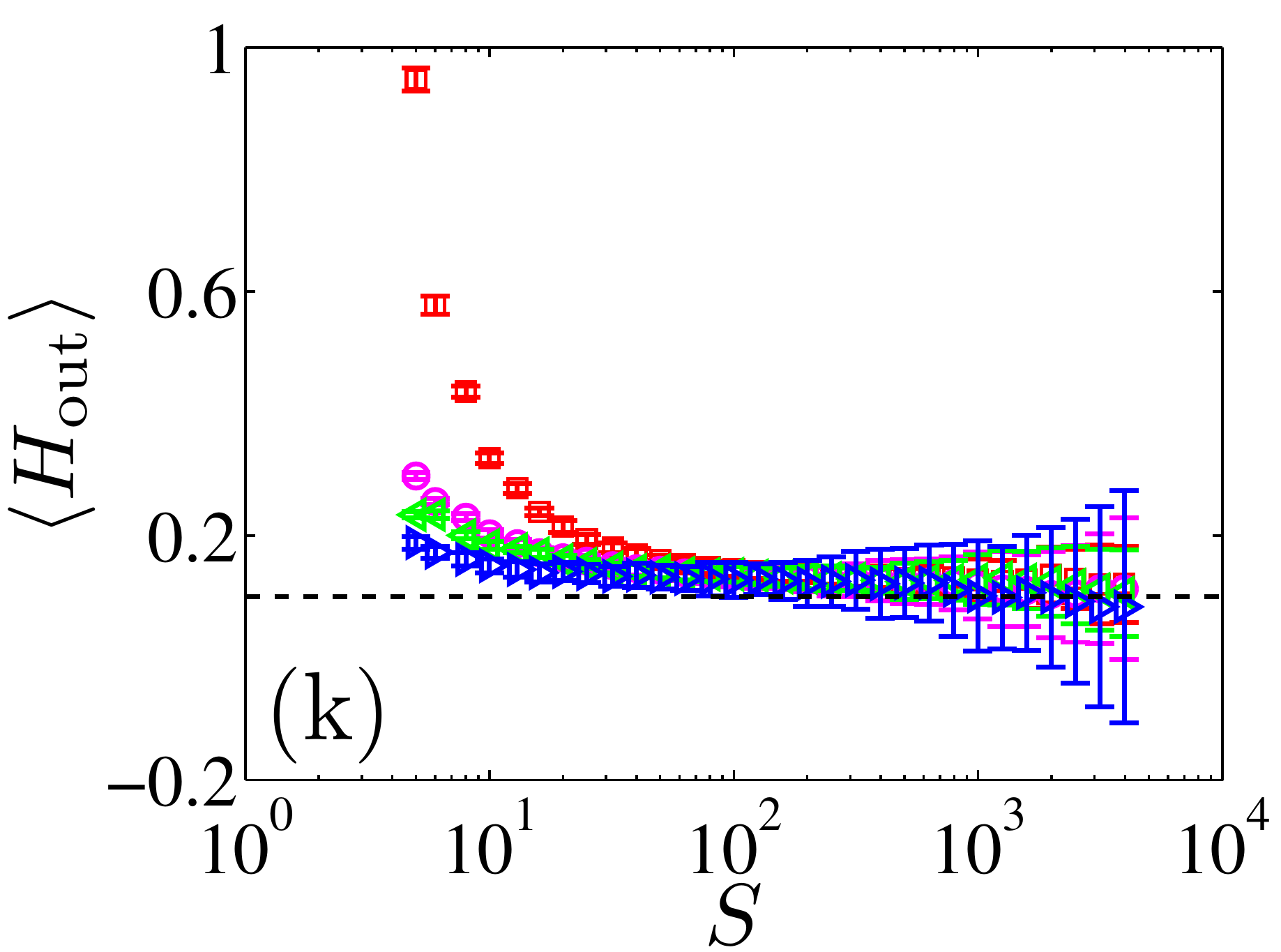}
\includegraphics[width=3.3cm]{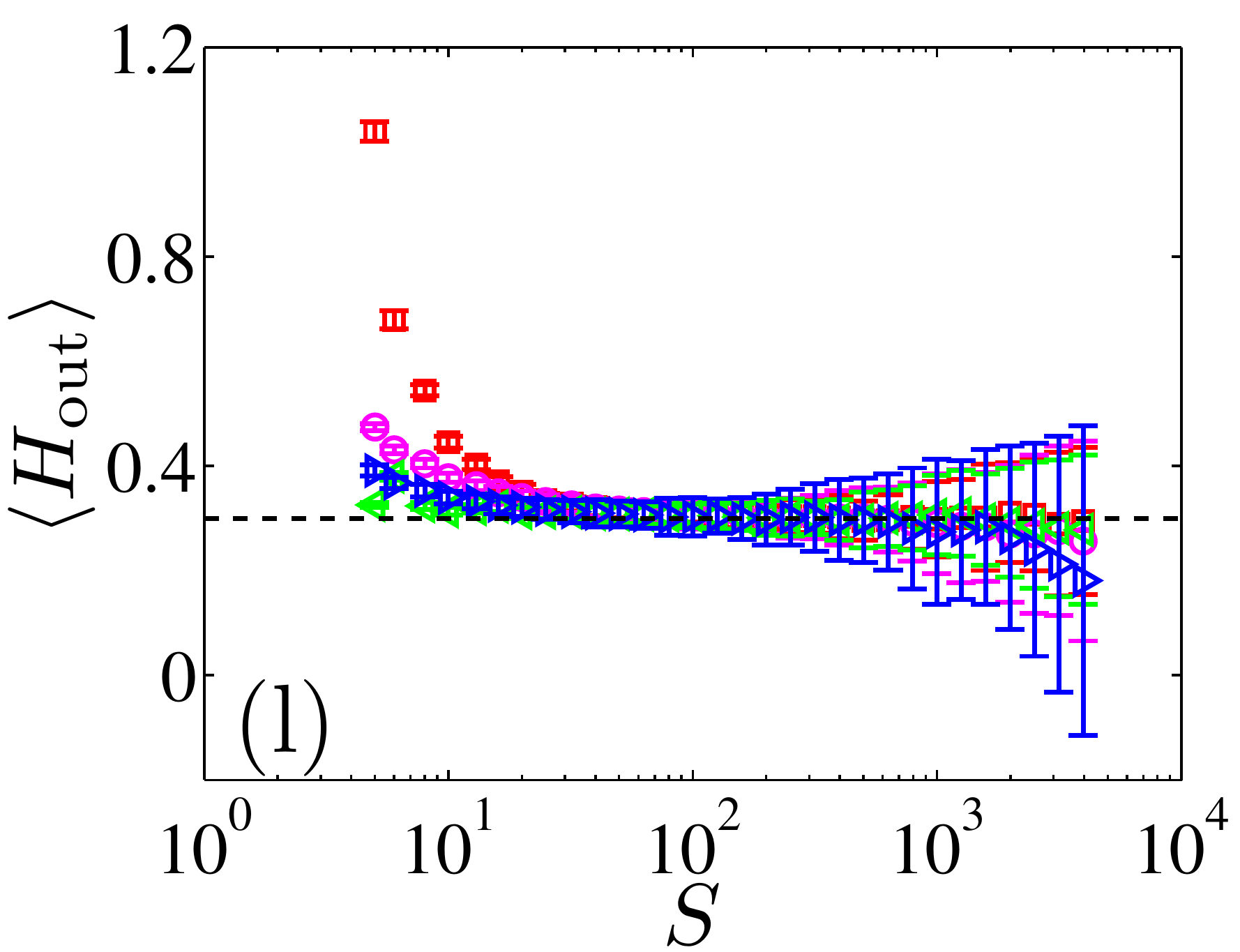}
\includegraphics[width=3.3cm]{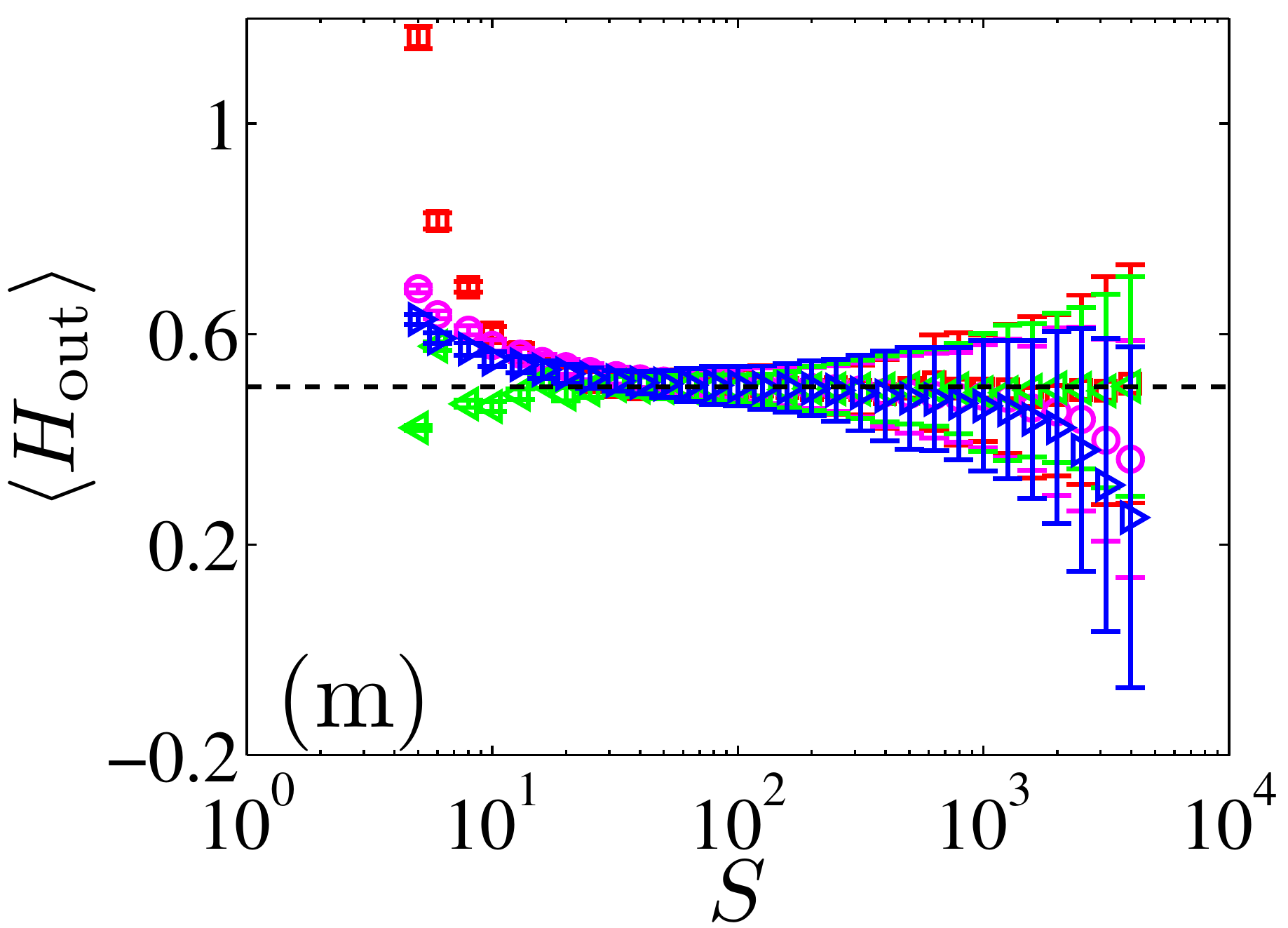}
\includegraphics[width=3.3cm]{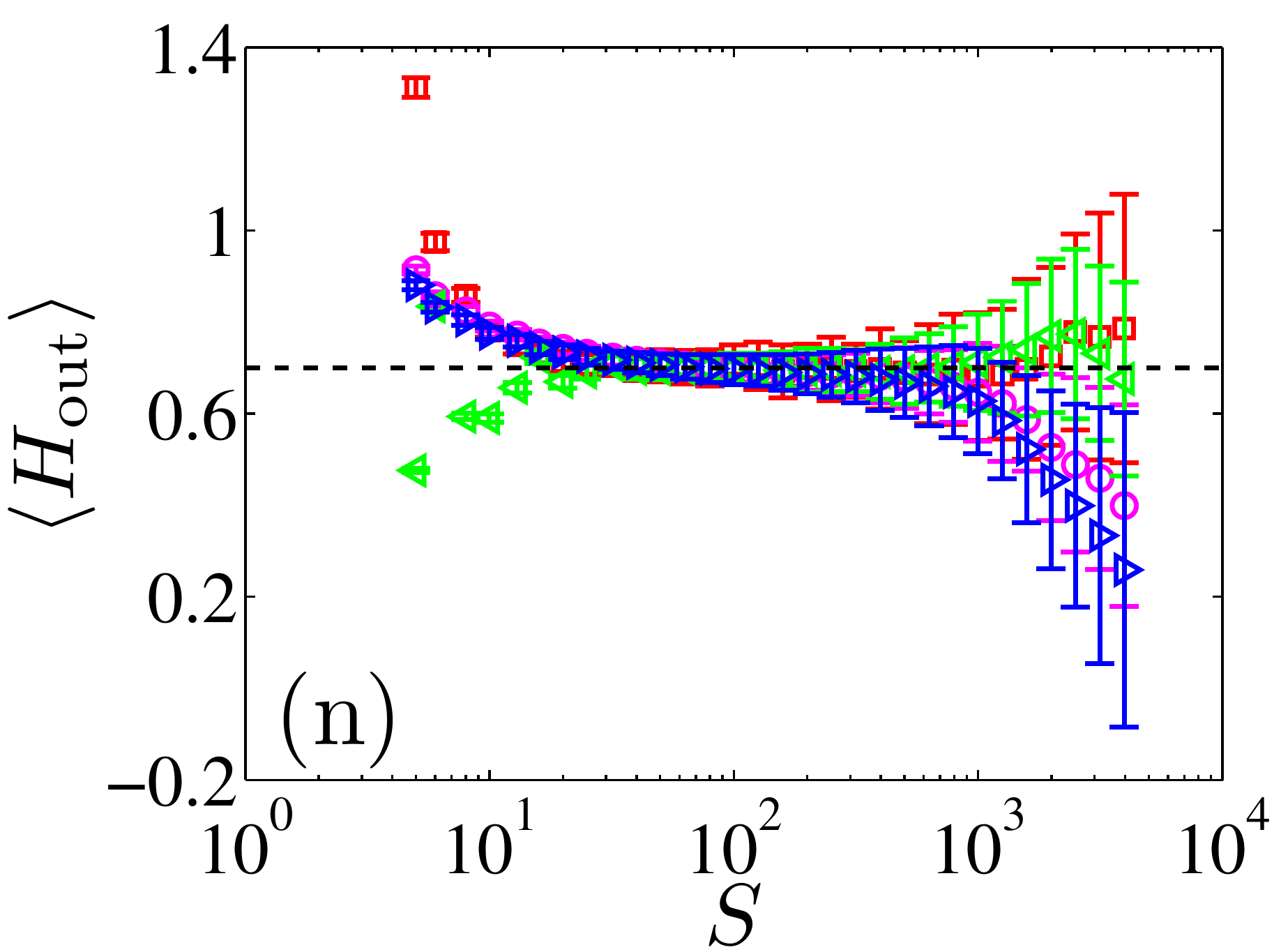}
\includegraphics[width=3.3cm]{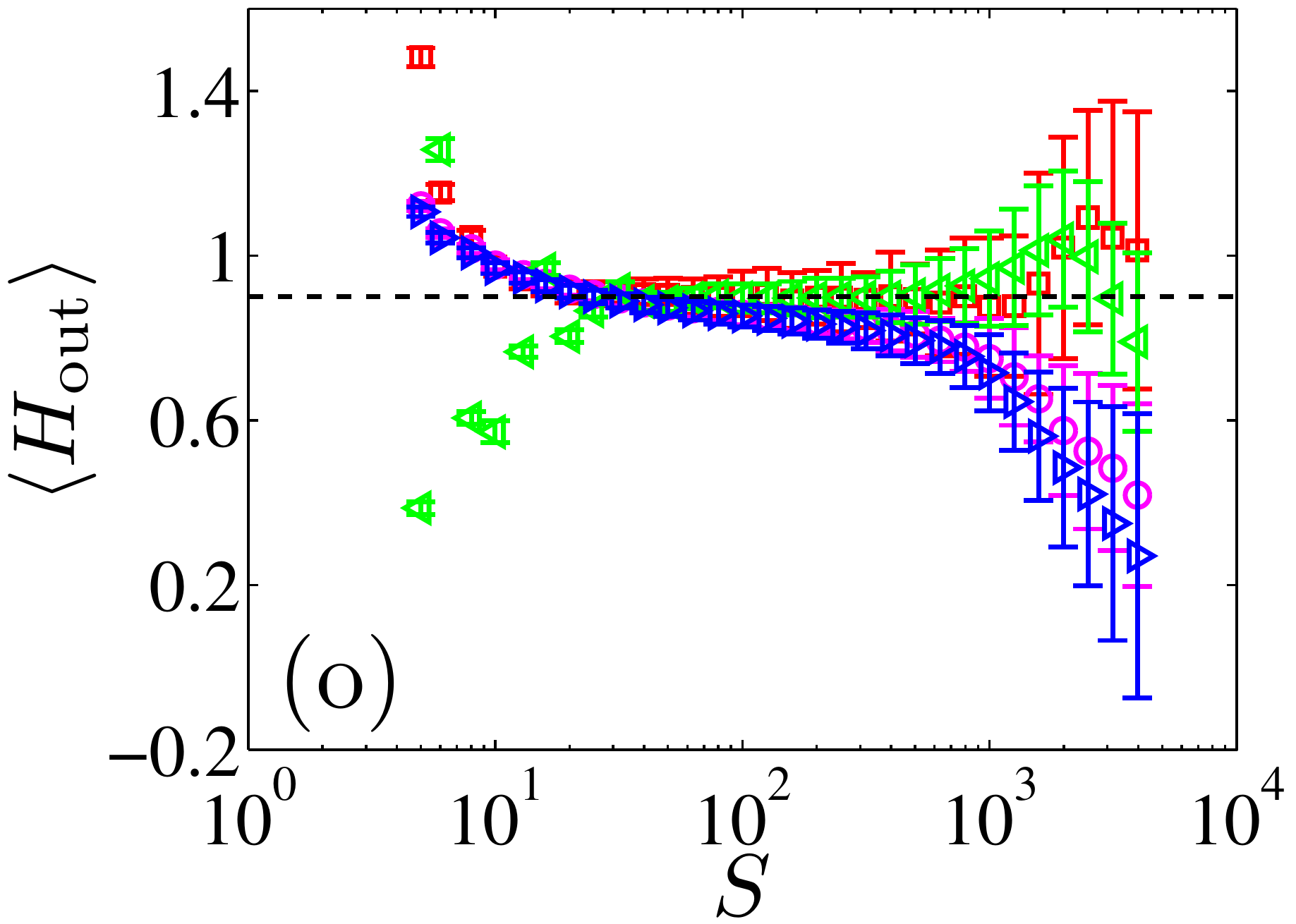}
\caption{{\textbf{Local slopes of the fluctuation functions.}} Each plot contains four curves obtained from four different scaling analysis methods (FA, BDMA, CDMA and DFA) and each curve represents a slope function averaged over 100 repeated simulated time series with the error bars showing the standard deviations. The three rows correspond to three generators (FGN-DH, FBM-RMD and WFBM from top to bottom). Each column corresponds to a fixed Hurst index ($H_{\mathrm{in}}=0.1,0.3,0.5,0.7$, and $0.9$ from left to right). The horizontal dashed lines
indicates the exact value of the Hurst index used to generate the synthetic time series.}
\label{Fig:LocalSlope}
\end{figure*}


\noindent{\textbf{Fluctuation functions.}} Figure \ref{Fig:FluctuationFunction} compares the fluctuation functions calculated with four different scaling analysis methods (FA, BDMA, CDMA, DFA) on time series generated using three different generators (FGN-DH, FBM-RMD and WFBM). We notice that panel (b) confirms the results in Ref.~\cite{Bryce-Sprague-2012-SR}, which compares the performances of FA and DFA on FGNs with $H_{\mathrm{in}}$.
One can also notice that the error bar increases with $s$ for each curve.

When the scale $s$ is small and the Hurst index $H_{\mathrm{in}}$ is small, the curvature of the fluctuation function for DFA is remarkable, while the FA curve looks quite straight. In addition, the DMA curves also exhibit some mild curvature. With the increase of the Hurst index $H_{\mathrm{in}}$ of the analysed time series, the curvature of the DFA and DMA curves decreases. We thus confirm that FA performs best in most cases and DFA performs worst at small scales.

However, the conclusions are very different at large scales. The DFA curves have the smallest error bars, the centred DMA curves show the second smallest error bars, and the FA curves exhibit the largest error bars. More significantly, the DFA and CDMA curves are very straight, while the FA and BDMA curves exhibit some clear curvature with the magnitude of the curvature becomes larger with the increase of the Hurst index $H_{\mathrm{in}}$.

These observations are qualitatively the same for different time series generators.


\noindent{\textbf{Local slopes.}} Figure \ref{Fig:LocalSlope} compares the local slopes,
which are the estimates of the Hurst exponent,  calculated with four different scaling analysis methods on the time series generated using three different generators. Comparing the three plots of each column, it is found that the relative performances are qualitatively the same for the three time series generators. For each scaling analysis method, the error bars become larger with the increase of the scale for each fixed Hurst index $H_{\mathrm{in}}$ or with the increase of the Hurst index $H_{\mathrm{in}}$ at fixed scale. Again, the error bars of the DFA curve are the
largest in each plot.

At large scales, we find that FA is the worst in the sense that the FA curves have the largest error bars and deviate the most from the theoretical line $\langle{H_{\mathrm{out}}}\rangle=H_{\mathrm{in}}$. In contrast, DFA and CDMA have comparable performances and perform best.

At small scales, the order of performance, as measured by the proximity of the
estimates of the scaling exponents to the true Hurst values and by the size of the error bars,
 is FA $\succ$ CDMA $\succ$ BDMA $\succ$ DFA for $H_{\mathrm{in}}=0.1$ in the first column, CDMA $\succ$ FA $\succ$ BDMA $\succ$ DFA for $H_{\mathrm{in}}=0.3$ in the second column, CDMA $\simeq$ FA $\succ$ BDMA $\succ$ DFA for $H_{\mathrm{in}}=0.5$ in the third column, FA $\succ$ BDMA $\succ$ CDMA $\succ$ DFA for $H_{\mathrm{in}}=0.7$ in the fourth column, and FA $\succ$ BDMA $\succ$ CDMA $\simeq$ DFA for $H_{\mathrm{in}}=0.9$ in the fifth column, where A $\succ$ B means that A is superior to B.


\noindent{\textbf{Effect of scaling range.}} In order to
perform the scaling analysis onto real systems using any of the above methods, it is of crucial importance to determine the scaling range. This is because the estimate of the
scaling exponent may vary dramatically if one changes the scaling range.
We now investigate the effect of
the scaling range on the estimation accuracy of the Hurst index
performed with the four scaling analysis methods applied to time series synthesized
by the three different generators.

\begin{figure*}
\centering
\includegraphics[width=5.5cm]{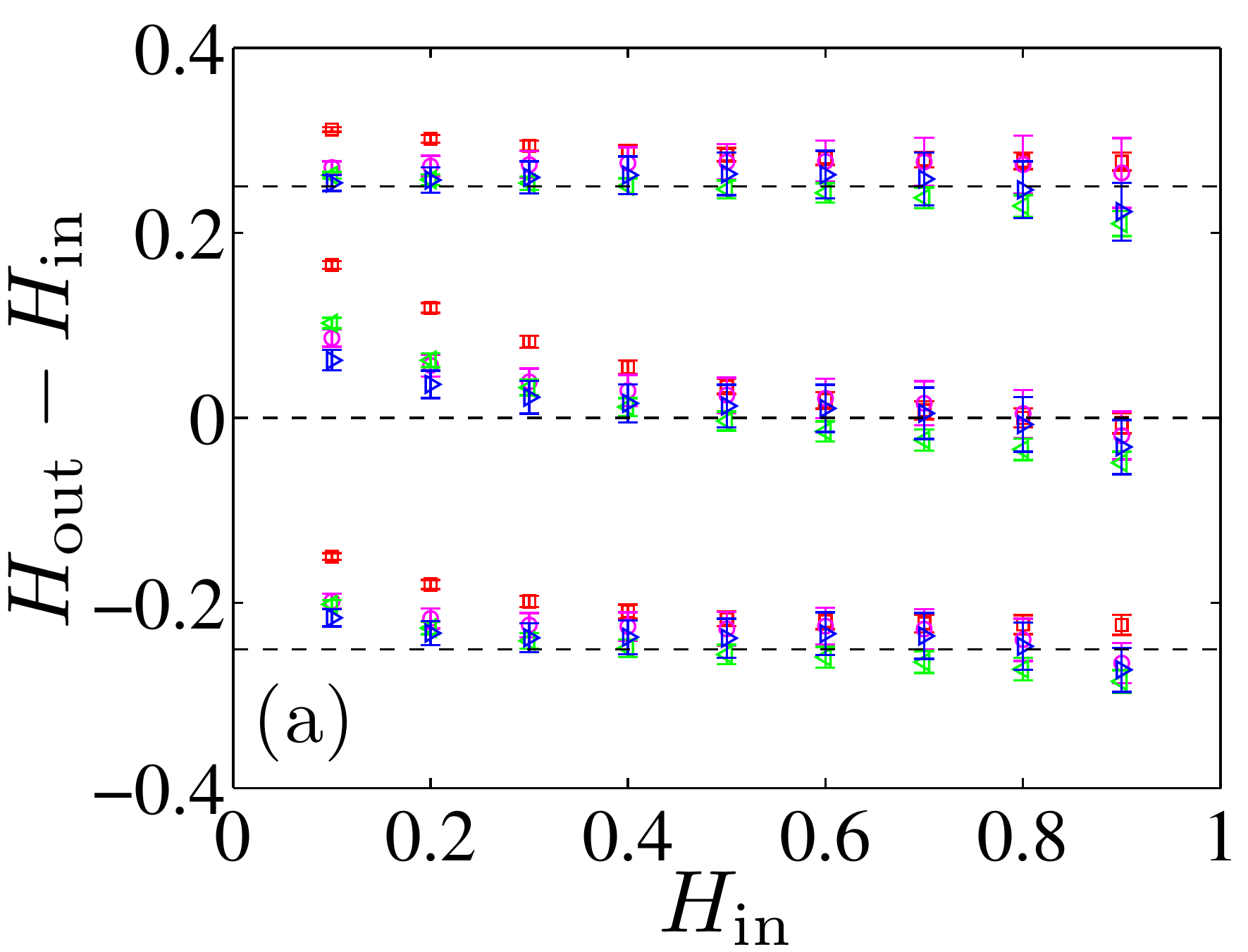}
\includegraphics[width=5.5cm]{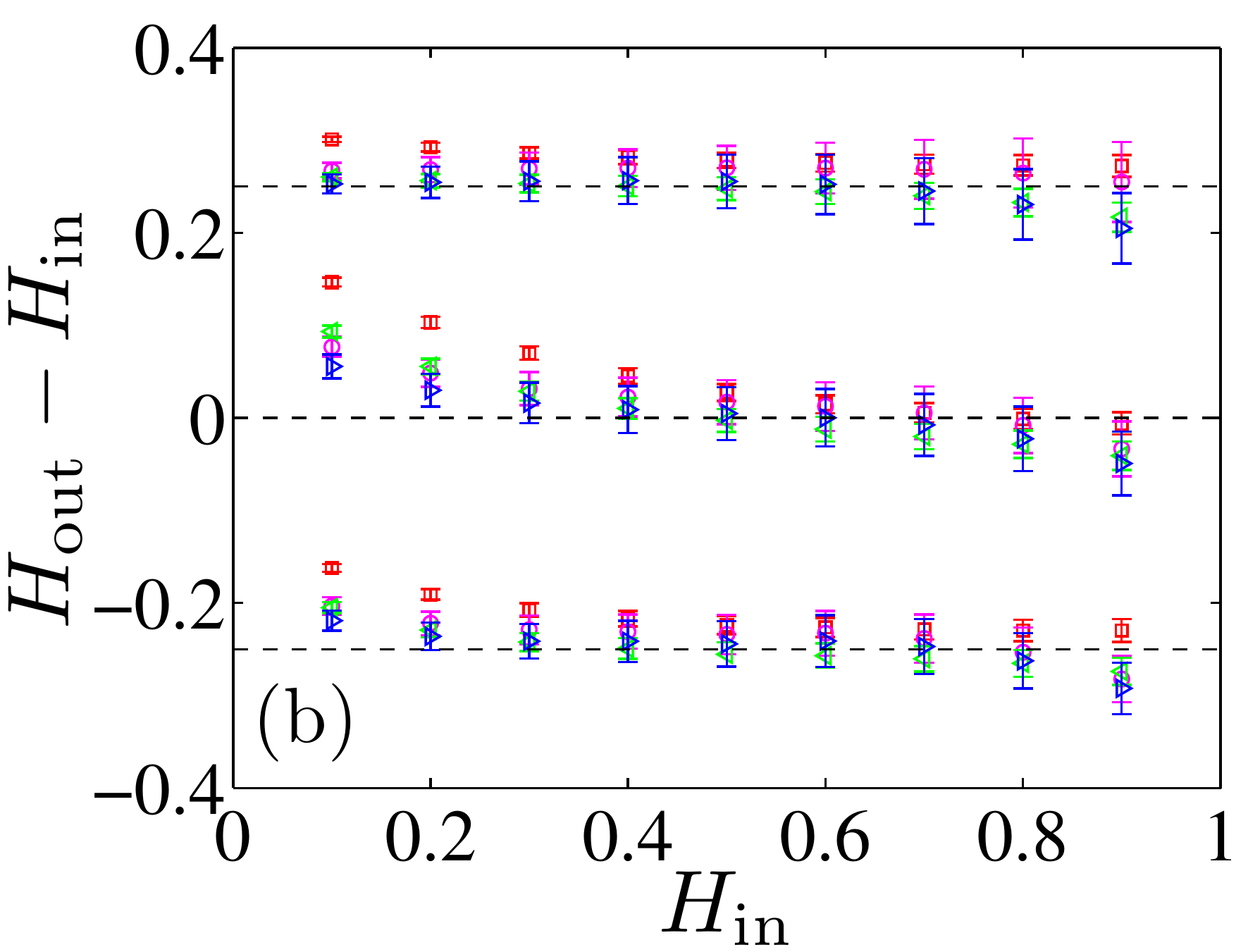}
\includegraphics[width=5.5cm]{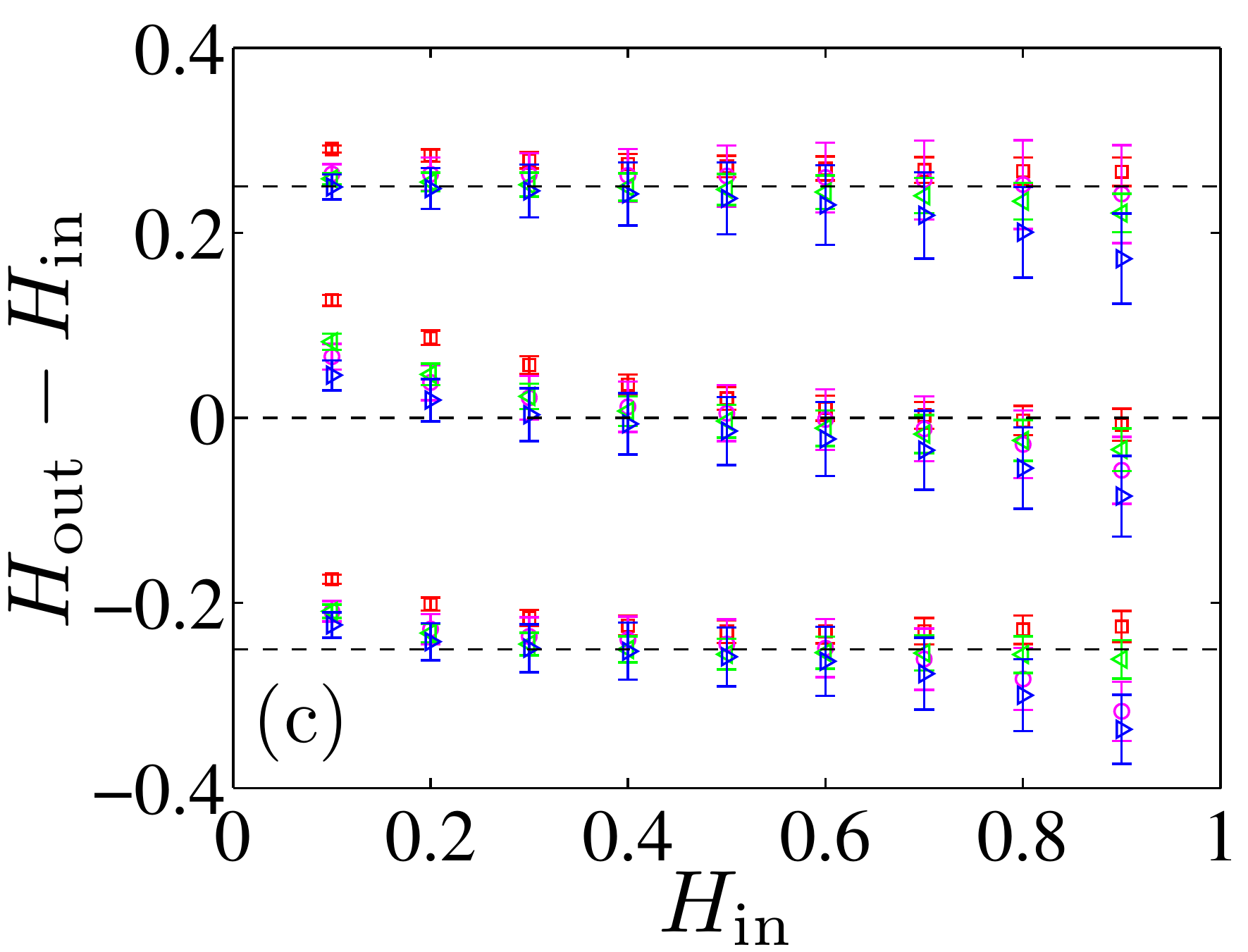}
\includegraphics[width=5.5cm]{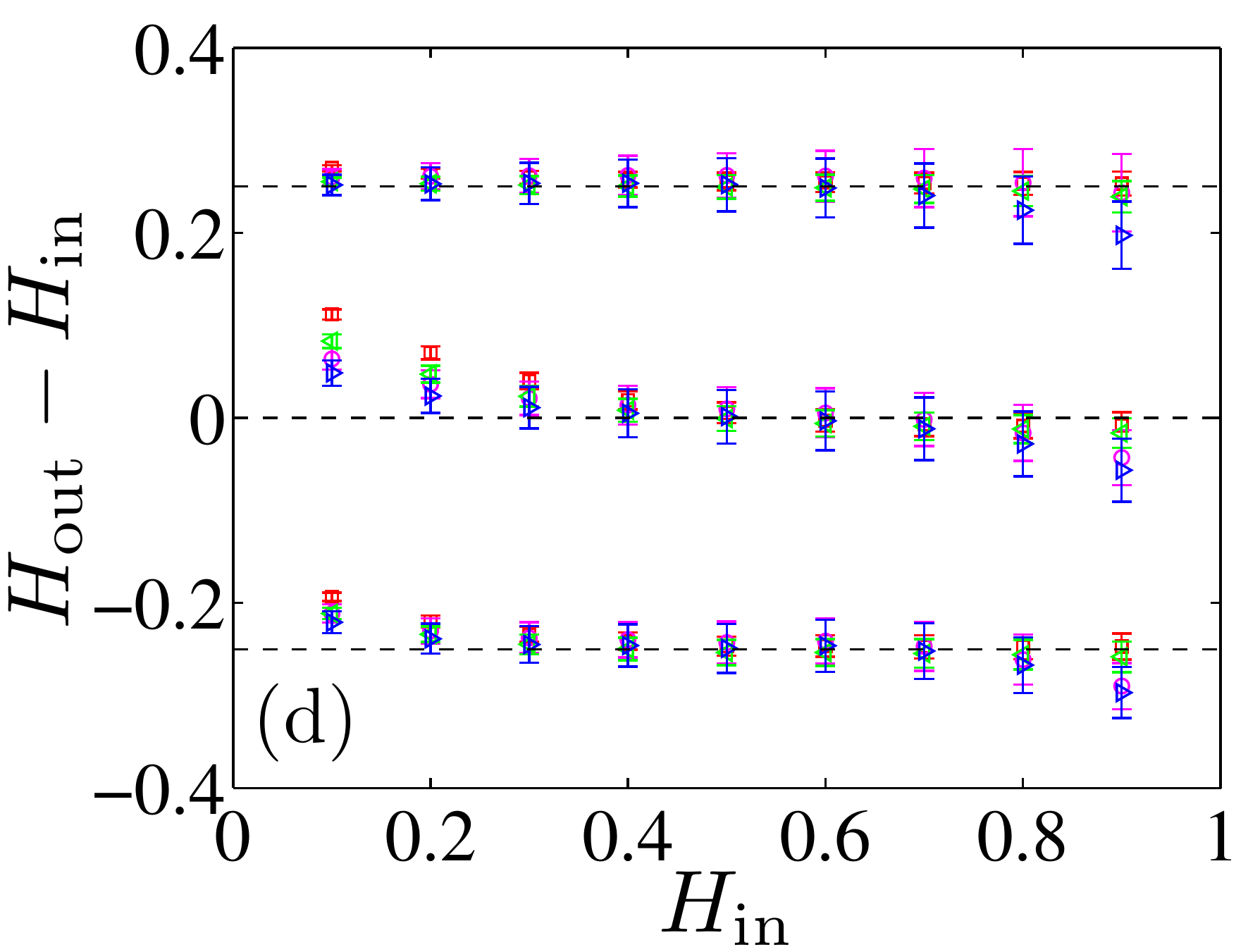}
\includegraphics[width=5.5cm]{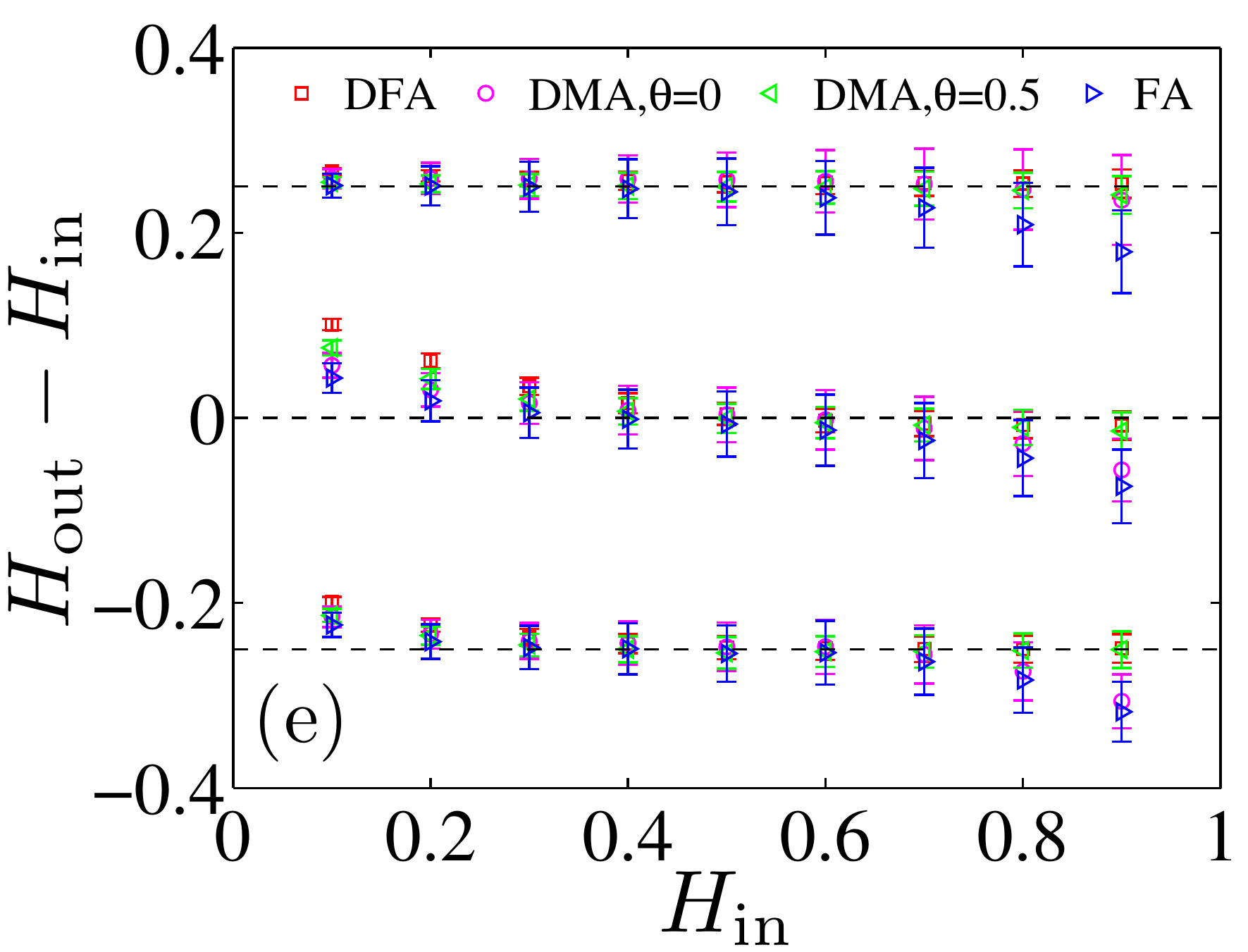}
\includegraphics[width=5.5cm]{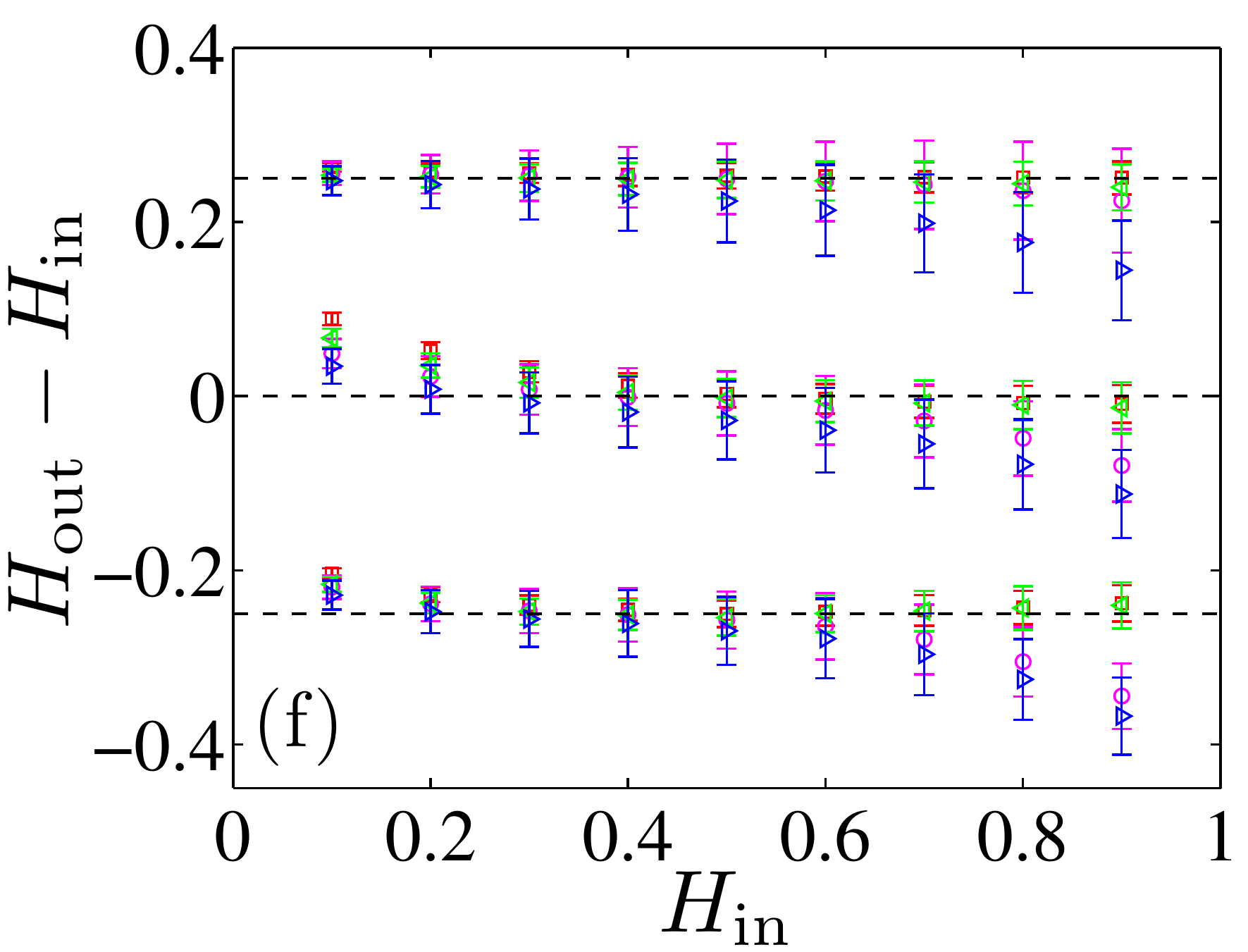}
\includegraphics[width=5.5cm]{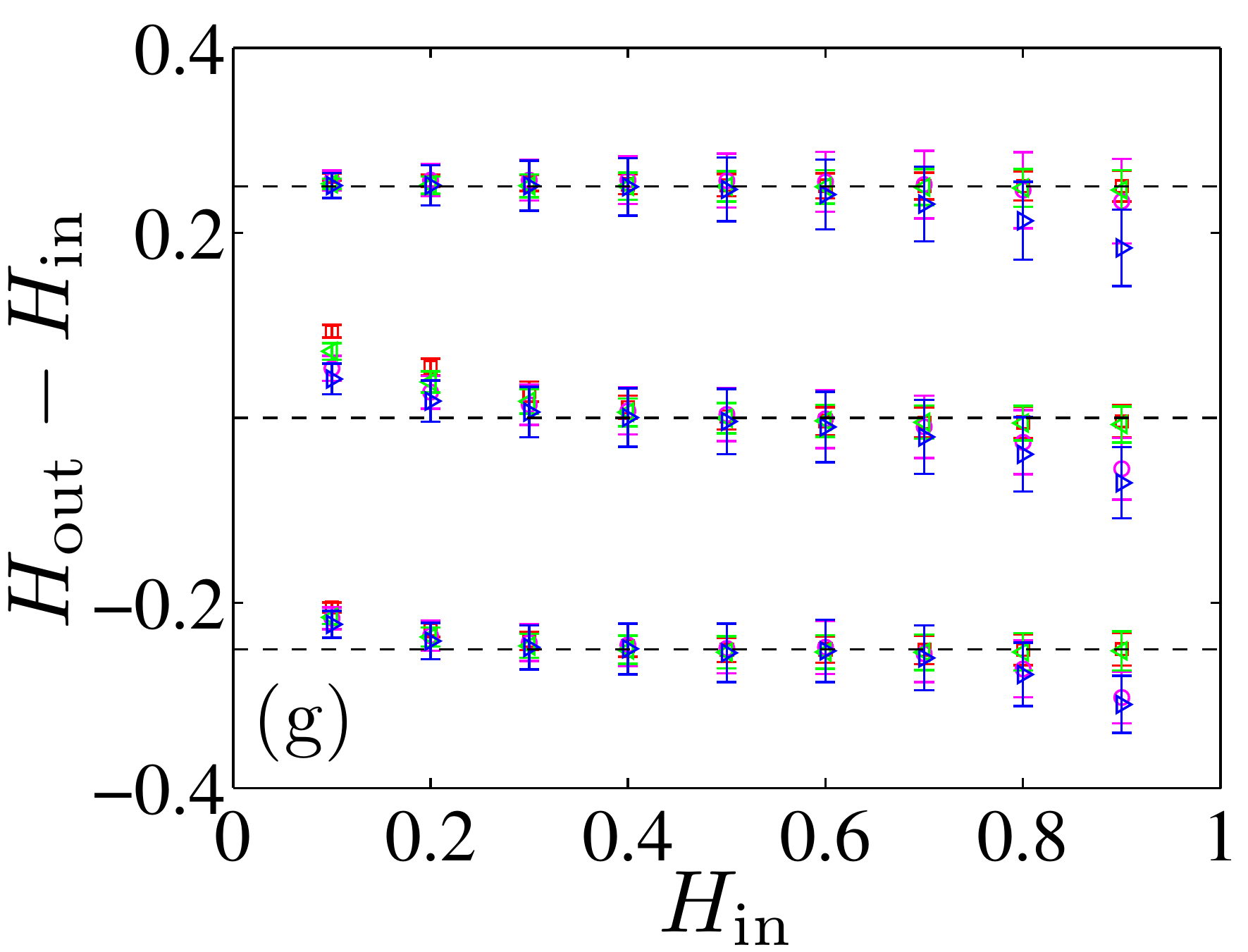}
\includegraphics[width=5.5cm]{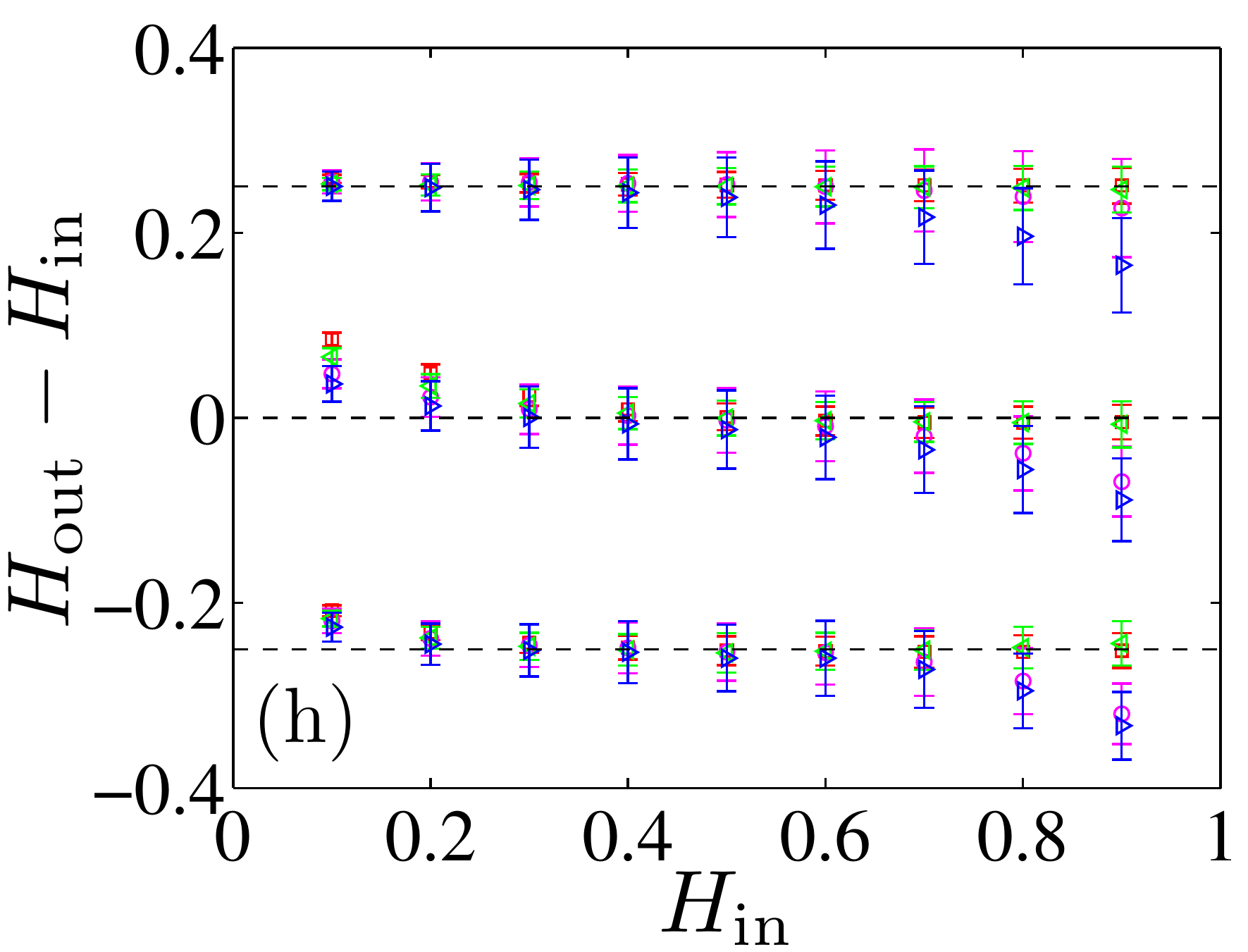}
\includegraphics[width=5.5cm]{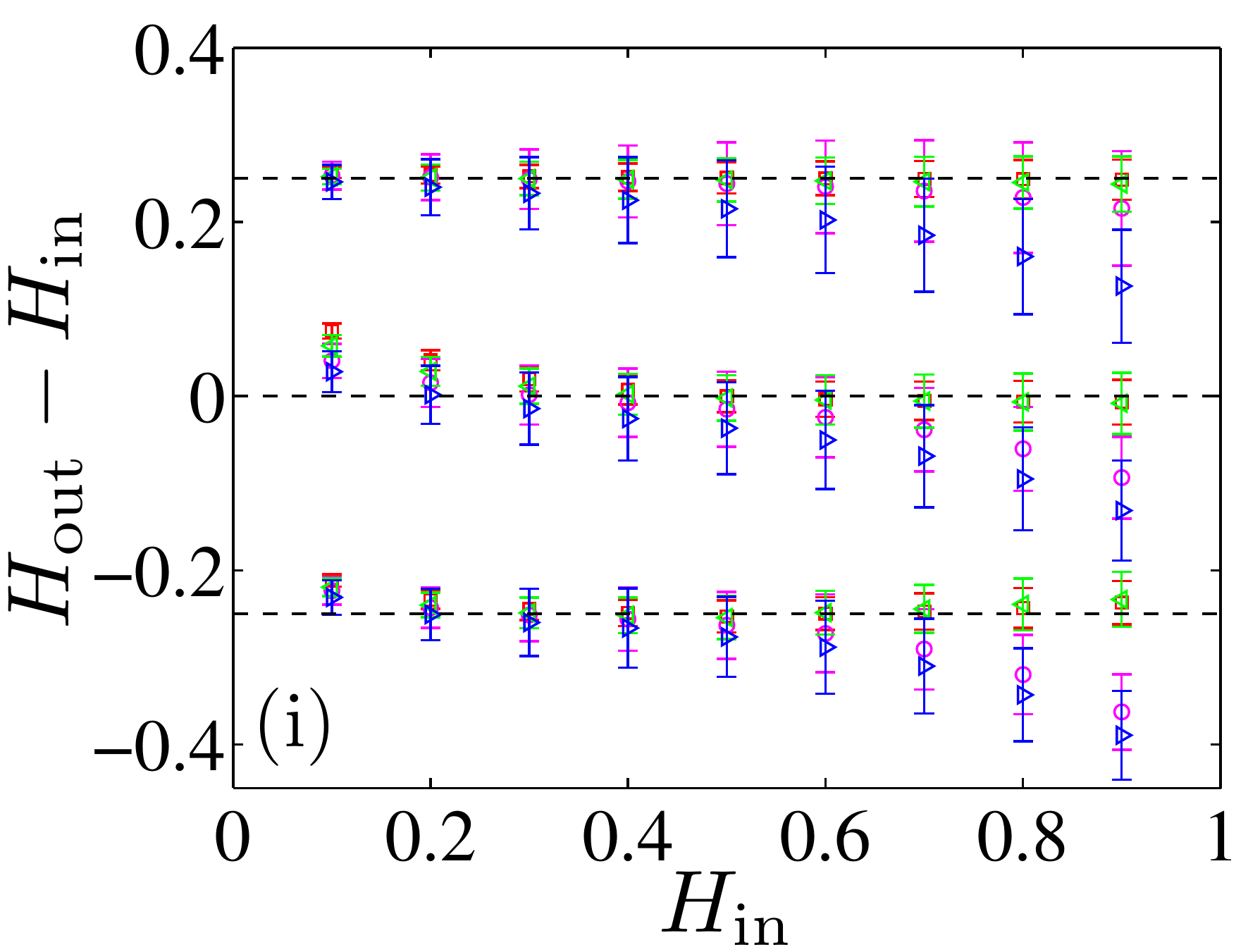}
\caption{{\textbf{Impacts of the scaling range on the Hurst index estimates.}} Each plot has a different scaling range $[s_{\mathrm{left}},s_{\mathrm{right}}$, where $s_{\mathrm{left}}=4,10,20$ from left column to right column and $s_{\mathrm{right}}=999,1992,5000$ from top row to bottom row. In each plot, there are three clusters of curves. Each cluster corresponds to the three generators (FGN-DH, FBM-RMD and WFBM from top to bottom). The top and bottom clusters have been shifted vertically by  +0.25 and -0.25 respectively for better visibility. In each clusters, there are four sets of points with their error bars that are obtained from four different analysis methods (FA, BDMA, CDMA and DFA). Each point shows the average slope of the Hurst index estimates over 100 simulated time series. The error bars show the standard deviations.}
\label{Fig:ScalingRange}
\end{figure*}

Let us first consider the FGNs. We find that the FA gives accurate estimates when $H_{\mathrm{in}}<0.5$,
while the estimated indexes deviate more and more from the theoretical values
when $H_{\mathrm{in}}$ increases in the persistent time series range, for all nine scaling ranges.
The DFA estimates are not accurate only when $s_{\mathrm{right}}=999$ (first row) and  $H_{\mathrm{in}}<0.5$ and DFA outperforms FA for all the other cases. More intriguingly, CDMA gives very accurate estimates of the Hurst indexes and performs the best almost in all situations. Overall, DFA outperforms BDMA and FA is the worst estimator.

For the time series generated with FBM-RMD and WFBM, the relative performances of the four scaling analysis methods are qualitatively the same. When $H_{\mathrm{in}}\ll 0.5$, FA $\succ$ BDMA $\succ$ CDMA $\succ$ DFA. For other situations, DFA and CDMA give very accurate estimates of the Hurst indexes and perform the best, while FA performs the worst.

Taking all these observations together, we conclude that CDMA has the best performance and DFA is slightly worse. When the scaling range is properly determined, DFA and CDMA have similar performances. In contrast, FA has the worst performance, especially in the sense that it cannot provide accurate estimations of the Hurst index for persistent time series.


\vspace{5mm}
\noindent{\Large\textbf{Discussion}}
\vspace{3mm}

We have investigated the performances of four estimators (FA, DFA, BDMA, and CDMA) for the characterization of long-range power-law correlated time series synthesized with three different generators (FGN-DH, FBM-RMD and WFBM). We have illustrated that, overall, CDMA and DFA are the best and exhibit comparable performances, while FA performs the worst. In particular, CDMA and DFA are less sensitive  than FA to the choice of the scaling range. We depart significantly from the conclusion of Ref.~\cite{Bryce-Sprague-2012-SR} that FA is superior to DFA, by showing that this statement holds only for very special cases (FGNs with $H_{\mathrm{in}}=0.3$) that cannot be extended to other situations.

An important issue is the effect of the length of time series on the results and conclusions, especially for short time series. We repeated the analysis by generating time series of length 2000, which corresponds to time windows of 8 years of trading at the daily scale, or less than a week of data sampled at the minute time scale. The analysis comparing the results for windows of 2000 time steps to those for windows of 20000 time steps is
presented in {\color{blue}{Supplementary Information}} and confirms that the conclusions remain unchanged, because the corresponding plots for the two cases with different time series lengths are almost indistinguishable, except that the results for shorter time series have larger fluctuations.


\vspace{5mm}
\noindent{\Large\textbf{Methods}}
\vspace{3mm}

{\bf Description and preprocessing of the data}: For each generator (FGN-DH, FBM-RMD or WFBM), we synthesize 100 time series of length 20000 for a given Hurst index $H_{\mathrm{in}}$. These time series are used in all the analyses. The discrete values of the fluctuation function $F(s)$ of each time series for each scaling analysis method are calculated at 32 $s$-values logarithmically sampled in the interval $[4,5000]$.

{\bf Figure 1 details}: Each point $(\langle{F(s)}\rangle,s)$ shows the average of 100 $F(s)$ values over the 100 time series for each $H_{\mathrm{in}}$ at scale $s$ for a given generator and a given estimator.

{\bf Figure 2 details}: For each time series, we calculate the local slope of $\ln{F(s)}$, which is the centred difference using two adjacent data points. Each point shows the average and the standard deviation estimated over the corresponding 100 local slopes.

{\bf Figure 3 details}: For each time series, we calculate the slope of $\ln{F(s)}$ using the data points within the chosen scaling range. Each point shows the average and the standard deviation over the corresponding 100 slopes.

\bibliography{E:/Papers/Auxiliary/Bibliography}


\vspace{5mm}
\noindent{\Large\textbf{Author contributions}}
\vspace{3mm}

ZQJ, WXZ and DS conceived the study, YHS, WXZ and DS designed the study, and YHS, GFG, ZQJ, WXZ and DS performed the study. WXZ and DS wrote the paper and reviewed the manuscript.

\begin{acknowledgments}
 This work was partially supported by the Natural Science Foundation of China (11075054), the Shanghai (Follow-up) Rising Star Program (11QH1400800), and the Fundamental Research Funds for the Central Universities.
\end{acknowledgments}

\begin{figure*}[t]
\centering
\includegraphics[width=3.3cm]{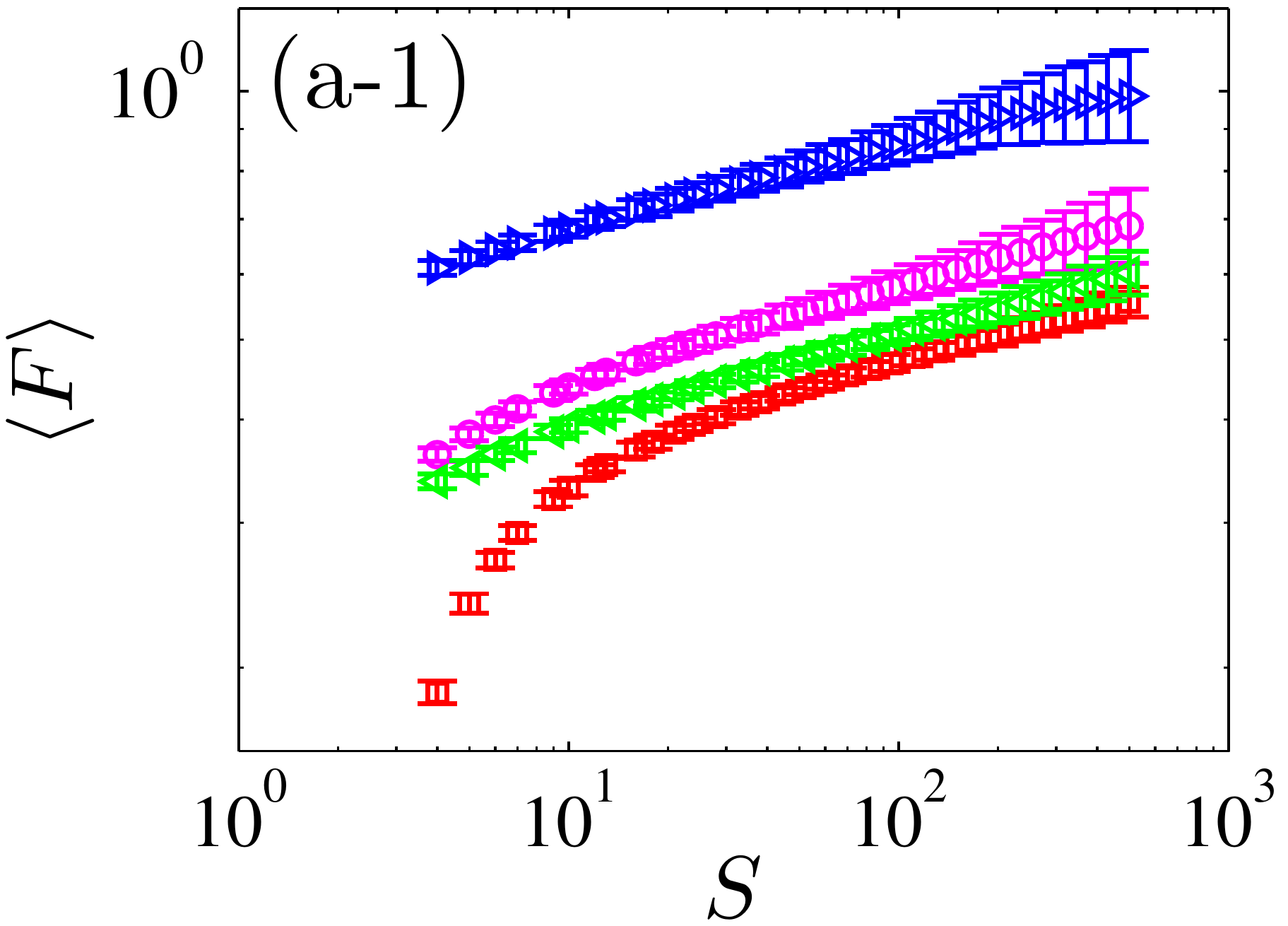}
\includegraphics[width=3.3cm]{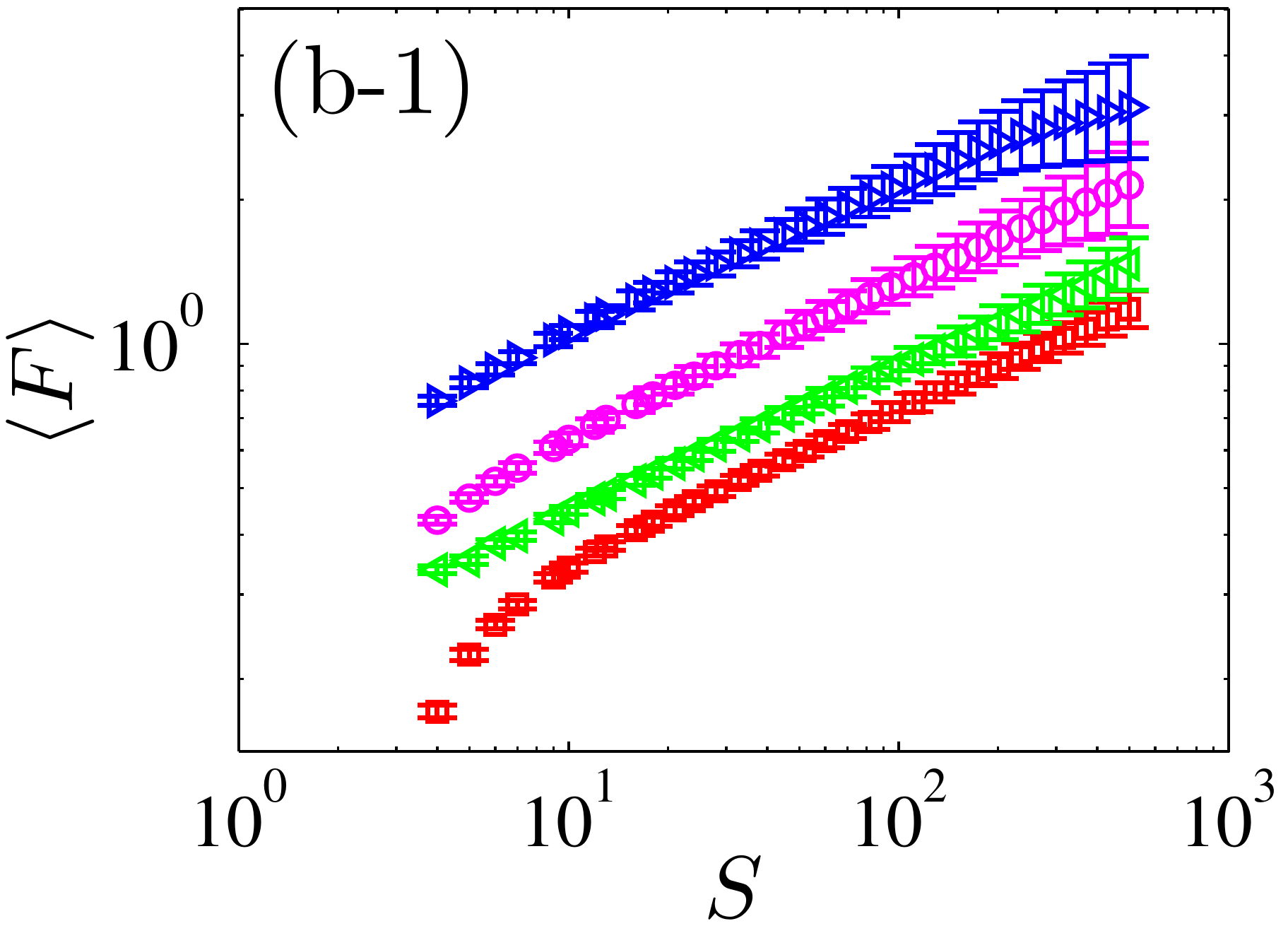}
\includegraphics[width=3.3cm]{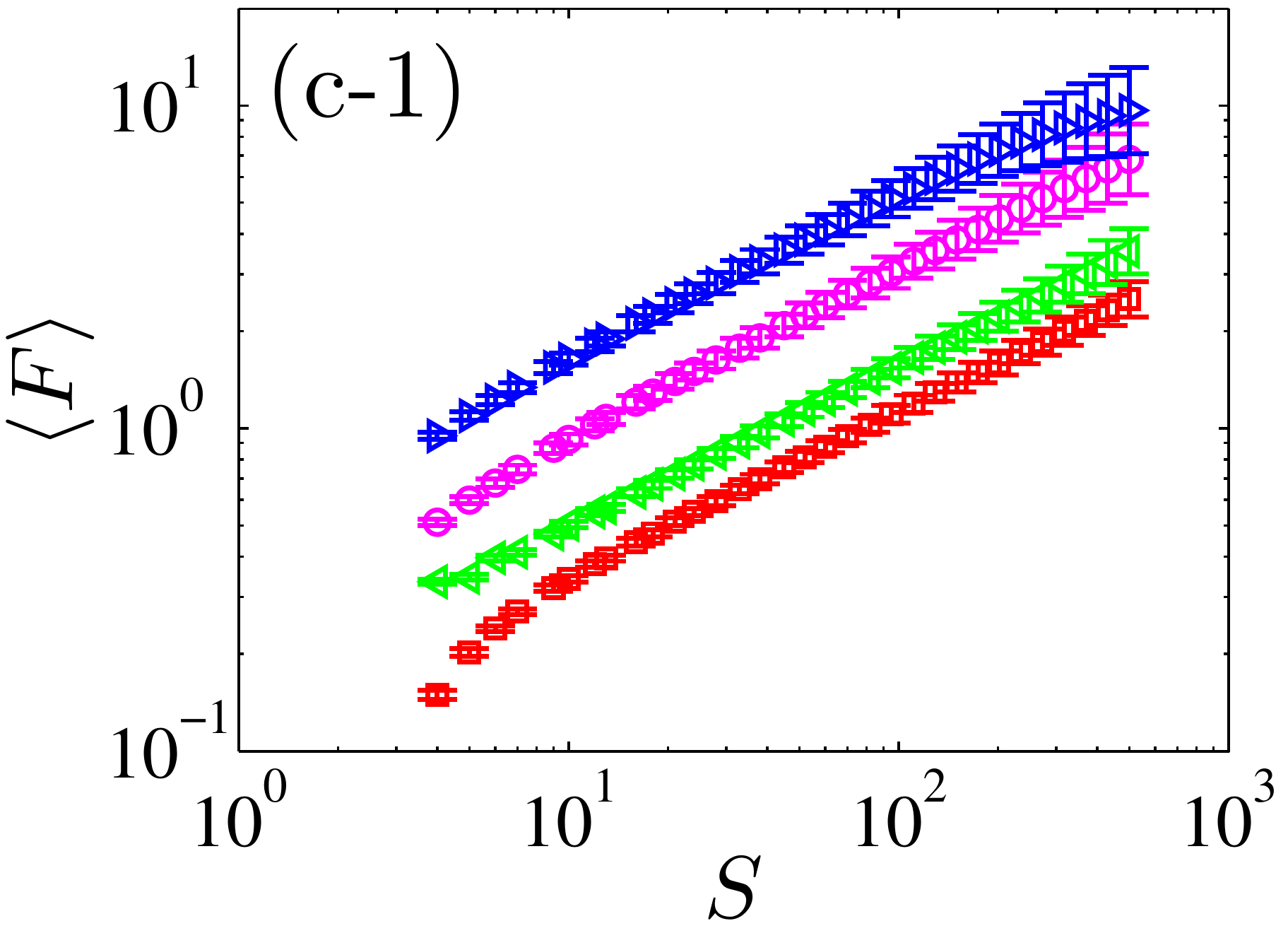}
\includegraphics[width=3.3cm]{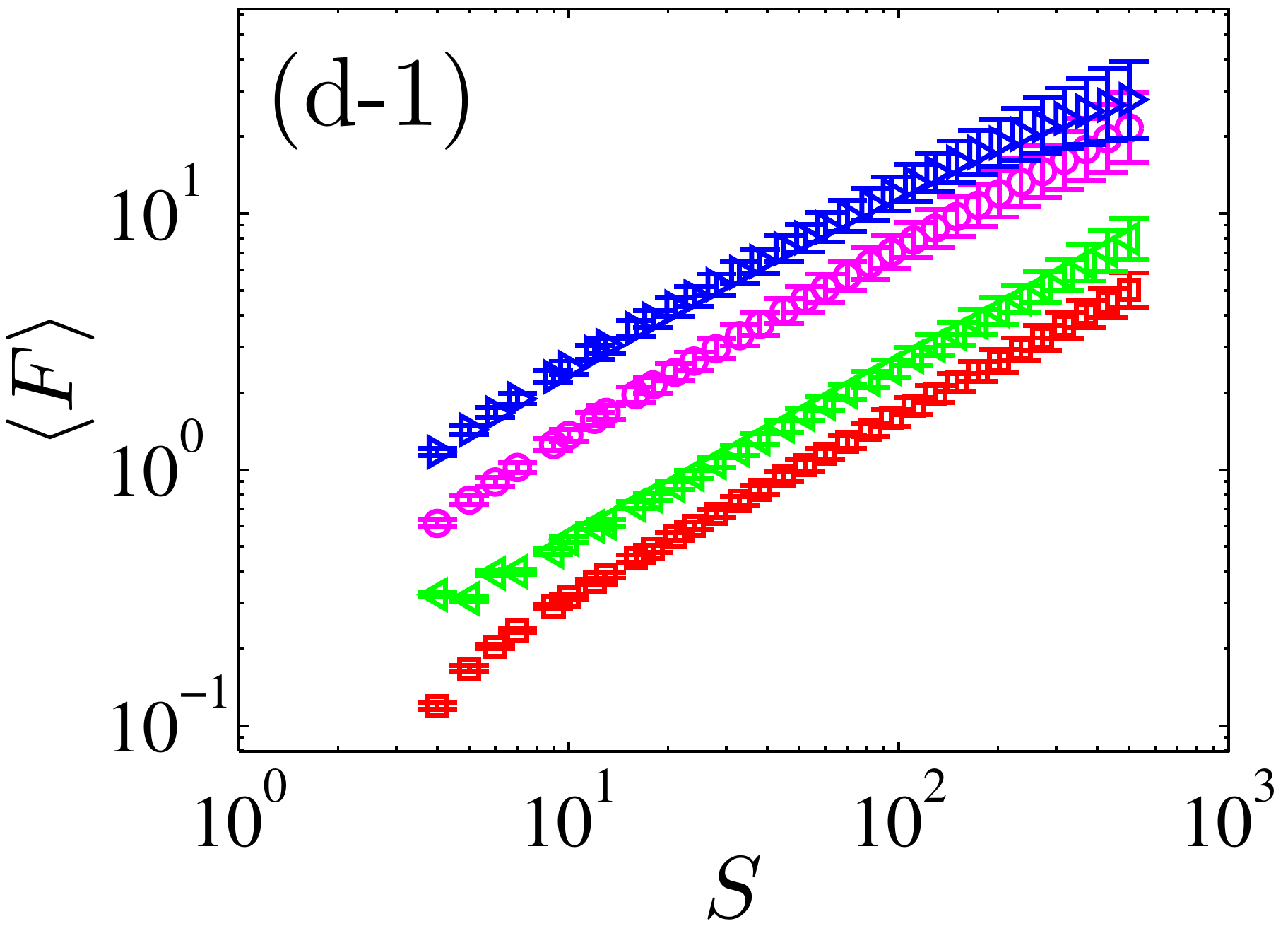}
\includegraphics[width=3.3cm]{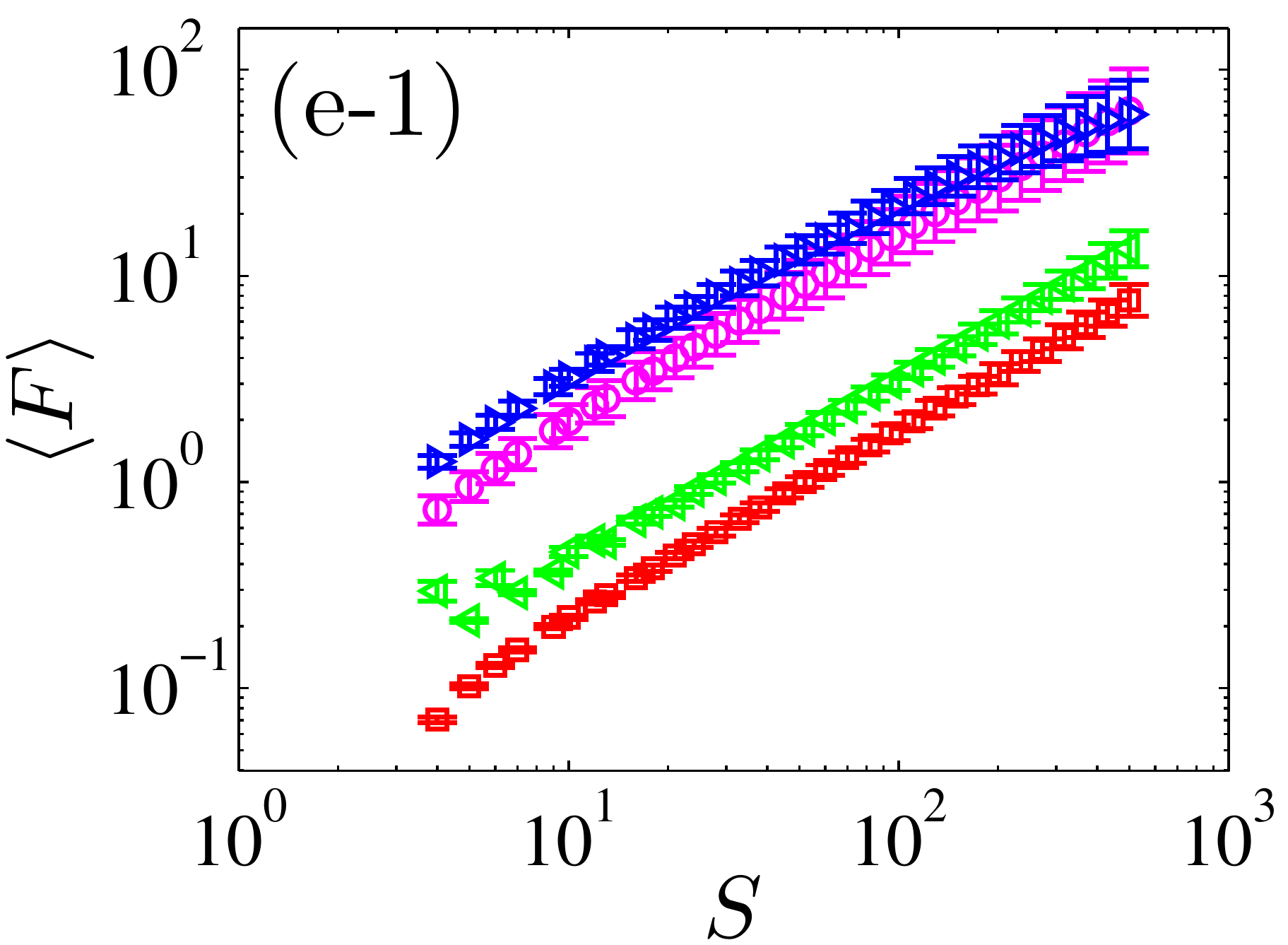}\\
\includegraphics[width=3.3cm]{Fig_fgn_01_FS_errorbar.pdf}
\includegraphics[width=3.3cm]{Fig_fgn_03_FS_errorbar.pdf}
\includegraphics[width=3.3cm]{Fig_fgn_05_FS_errorbar.pdf}
\includegraphics[width=3.3cm]{Fig_fgn_07_FS_errorbar.pdf}
\includegraphics[width=3.3cm]{Fig_fgn_09_FS_errorbar.pdf}\\
\includegraphics[width=3.3cm]{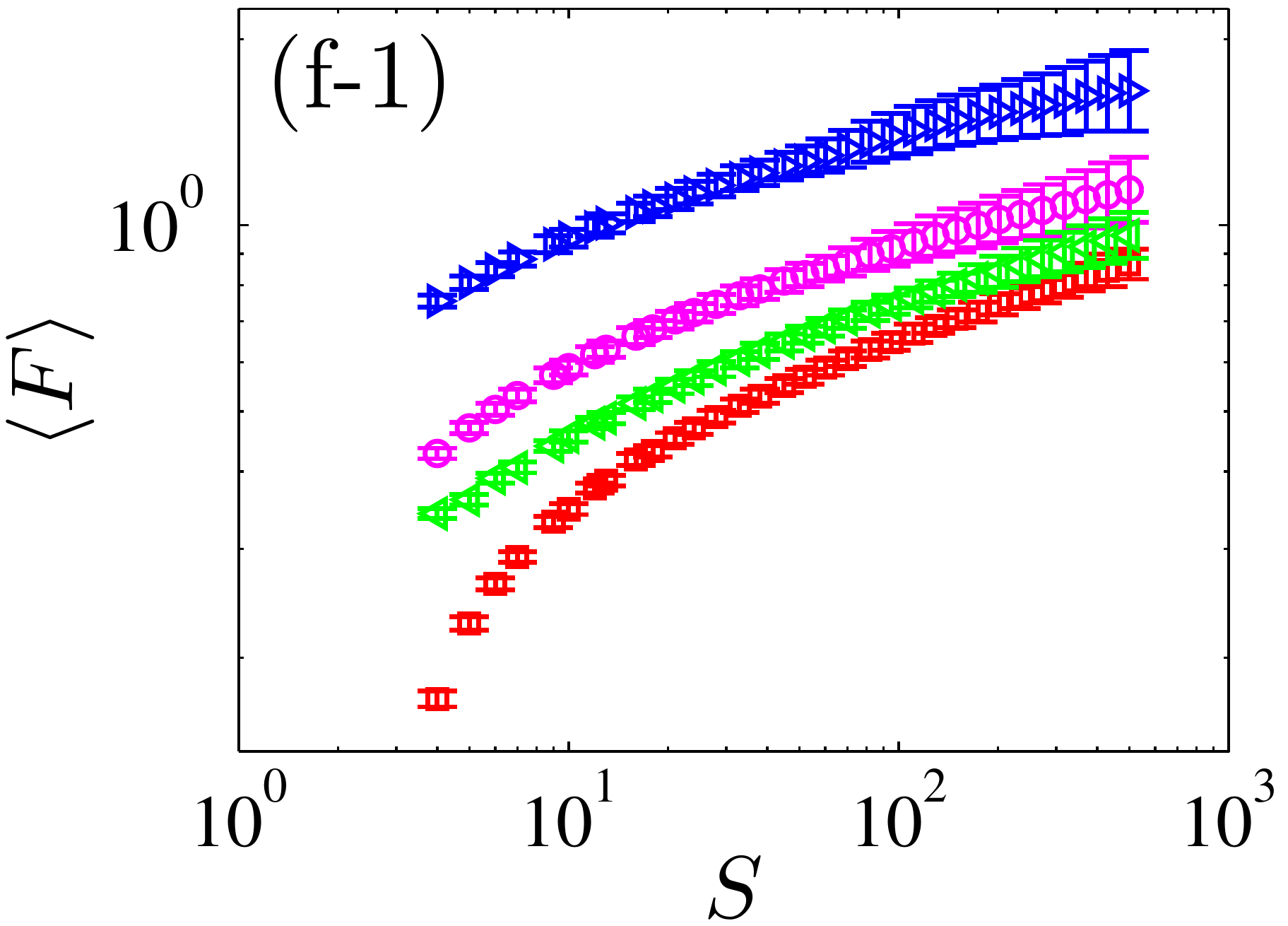}
\includegraphics[width=3.3cm]{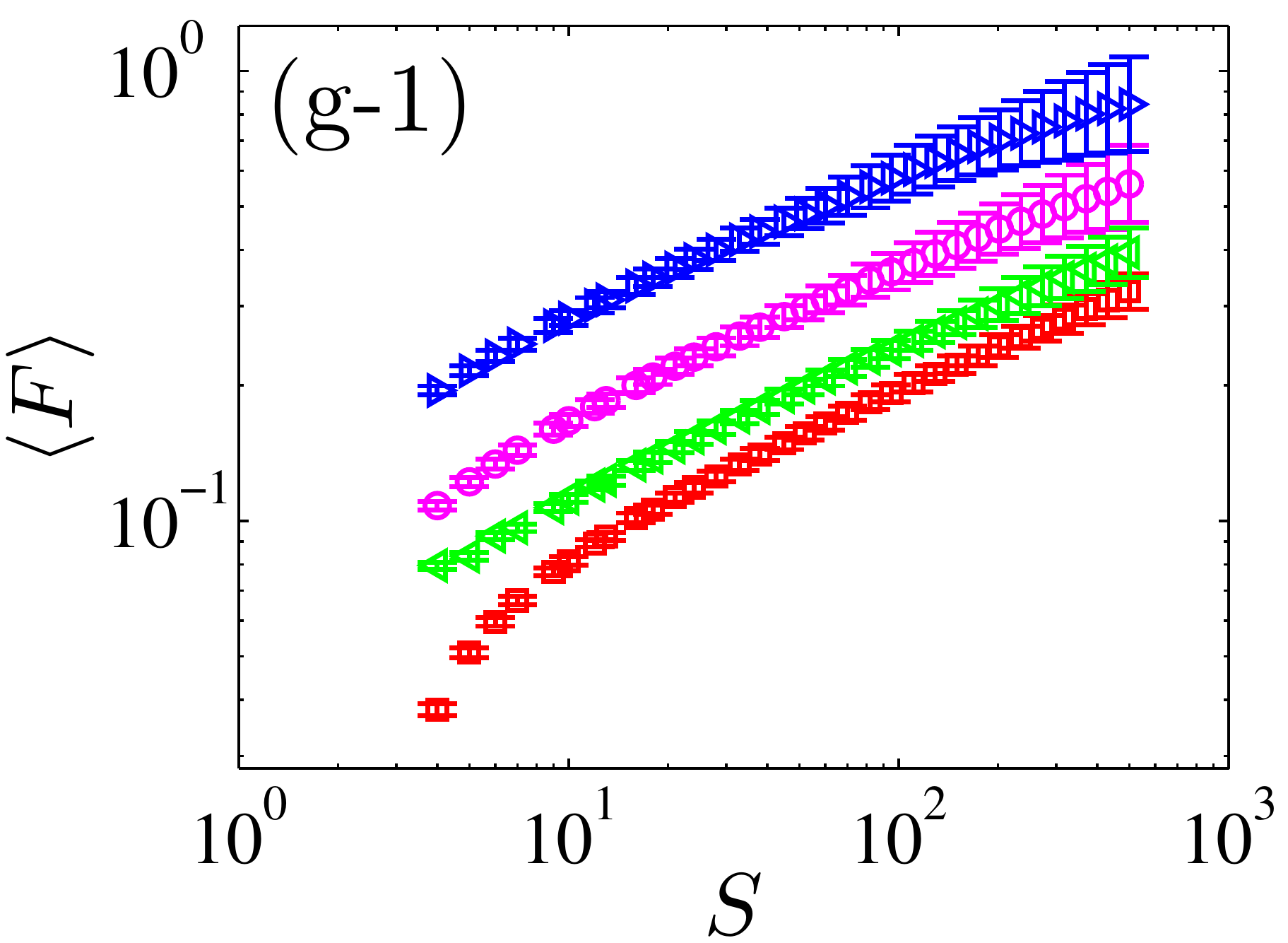}
\includegraphics[width=3.3cm]{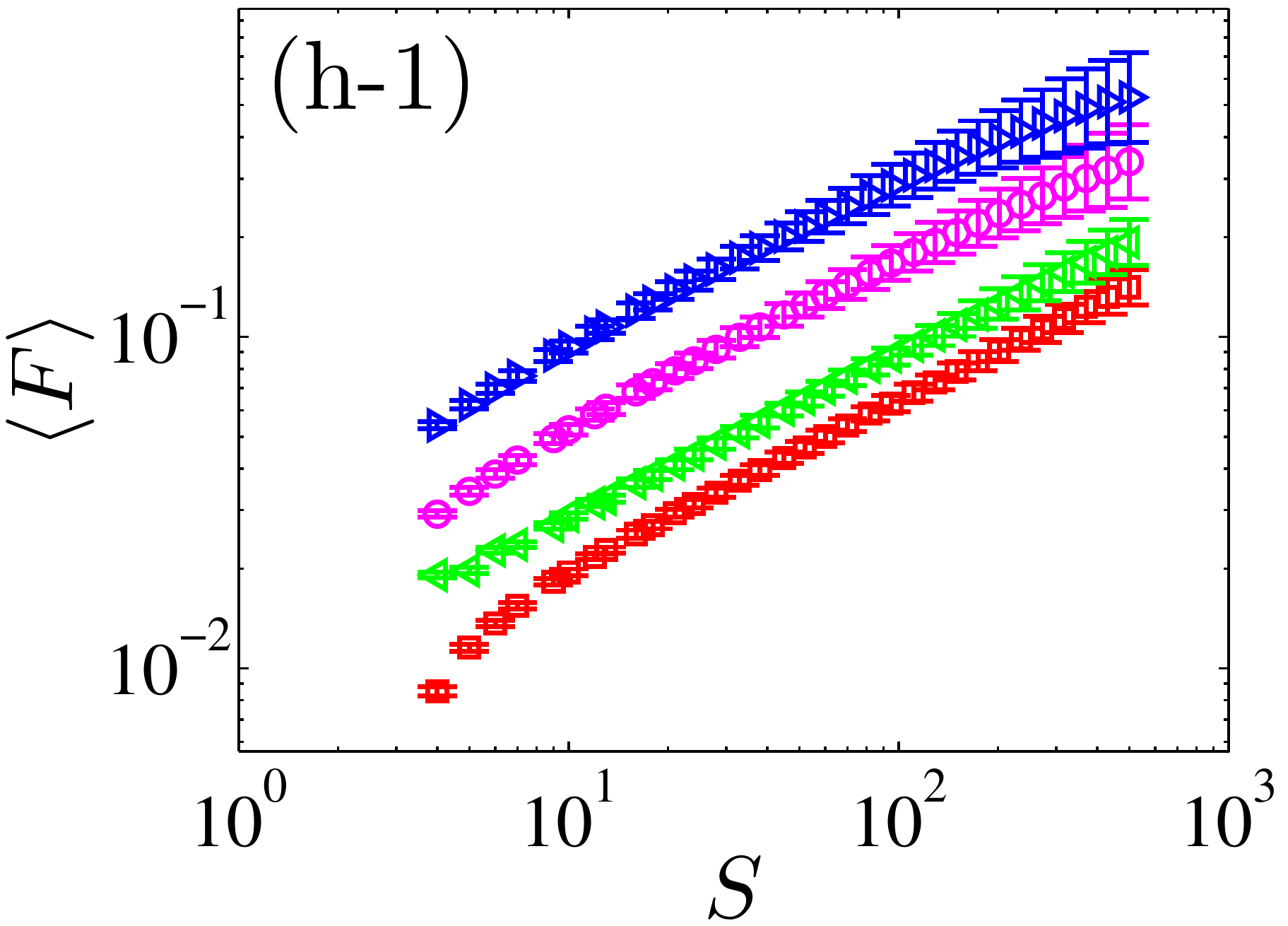}
\includegraphics[width=3.3cm]{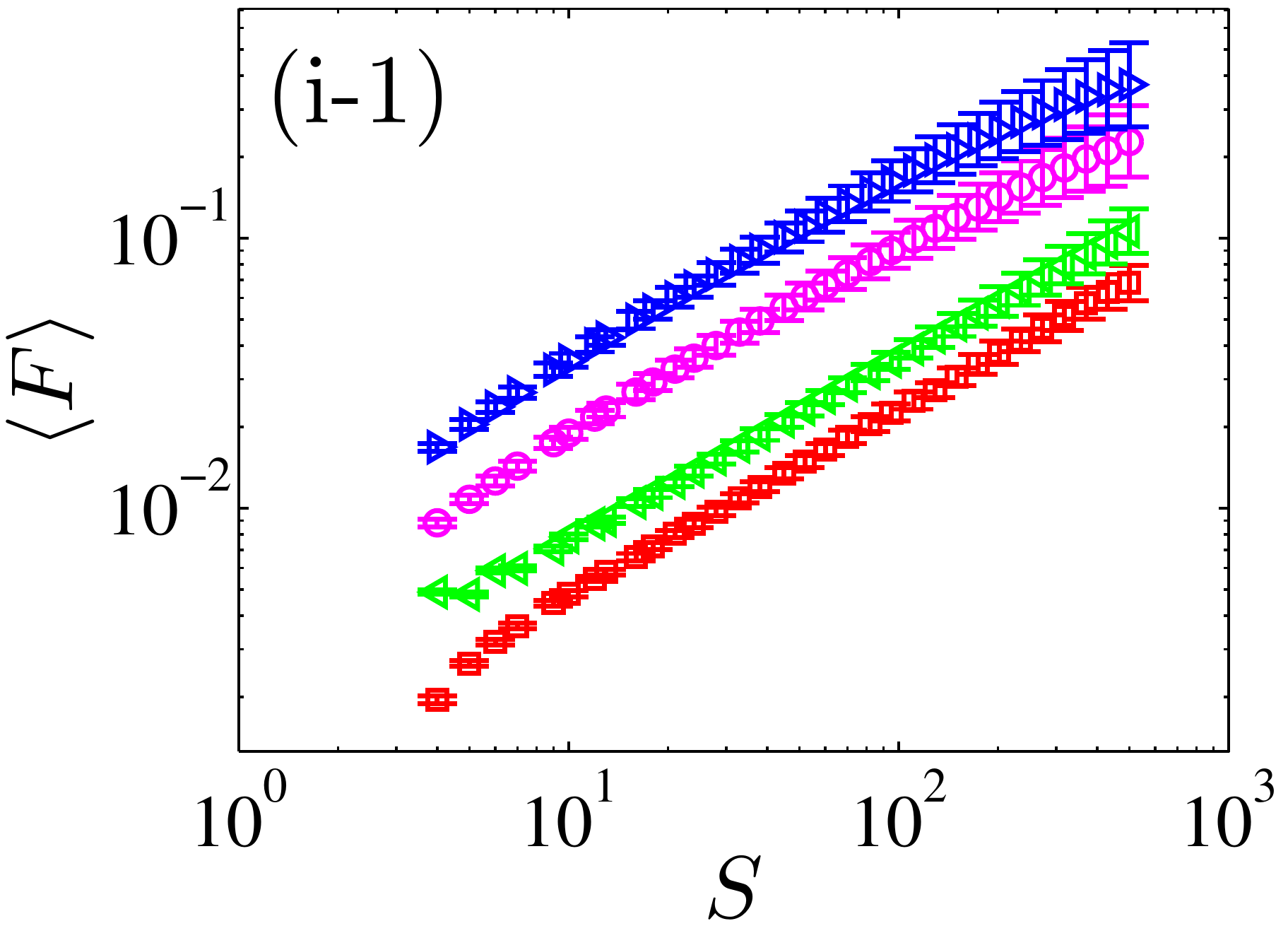}
\includegraphics[width=3.3cm]{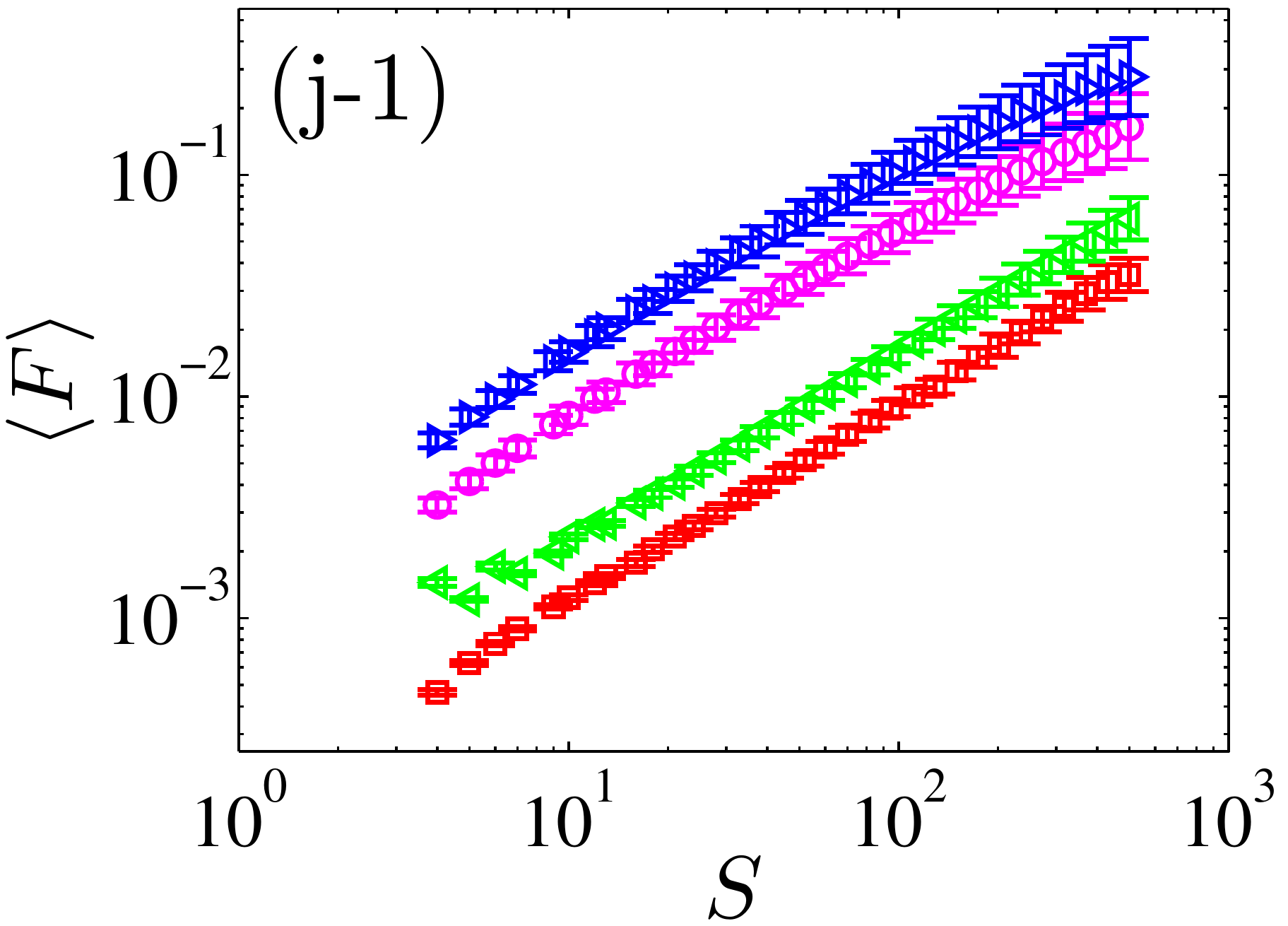}\\
\includegraphics[width=3.3cm]{Fig_rmd_01_FS_errorbar.pdf}
\includegraphics[width=3.3cm]{Fig_rmd_03_FS_errorbar.pdf}
\includegraphics[width=3.3cm]{Fig_rmd_05_FS_errorbar.pdf}
\includegraphics[width=3.3cm]{Fig_rmd_07_FS_errorbar.pdf}
\includegraphics[width=3.3cm]{Fig_rmd_09_FS_errorbar.pdf}\\
\includegraphics[width=3.3cm]{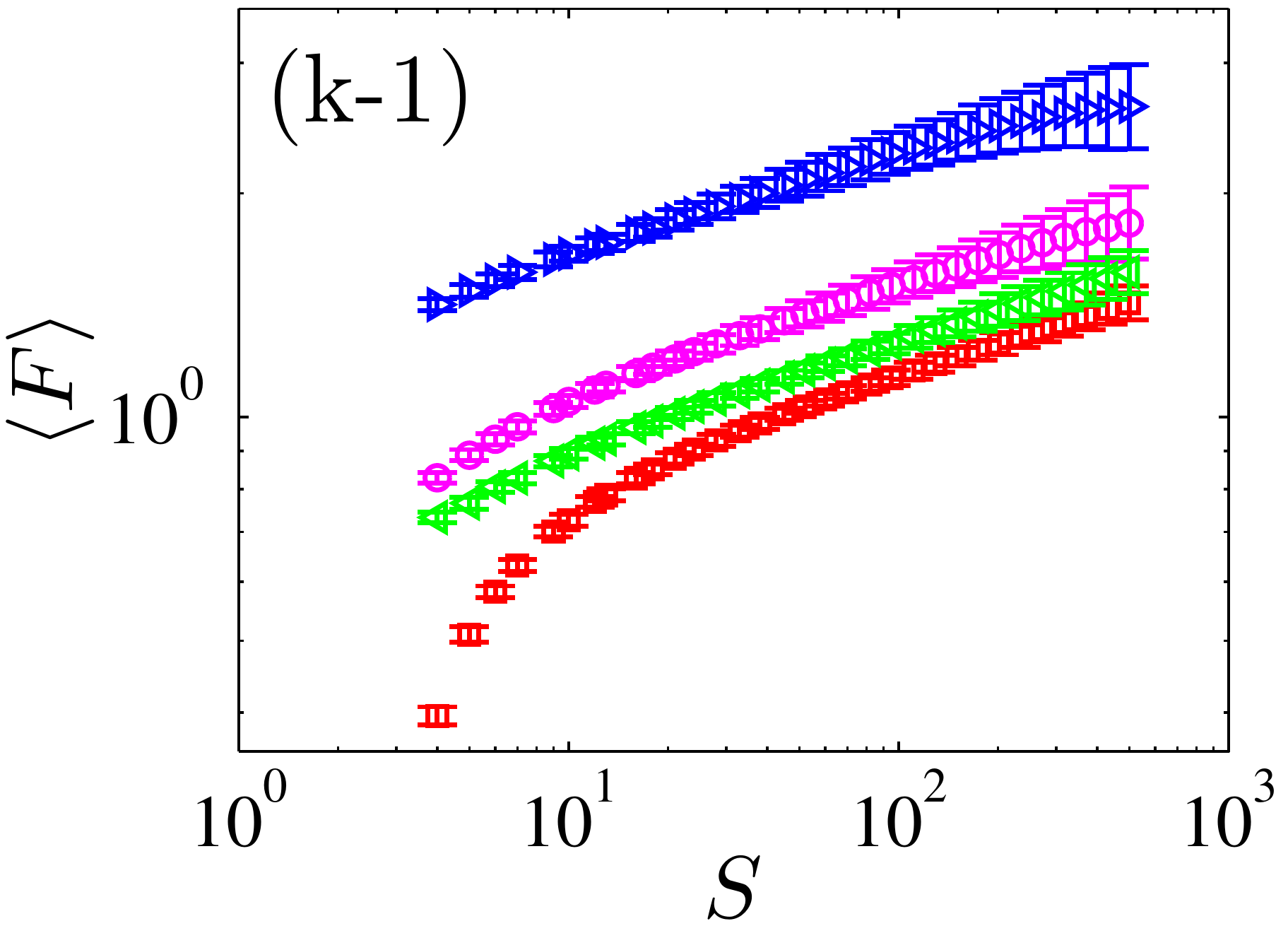}
\includegraphics[width=3.3cm]{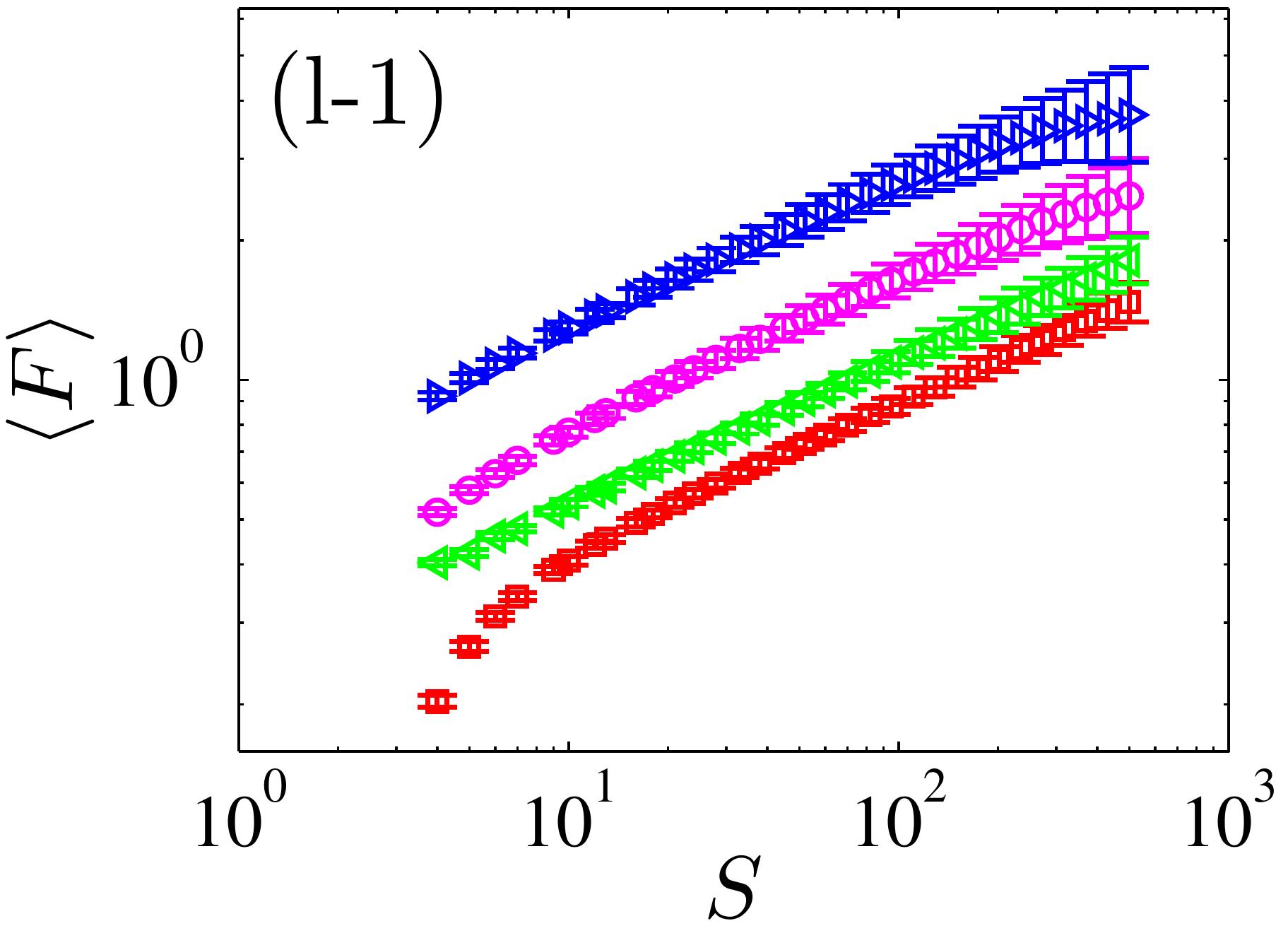}
\includegraphics[width=3.3cm]{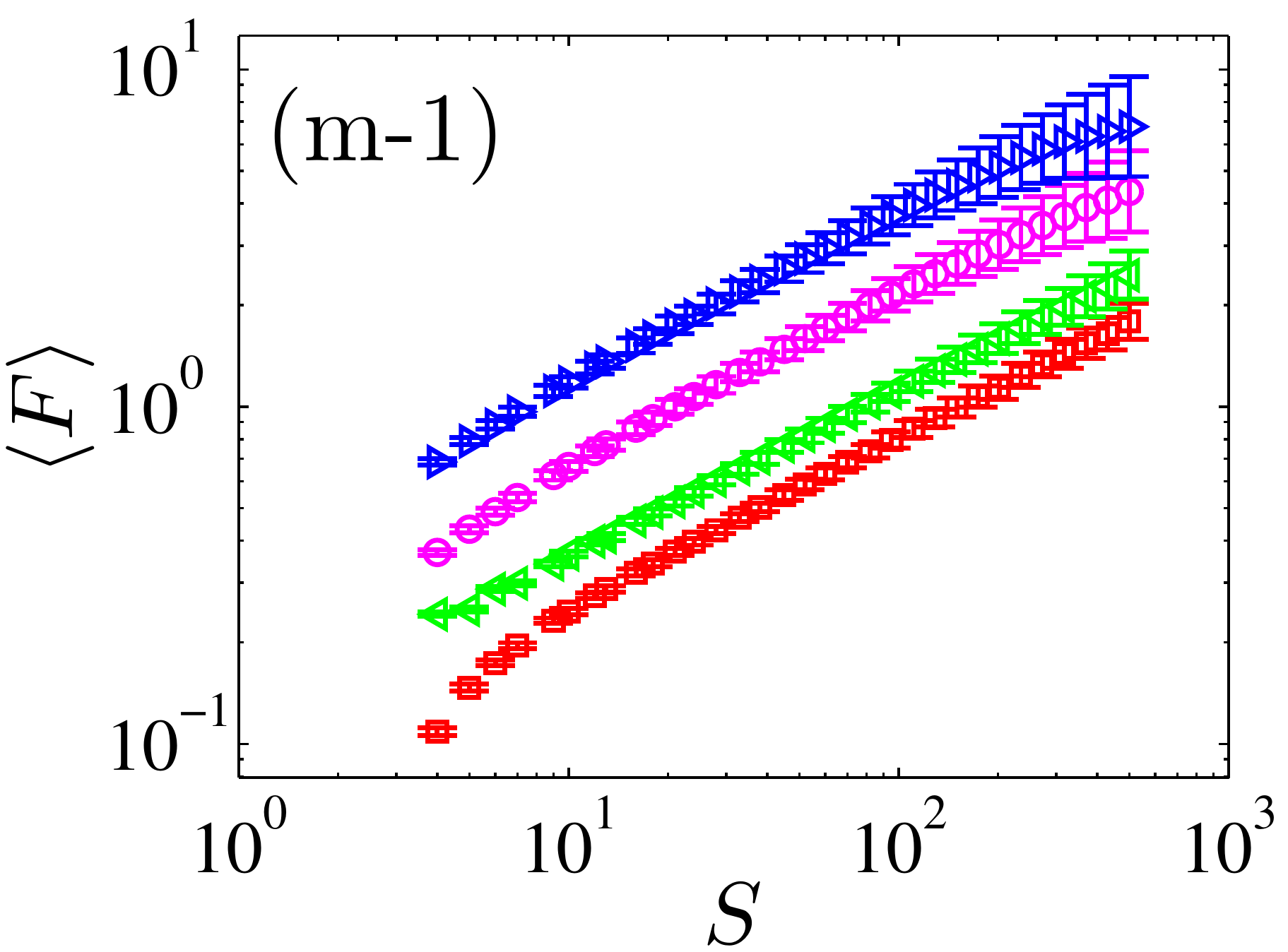}
\includegraphics[width=3.3cm]{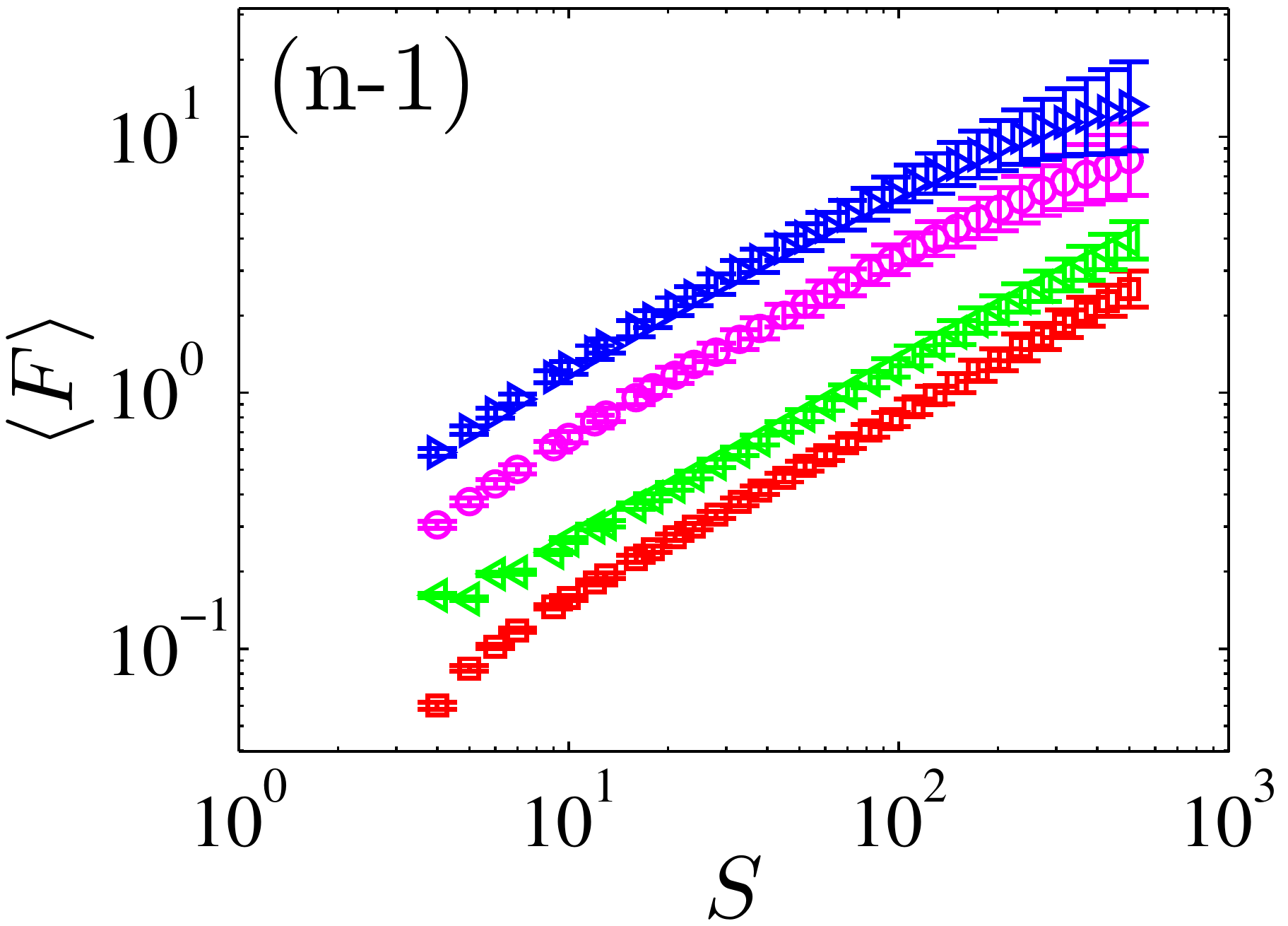}
\includegraphics[width=3.3cm]{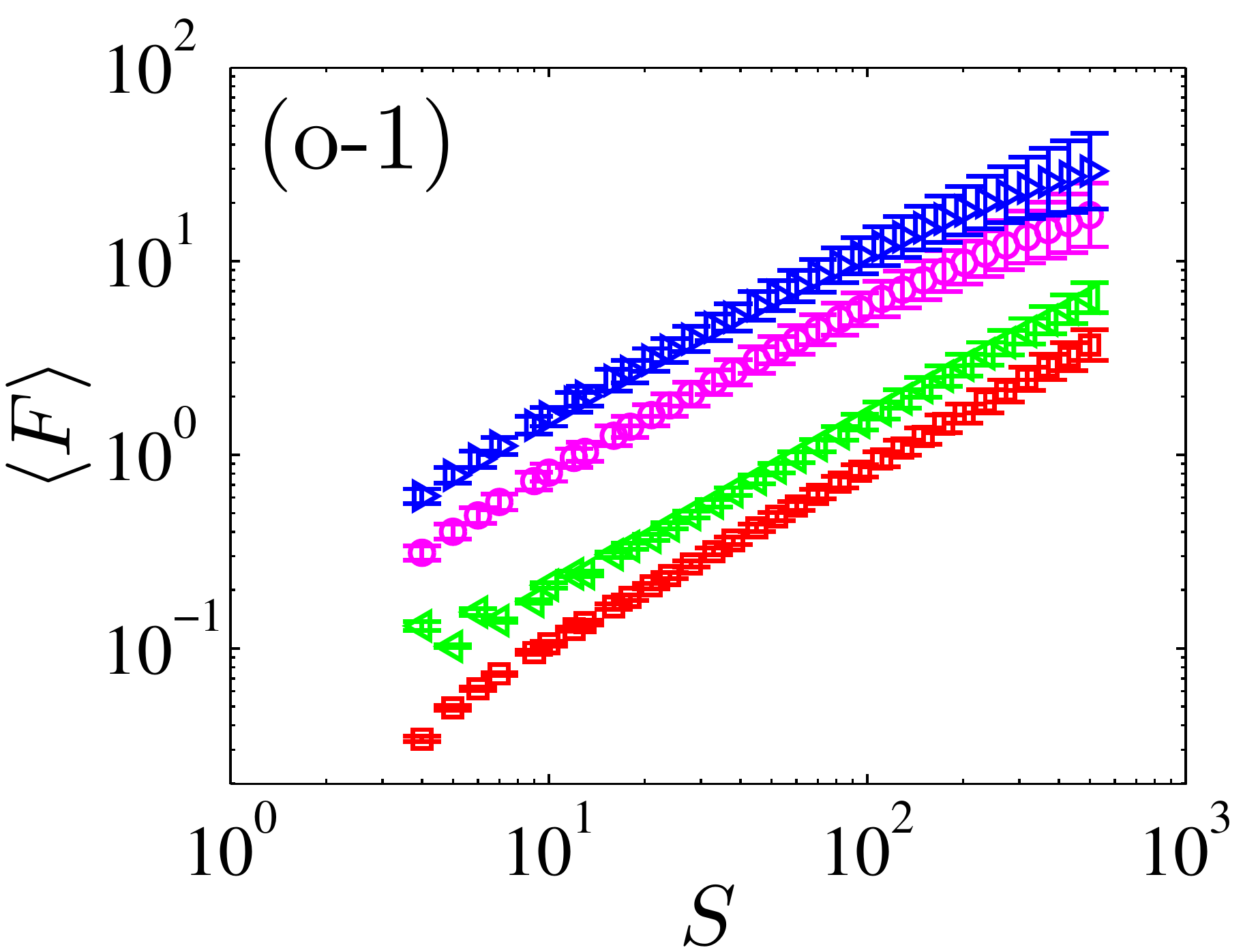}\\
\includegraphics[width=3.3cm]{Fig_wfbm_01_FS_errorbar.pdf}
\includegraphics[width=3.3cm]{Fig_wfbm_03_FS_errorbar.pdf}
\includegraphics[width=3.3cm]{Fig_wfbm_05_FS_errorbar.pdf}
\includegraphics[width=3.3cm]{Fig_wfbm_07_FS_errorbar.pdf}
\includegraphics[width=3.3cm]{Fig_wfbm_09_FS_errorbar.pdf}
\caption{Comparing plots of $\langle{F}\rangle$ against $s$. The plots labeled (a)-(o) are the same as in the paper where the length of time series is 20000, while the plots labeled with (a-1) to (o-1) are the results where the length of time series is 2000.}
\label{SFig:FluctuationFunction}
\end{figure*}

\begin{figure*}
\centering
\includegraphics[width=3.3cm]{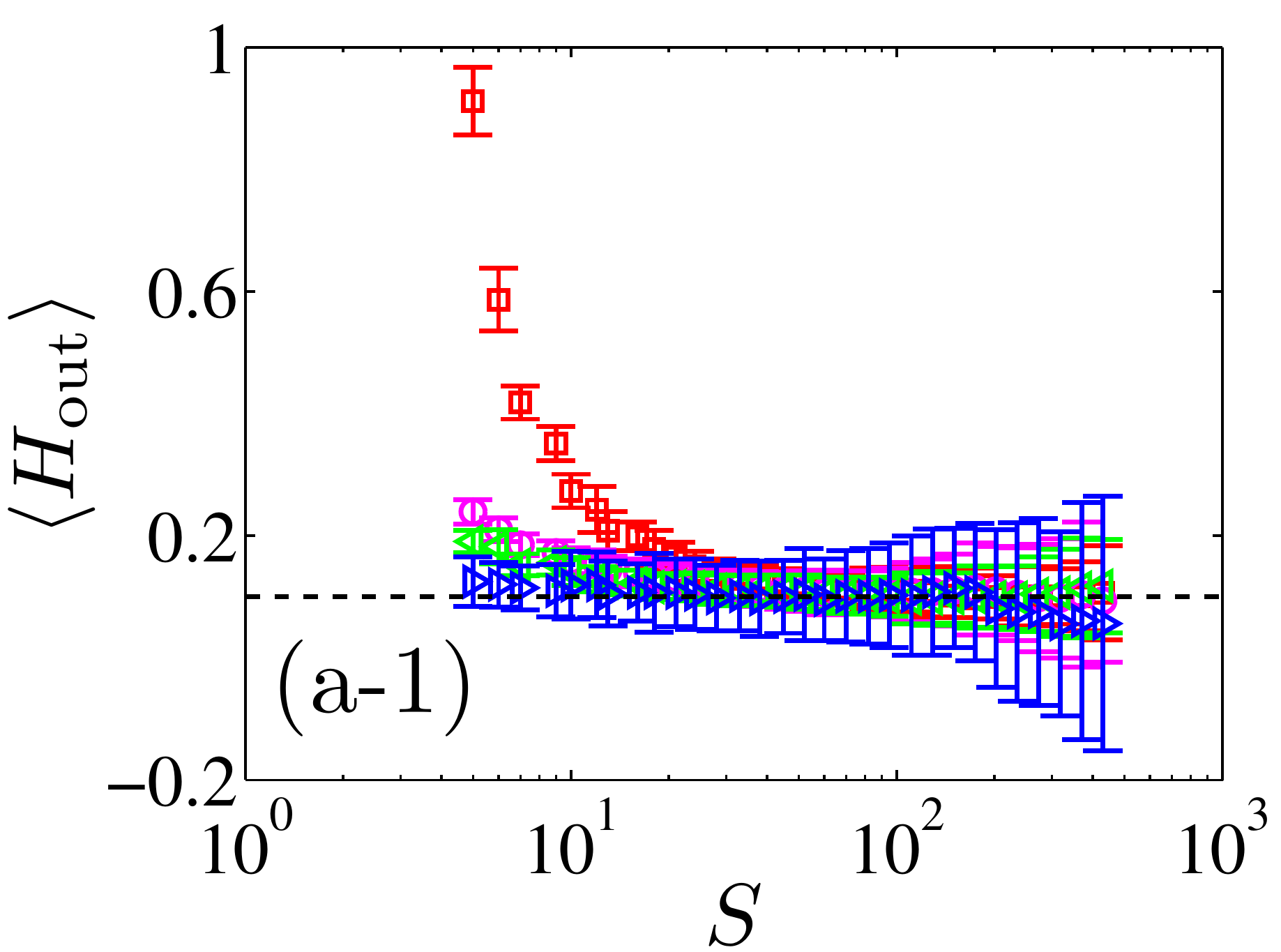}
\includegraphics[width=3.3cm]{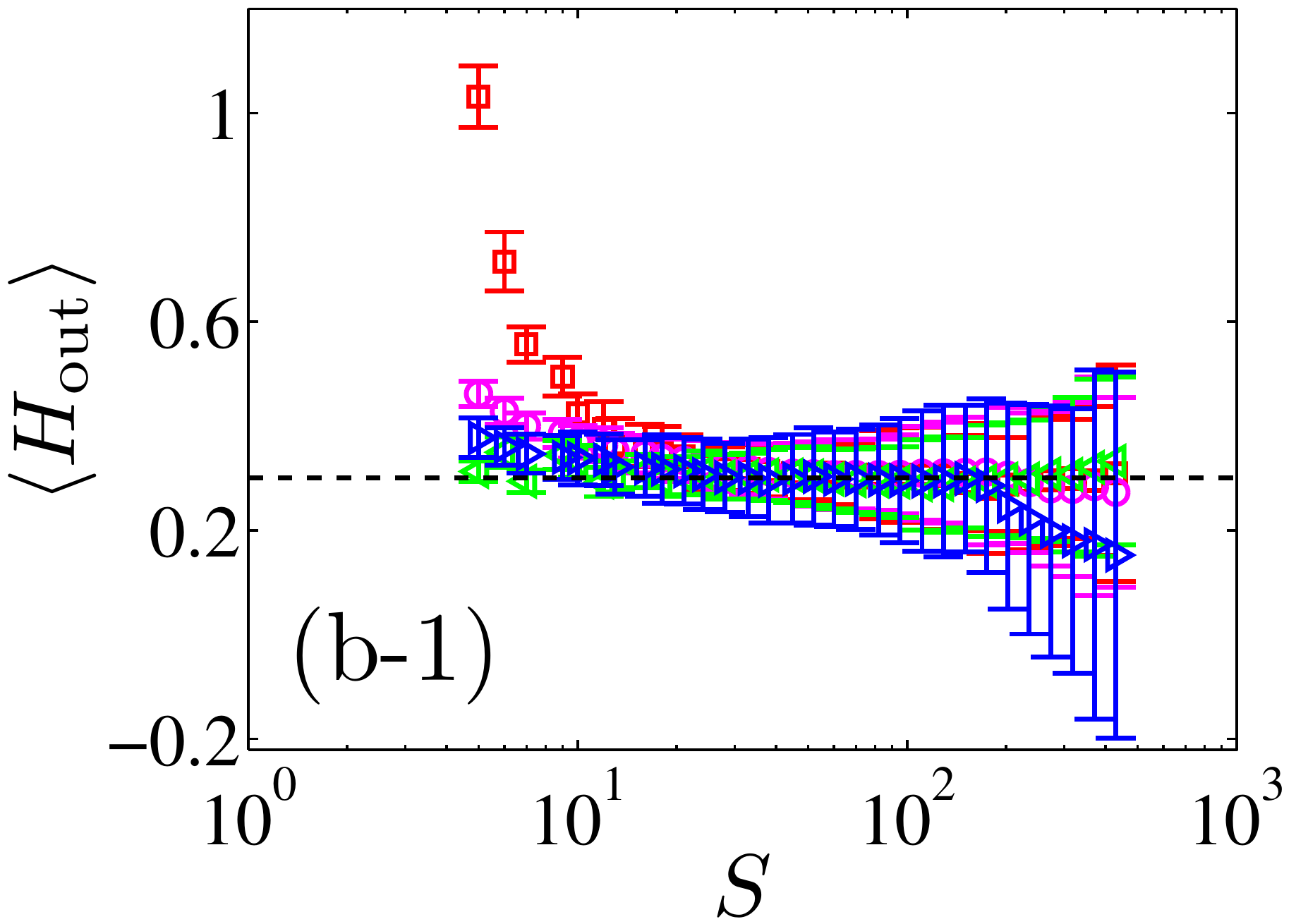}
\includegraphics[width=3.3cm]{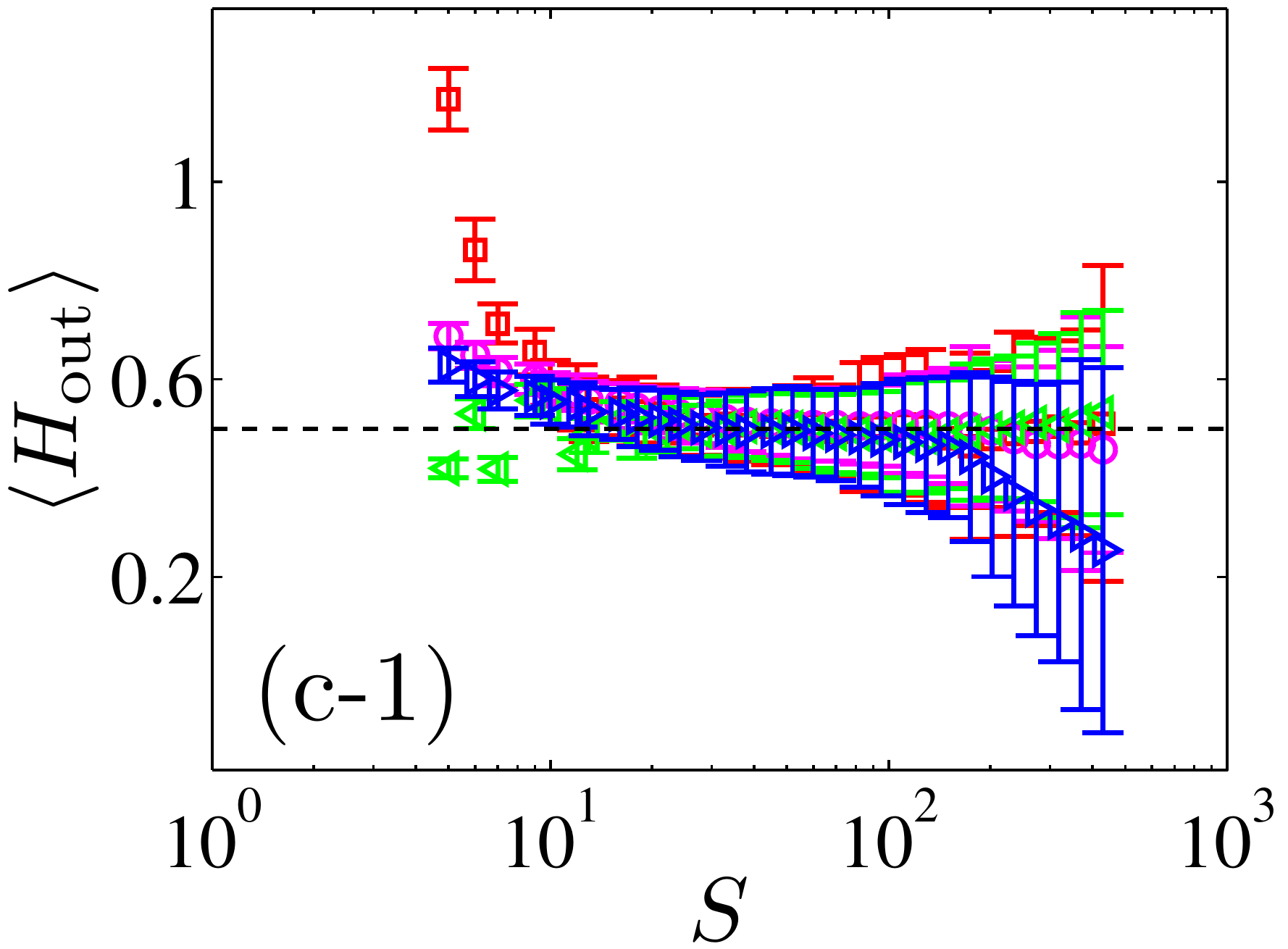}
\includegraphics[width=3.3cm]{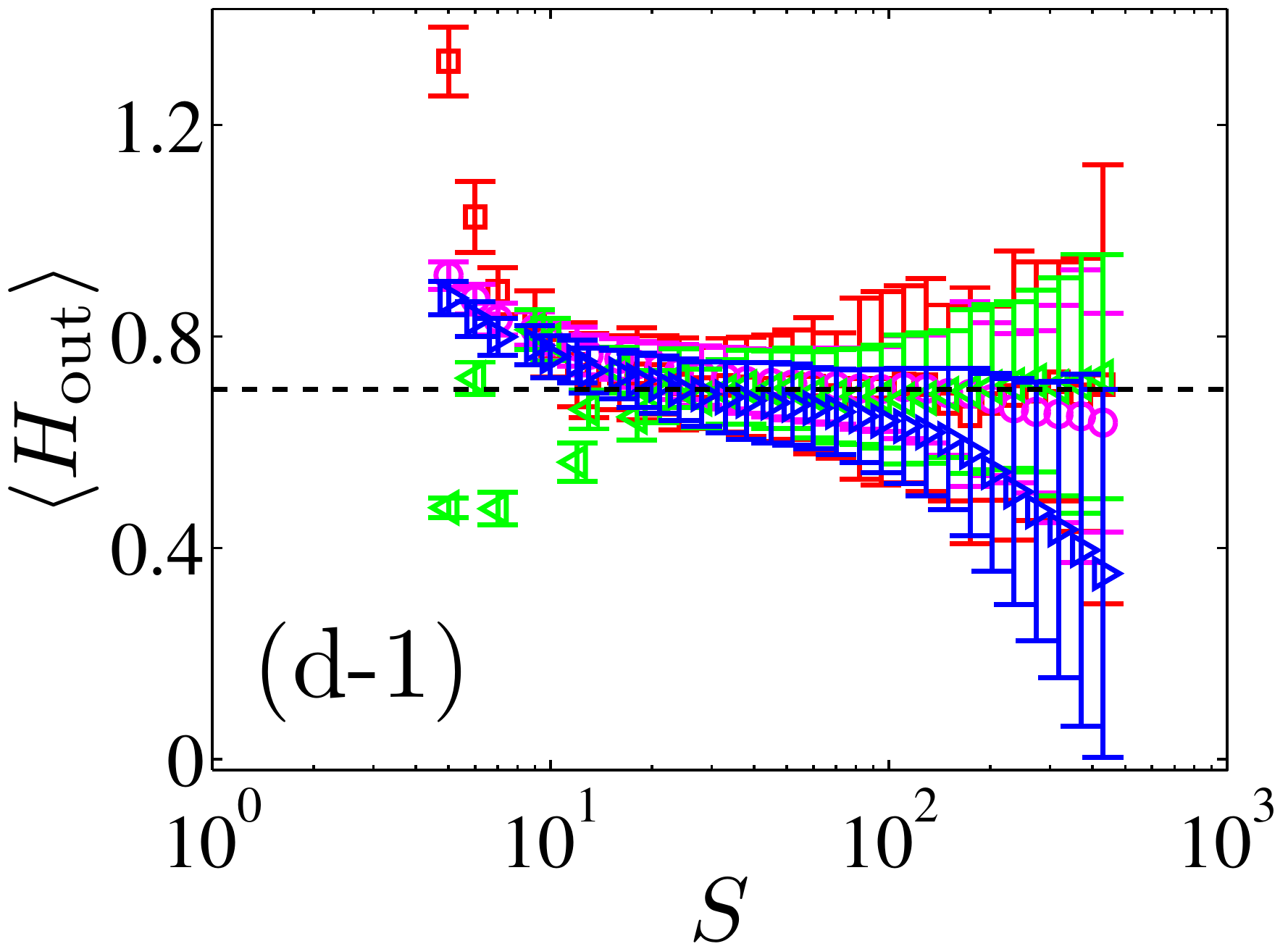}
\includegraphics[width=3.3cm]{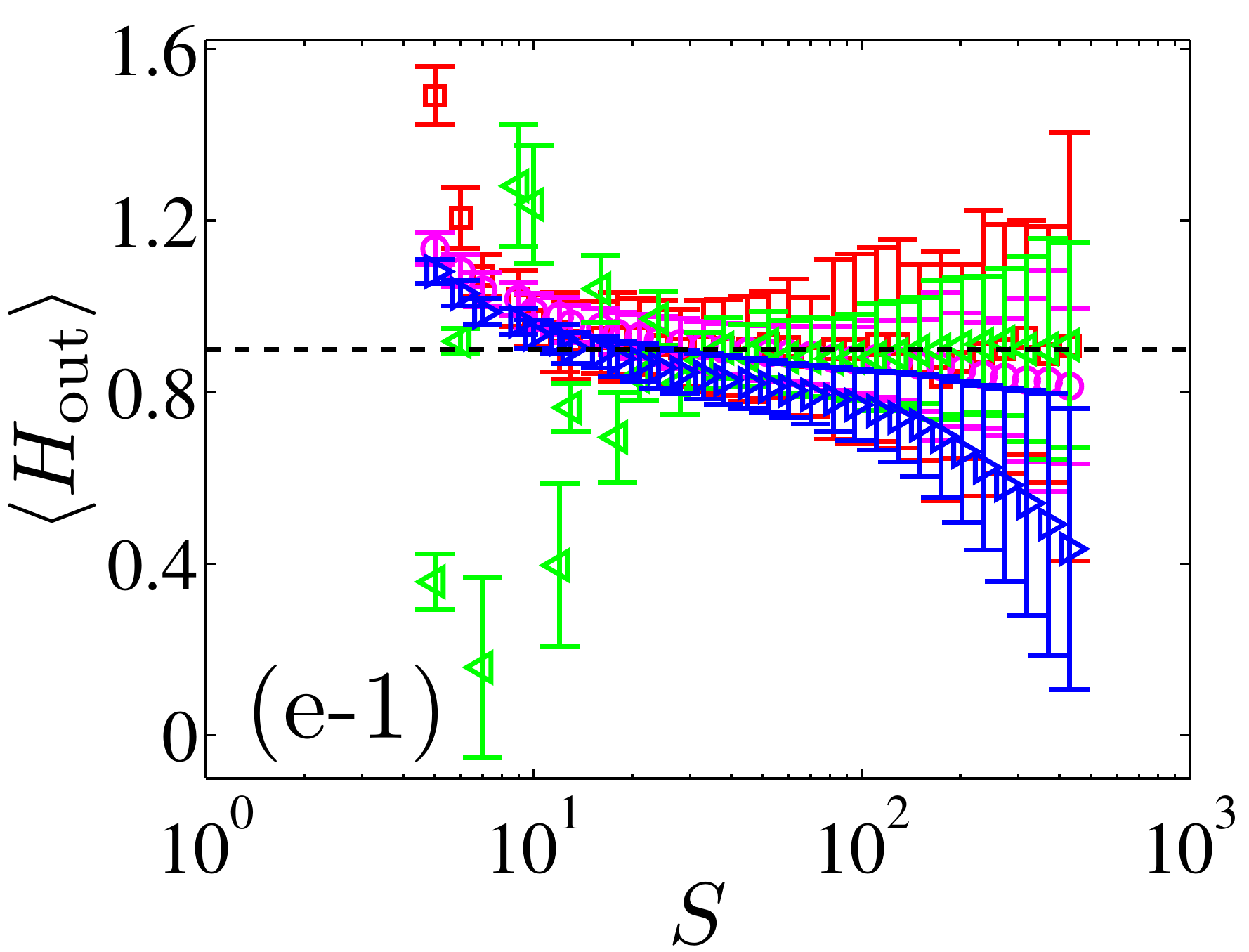}
\includegraphics[width=3.3cm]{Fig_fgn_01_h_errorbar.pdf}
\includegraphics[width=3.3cm]{Fig_fgn_03_h_errorbar.pdf}
\includegraphics[width=3.3cm]{Fig_fgn_05_h_errorbar.pdf}
\includegraphics[width=3.3cm]{Fig_fgn_07_h_errorbar.pdf}
\includegraphics[width=3.3cm]{Fig_fgn_09_h_errorbar.pdf}\\
\includegraphics[width=3.3cm]{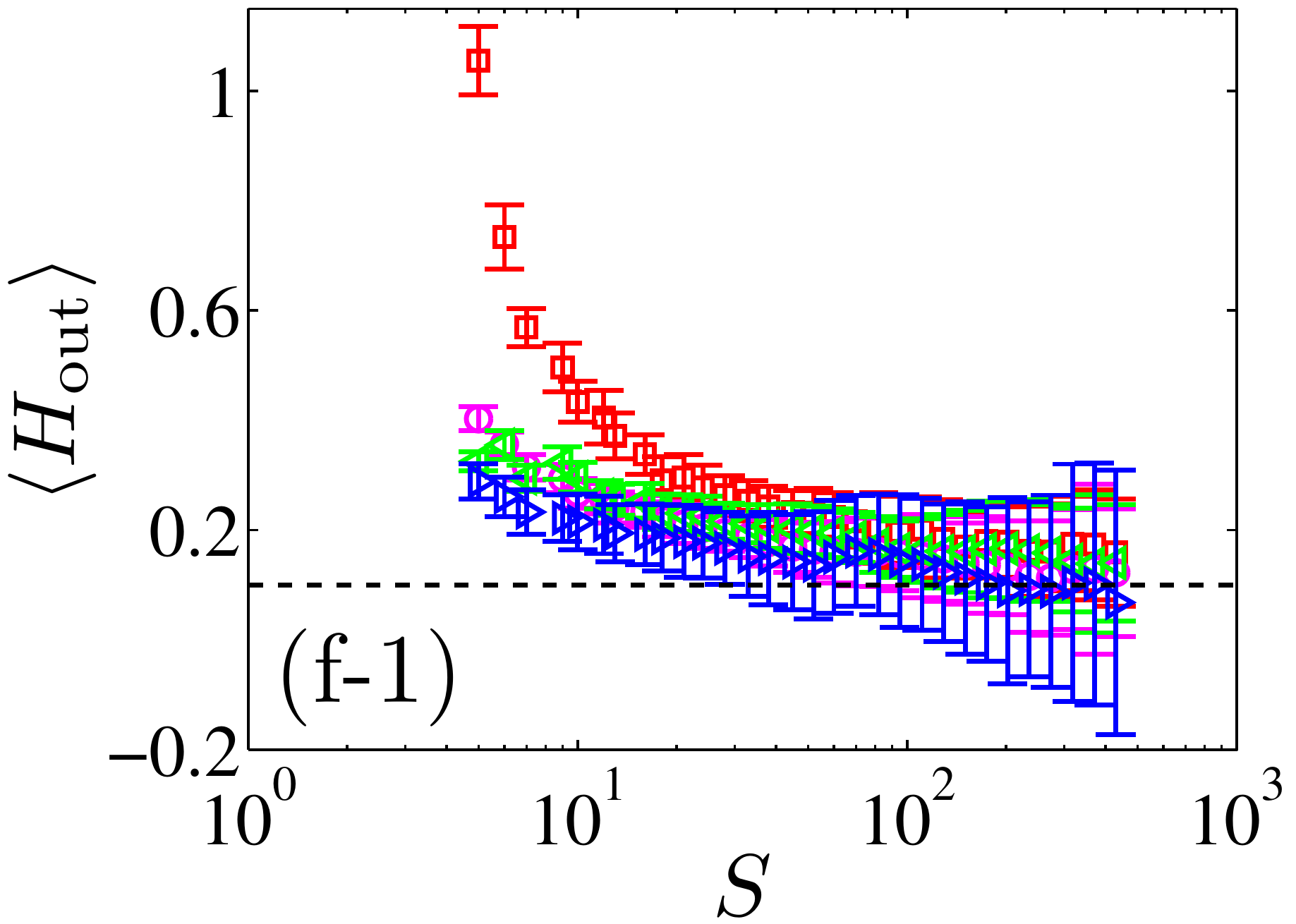}
\includegraphics[width=3.3cm]{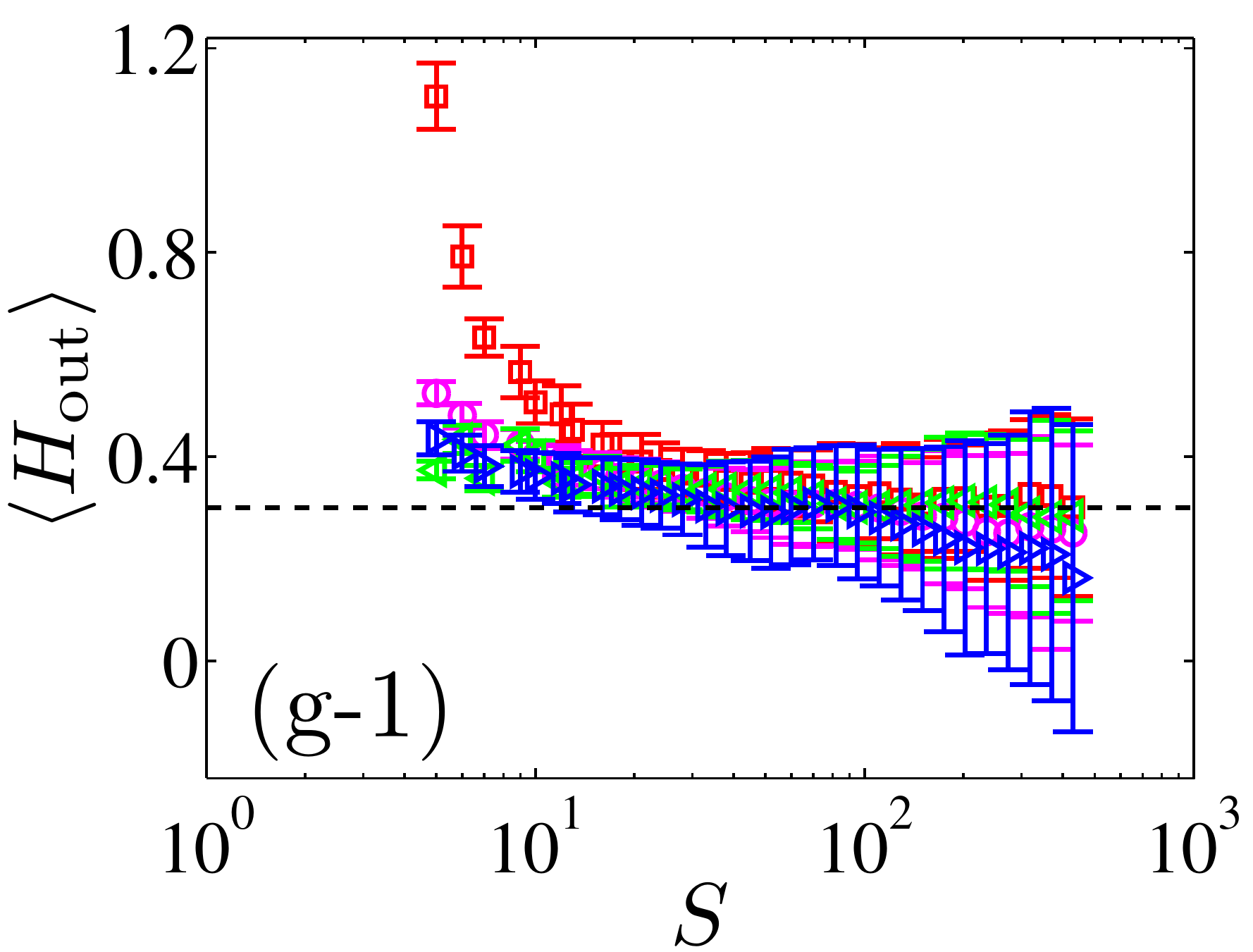}
\includegraphics[width=3.3cm]{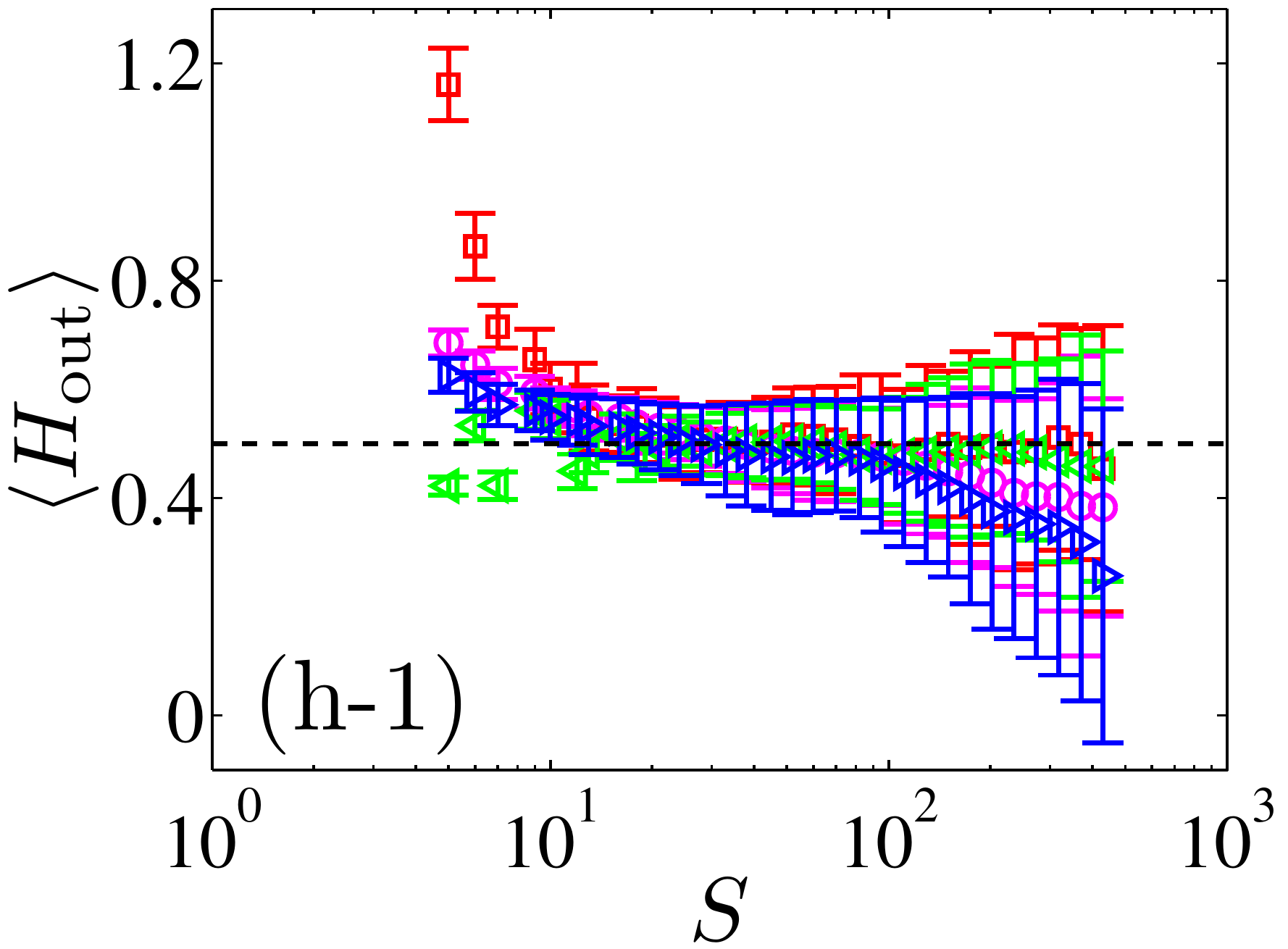}
\includegraphics[width=3.3cm]{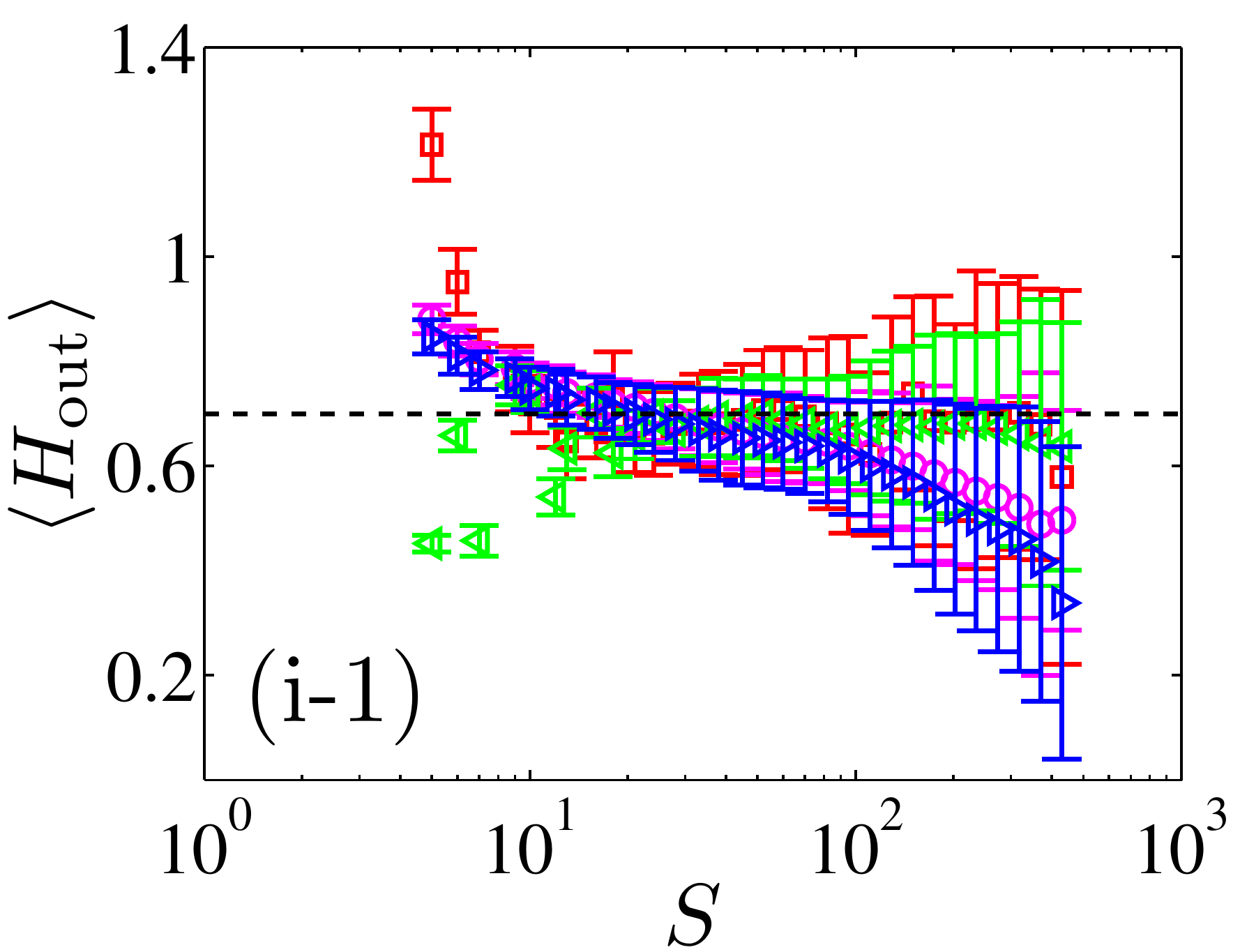}
\includegraphics[width=3.3cm]{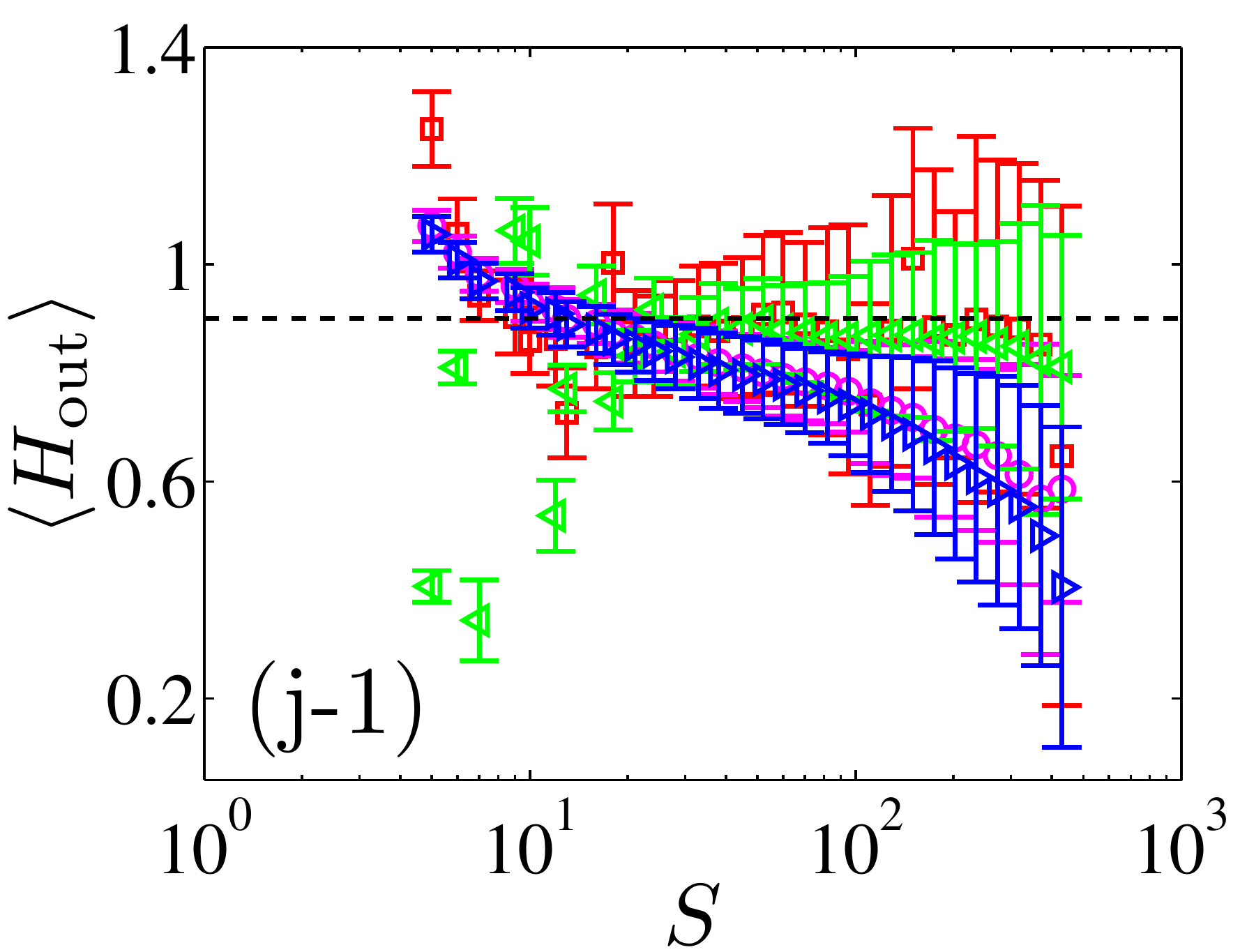}
\includegraphics[width=3.3cm]{Fig_rmd_01_h_errorbar.pdf}
\includegraphics[width=3.3cm]{Fig_rmd_03_h_errorbar.pdf}
\includegraphics[width=3.3cm]{Fig_rmd_05_h_errorbar.pdf}
\includegraphics[width=3.3cm]{Fig_rmd_07_h_errorbar.pdf}
\includegraphics[width=3.3cm]{Fig_rmd_09_h_errorbar.pdf}\\
\includegraphics[width=3.3cm]{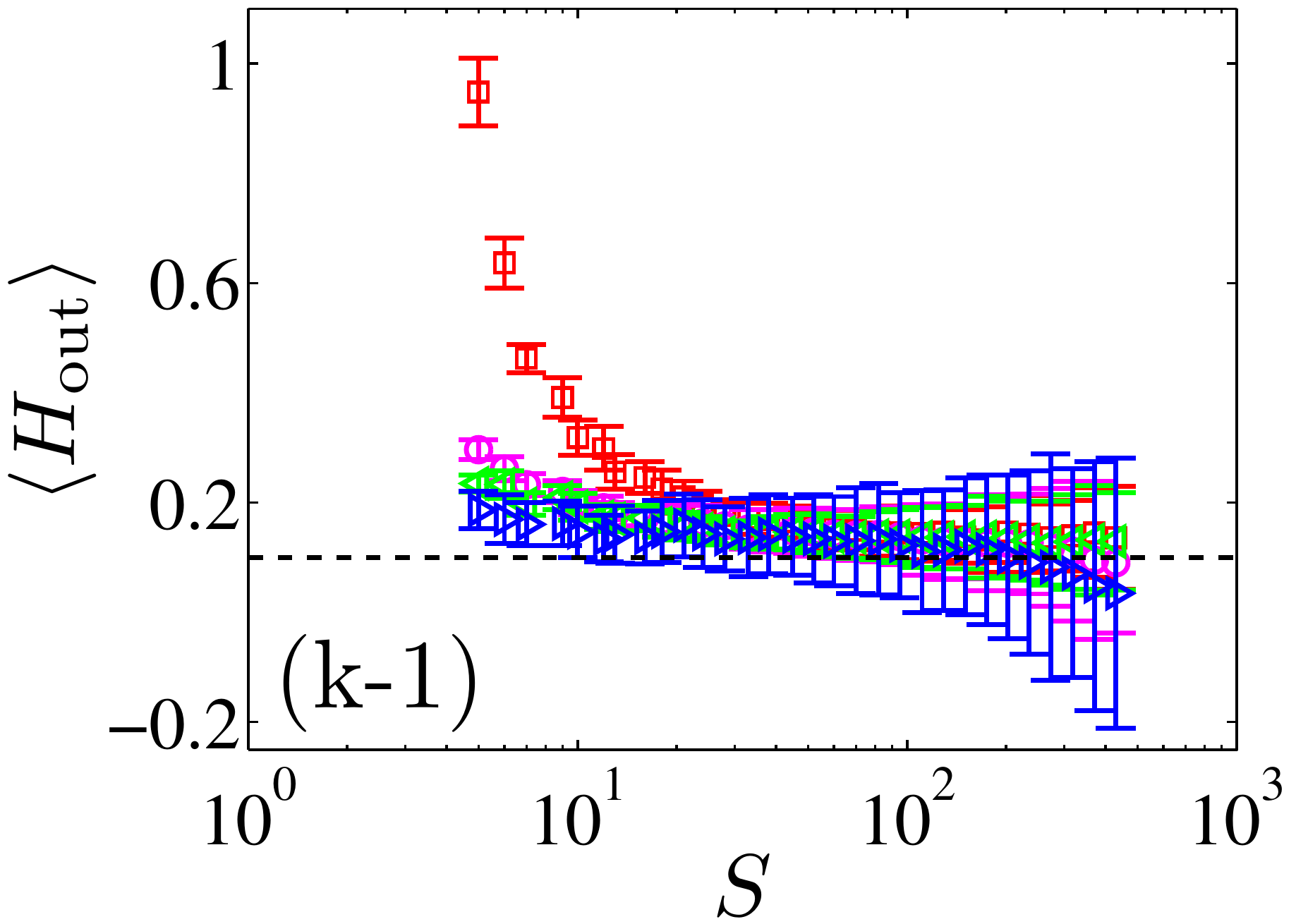}
\includegraphics[width=3.3cm]{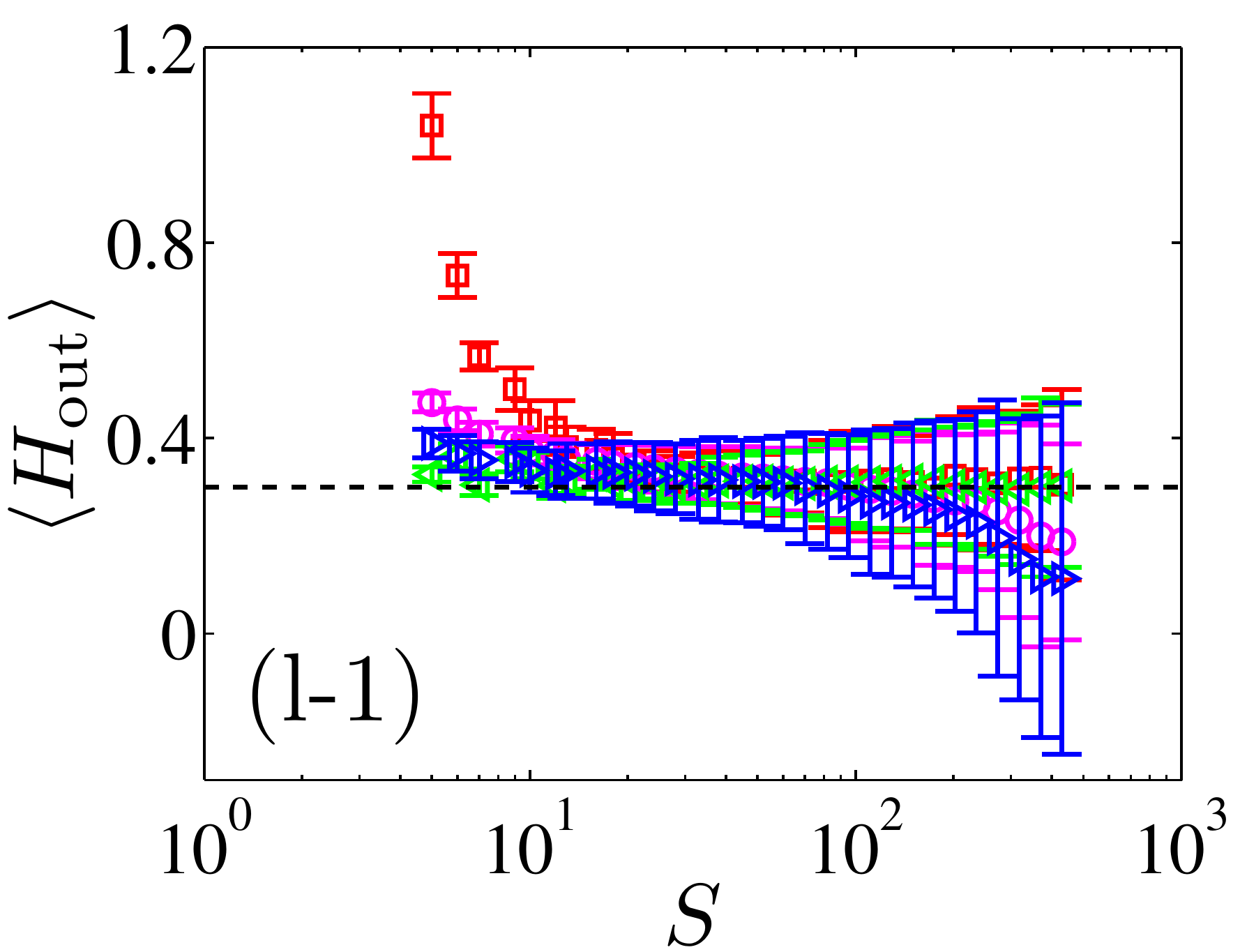}
\includegraphics[width=3.3cm]{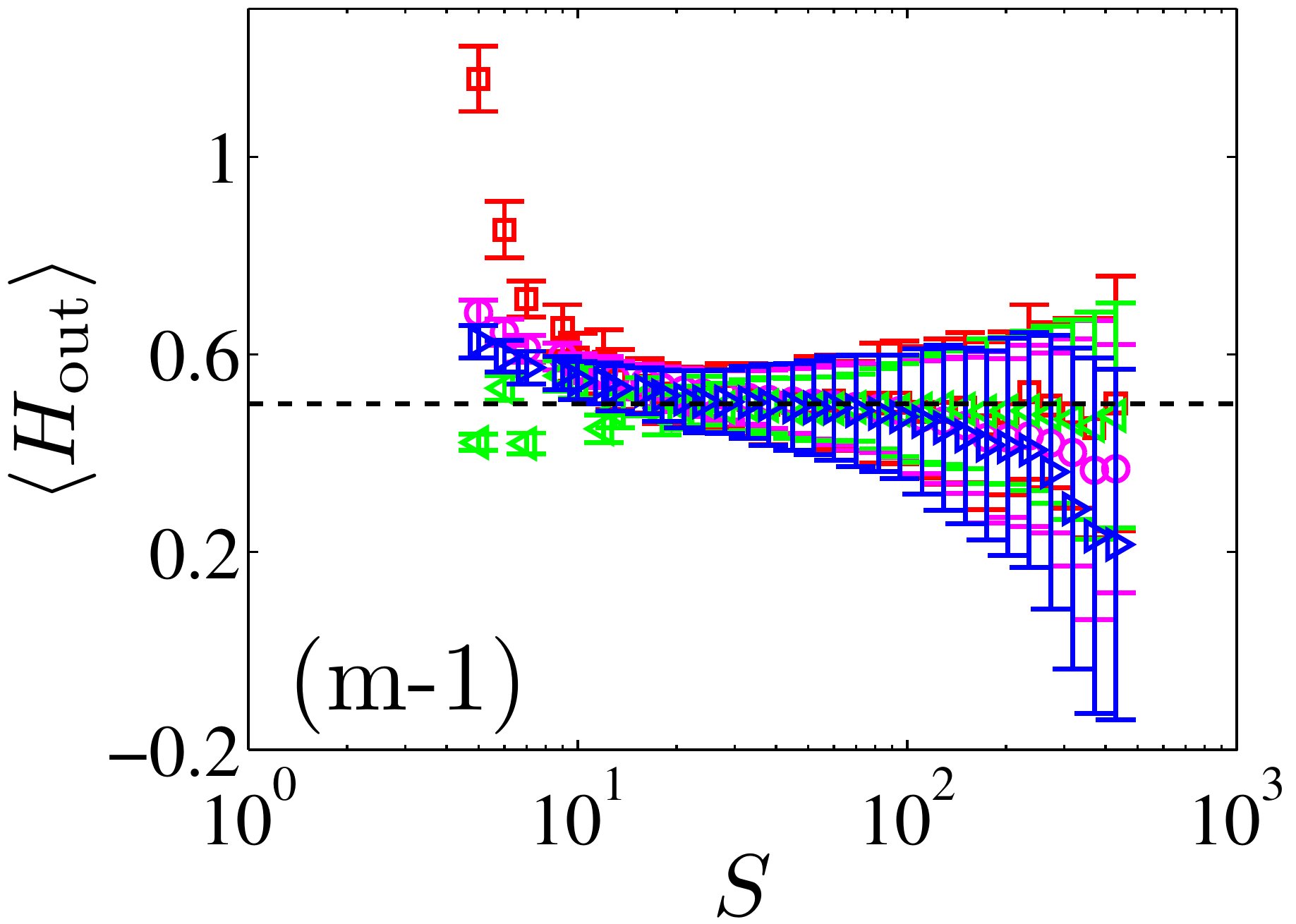}
\includegraphics[width=3.3cm]{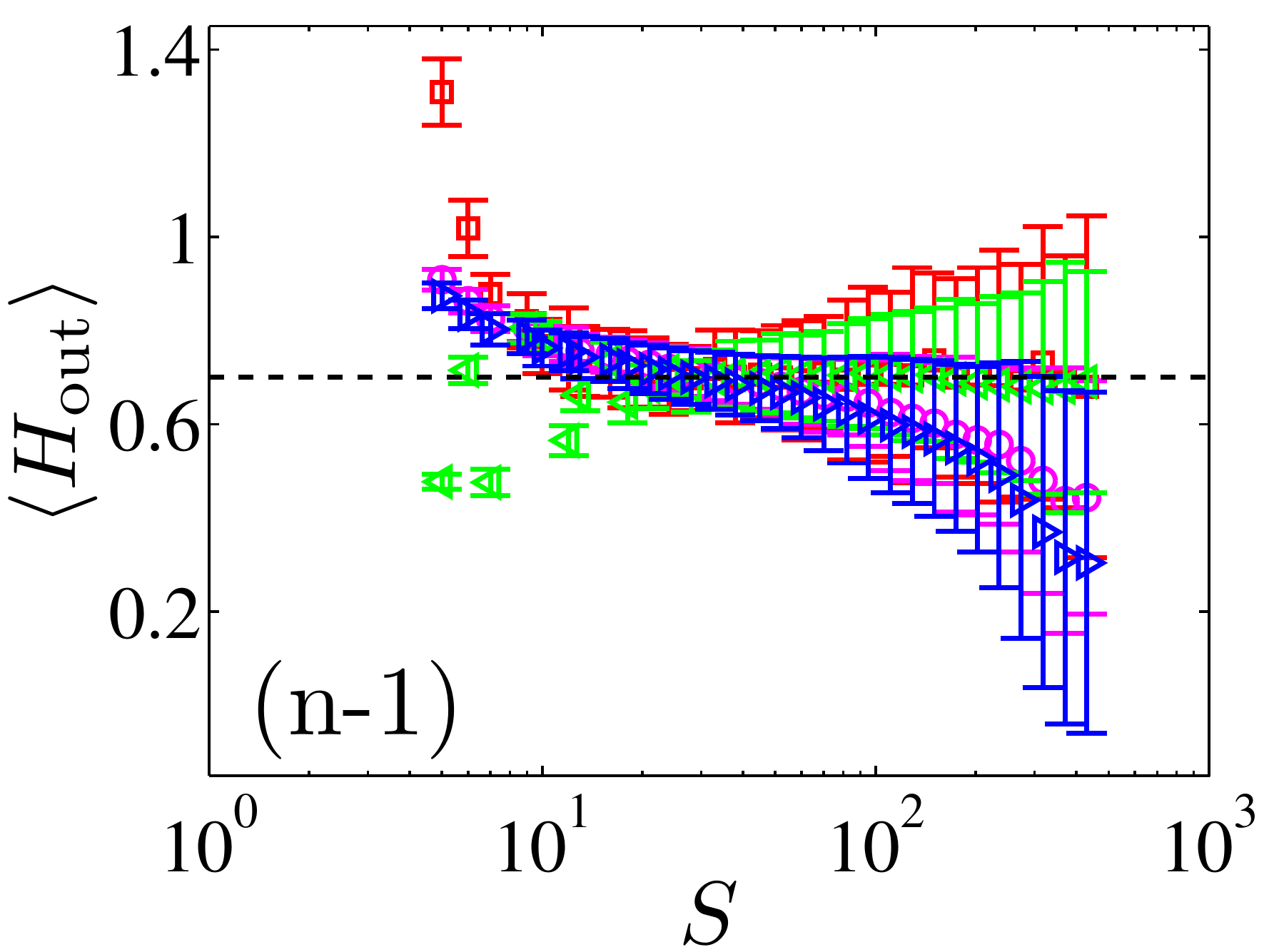}
\includegraphics[width=3.3cm]{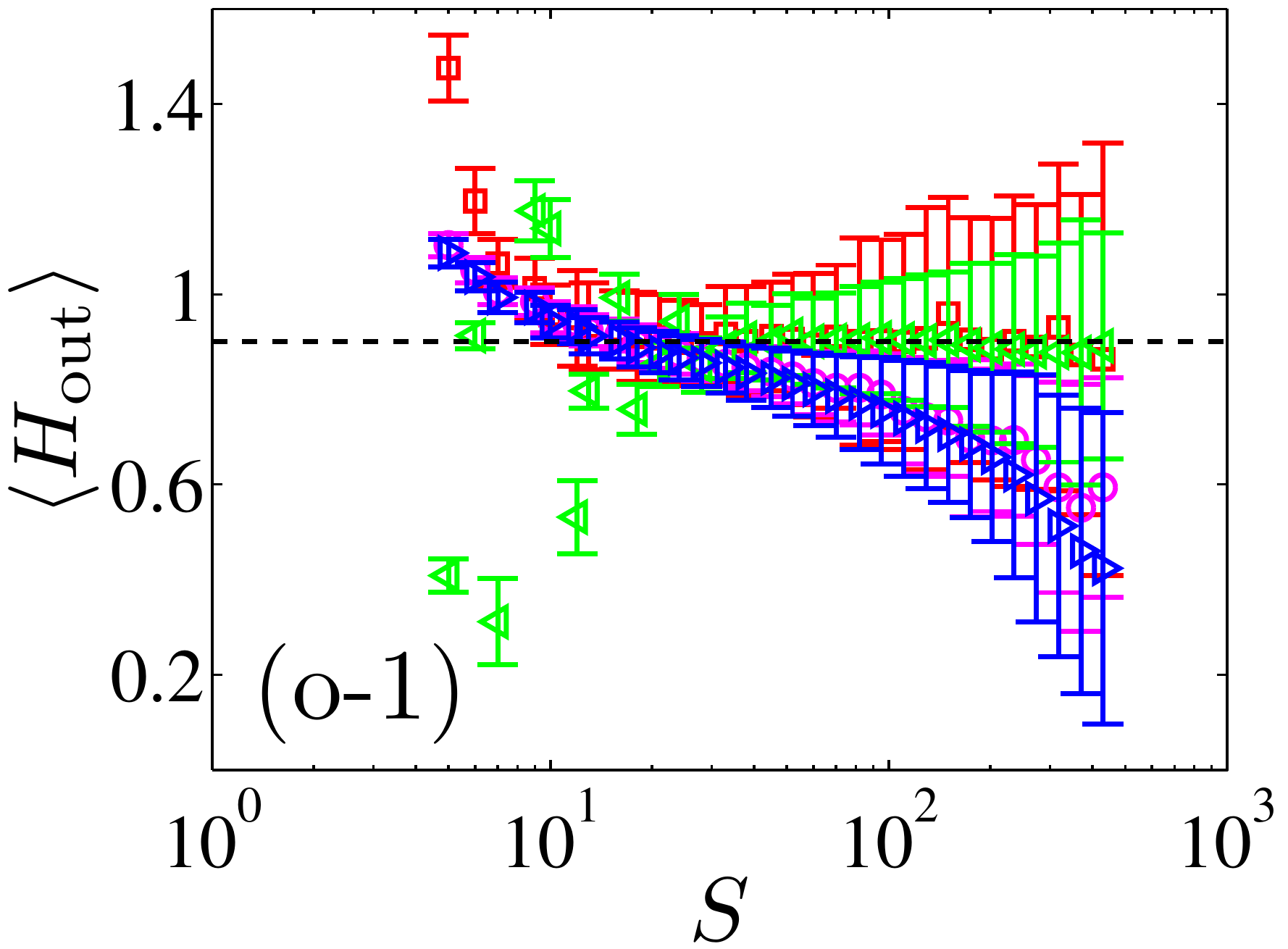}
\includegraphics[width=3.3cm]{Fig_wfbm_01_h_errorbar.pdf}
\includegraphics[width=3.3cm]{Fig_wfbm_03_h_errorbar.pdf}
\includegraphics[width=3.3cm]{Fig_wfbm_05_h_errorbar.pdf}
\includegraphics[width=3.3cm]{Fig_wfbm_07_h_errorbar.pdf}
\includegraphics[width=3.3cm]{Fig_wfbm_09_h_errorbar.pdf}
\caption{{\textbf{Comparing local slopes of the fluctuation functions.}} The plots labeled (a)-(o) are the same as in the paper where the length of time series is 20000, while the plots labeled with (a-1) to (o-1) are the results where the length of time series is 2000.}
\label{SFig:LocalSlope}
\end{figure*}

\begin{figure*}
\centering
\includegraphics[width=5.5cm,height=4cm]{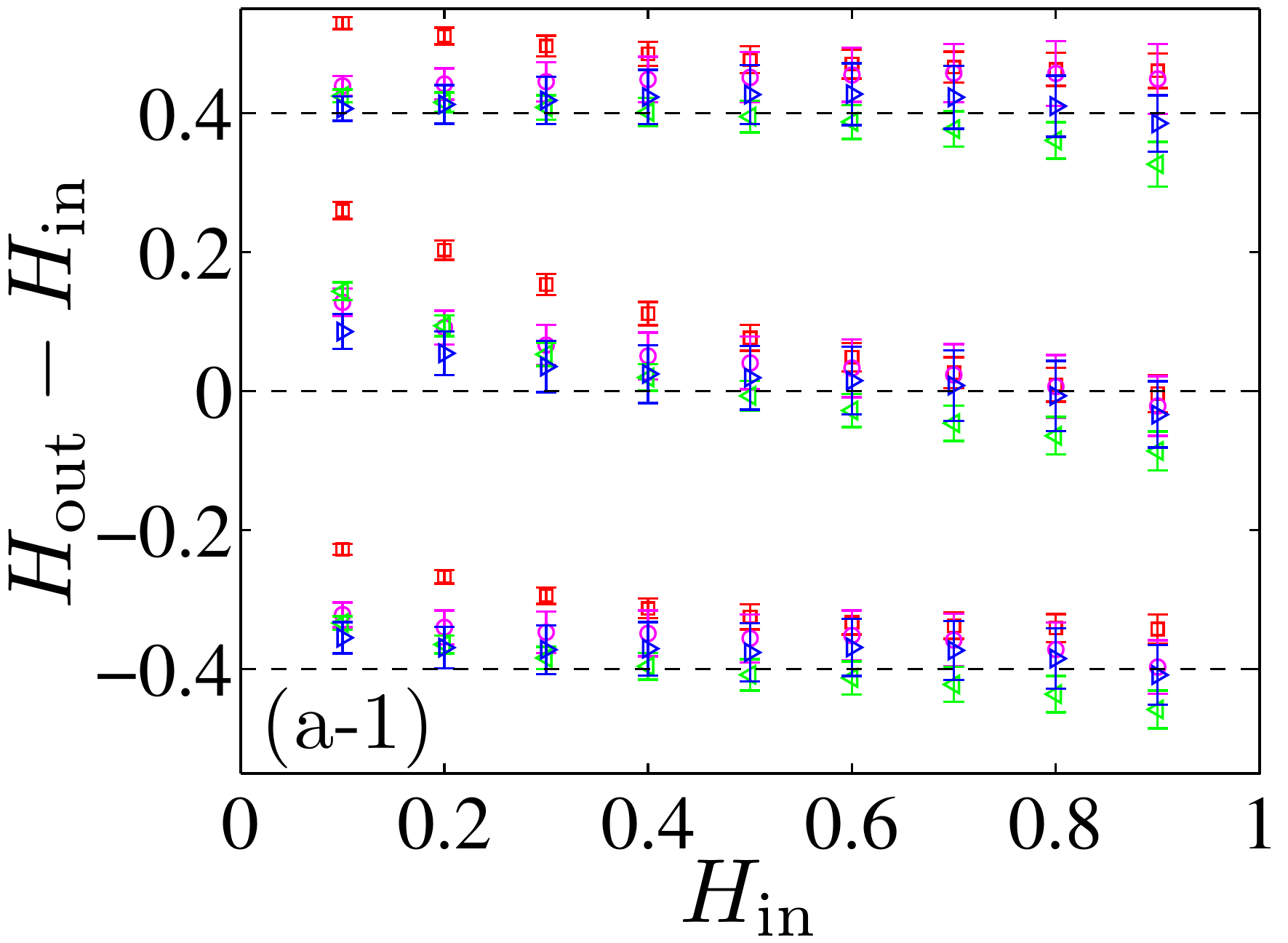}
\includegraphics[width=5.5cm,height=4cm]{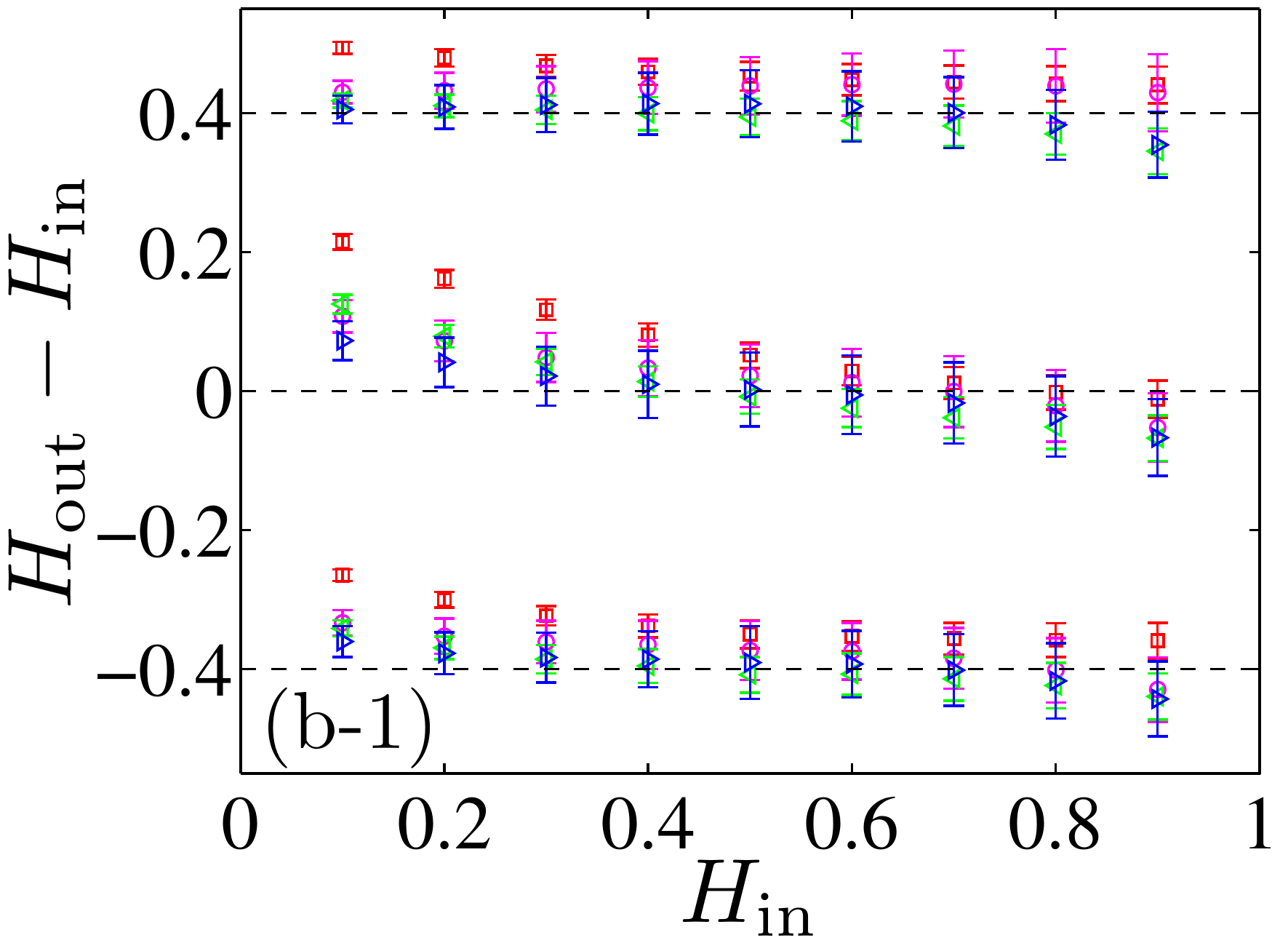}
\includegraphics[width=5.5cm,height=4cm]{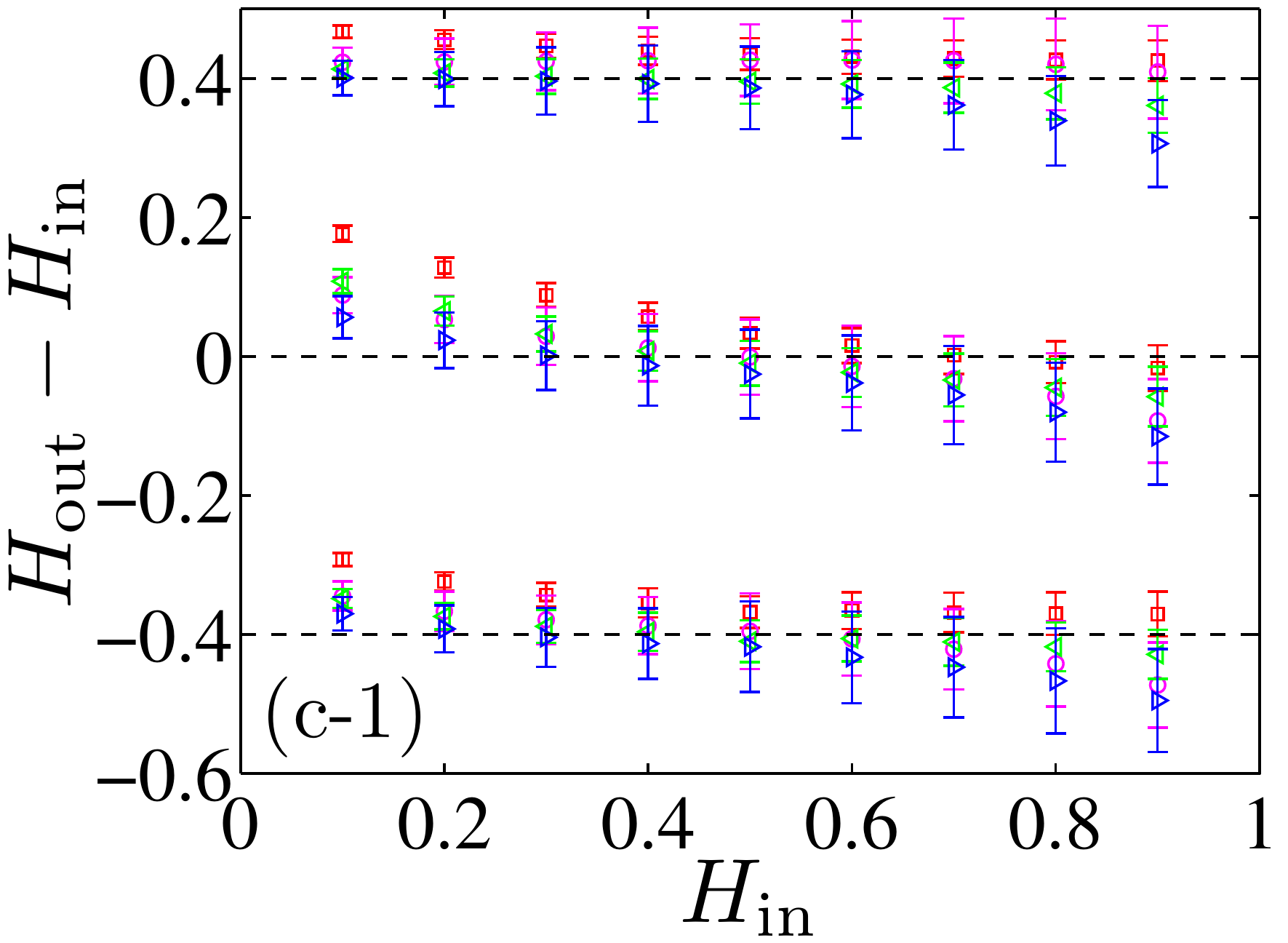}
\includegraphics[width=5.5cm,height=4cm]{Fig_H_4_999.pdf}
\includegraphics[width=5.5cm,height=4cm]{Fig_H_4_1992.pdf}
\includegraphics[width=5.5cm,height=4cm]{Fig_H_4_5000.pdf}\\
\includegraphics[width=5.5cm,height=4cm]{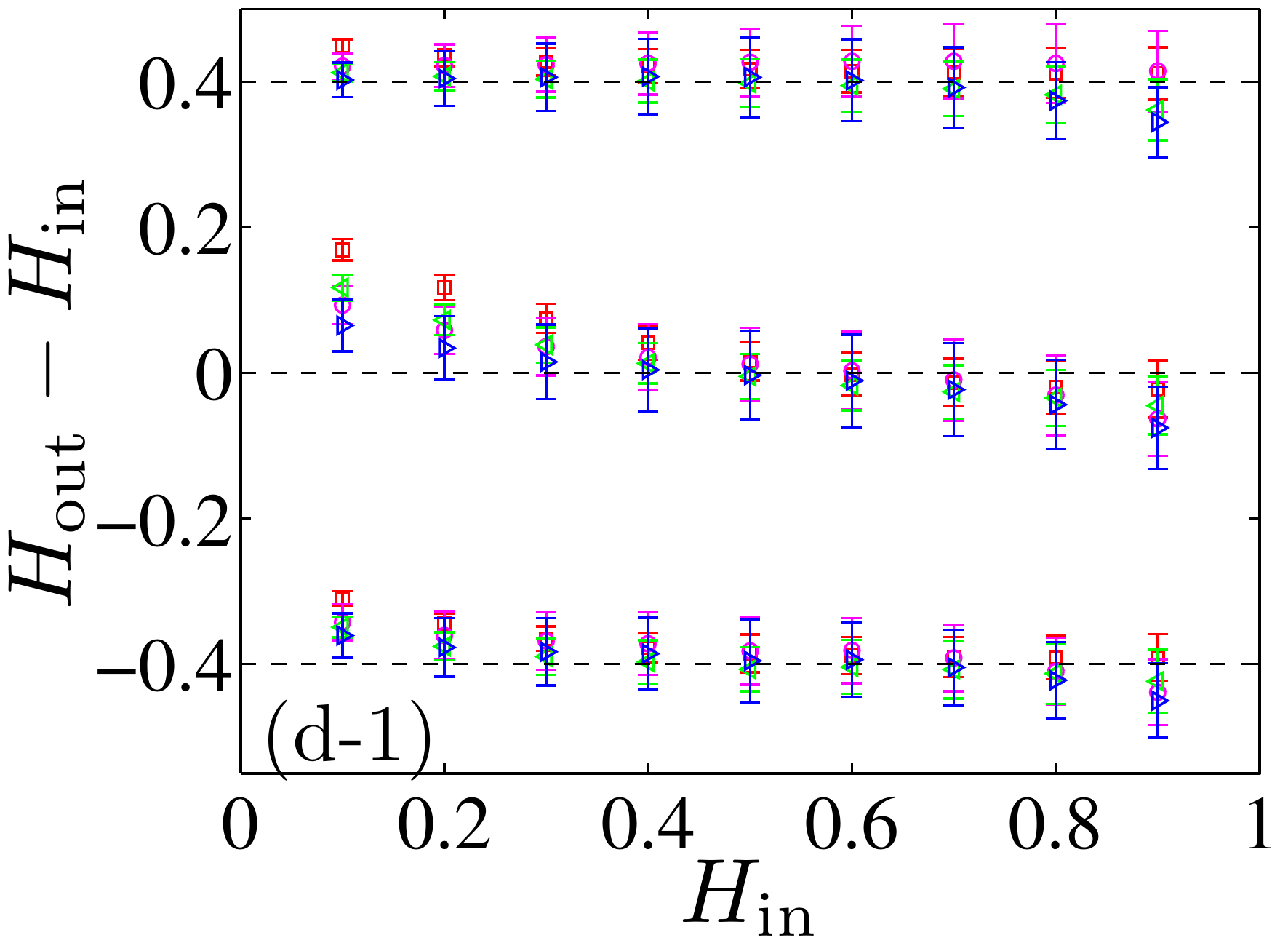}
\includegraphics[width=5.5cm,height=4cm]{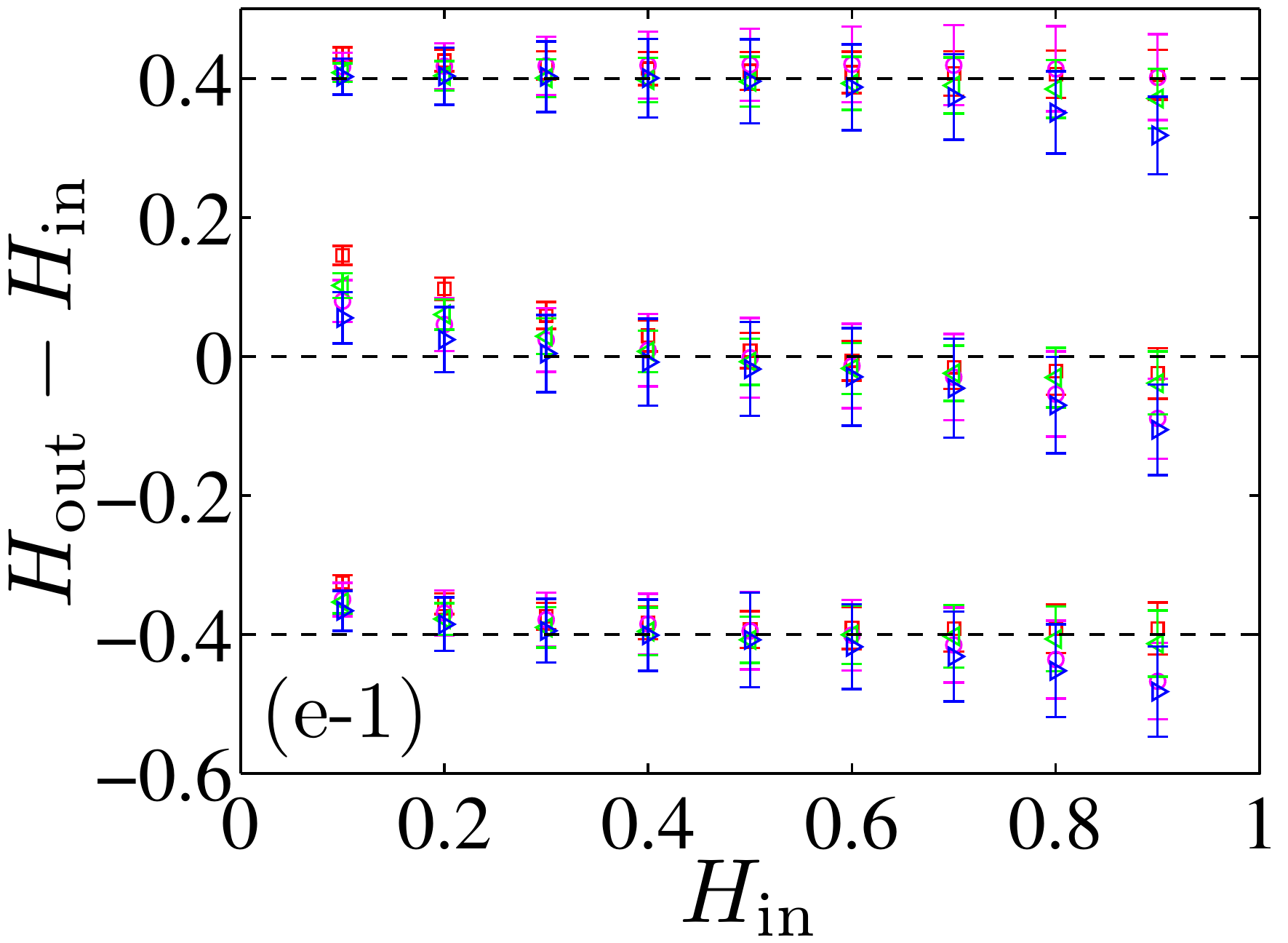}
\includegraphics[width=5.5cm,height=4cm]{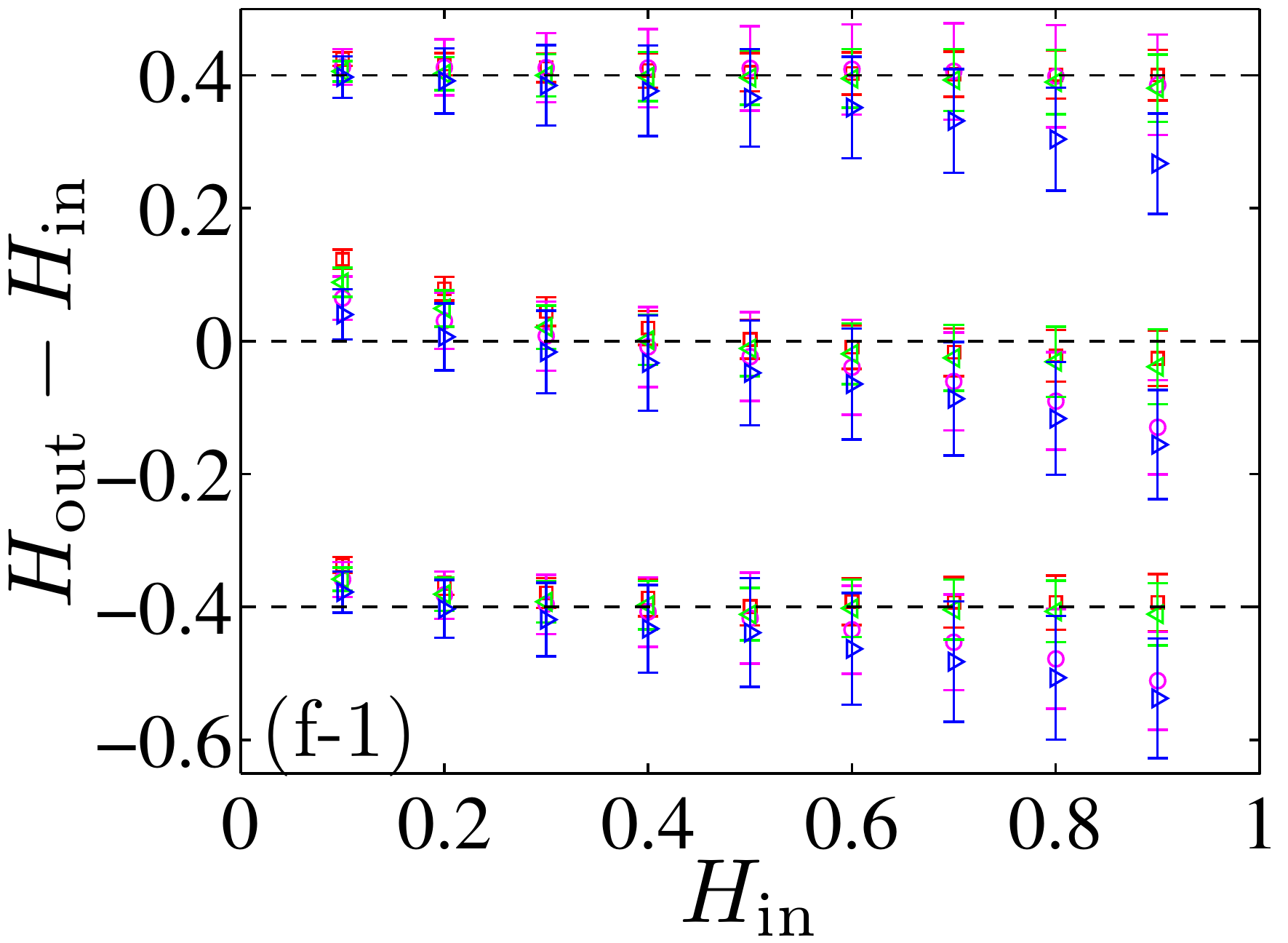}
\includegraphics[width=5.5cm,height=4cm]{Fig_H_10_999.pdf}
\includegraphics[width=5.5cm,height=4cm]{Fig_H_10_1992.pdf}
\includegraphics[width=5.5cm,height=4cm]{Fig_H_10_5000.pdf}\\
\includegraphics[width=5.5cm,height=4cm]{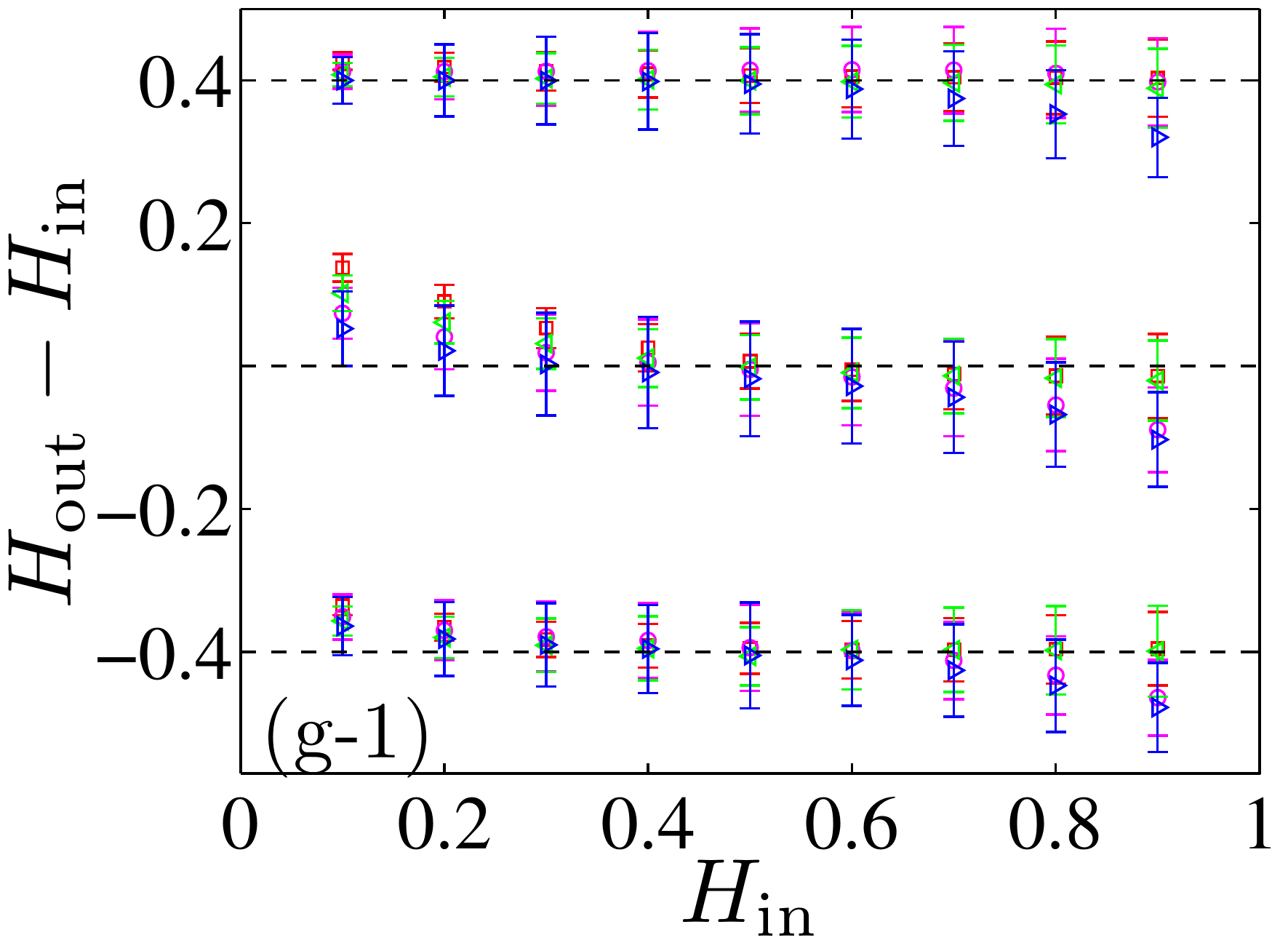}
\includegraphics[width=5.5cm,height=4cm]{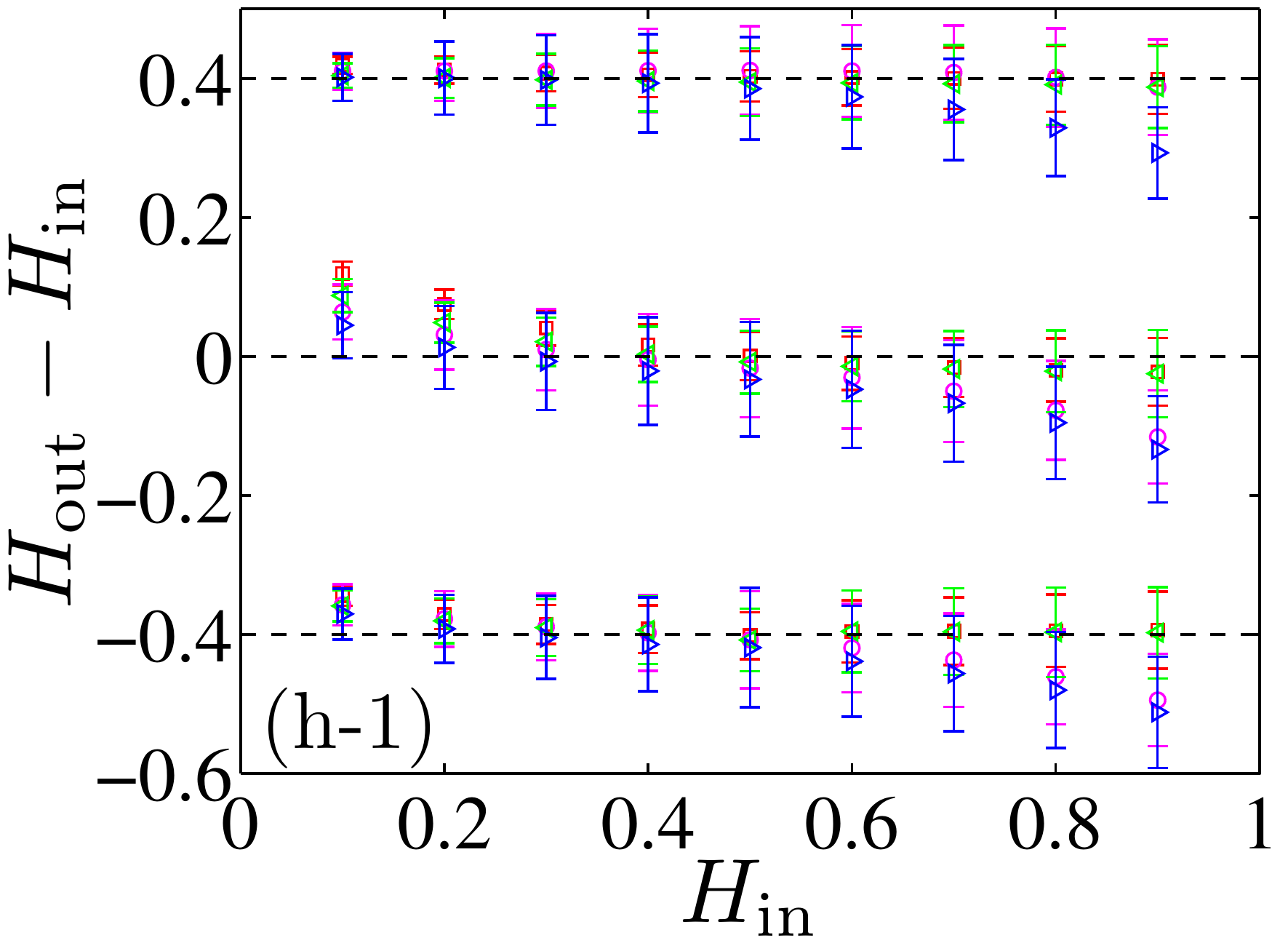}
\includegraphics[width=5.5cm,height=4cm]{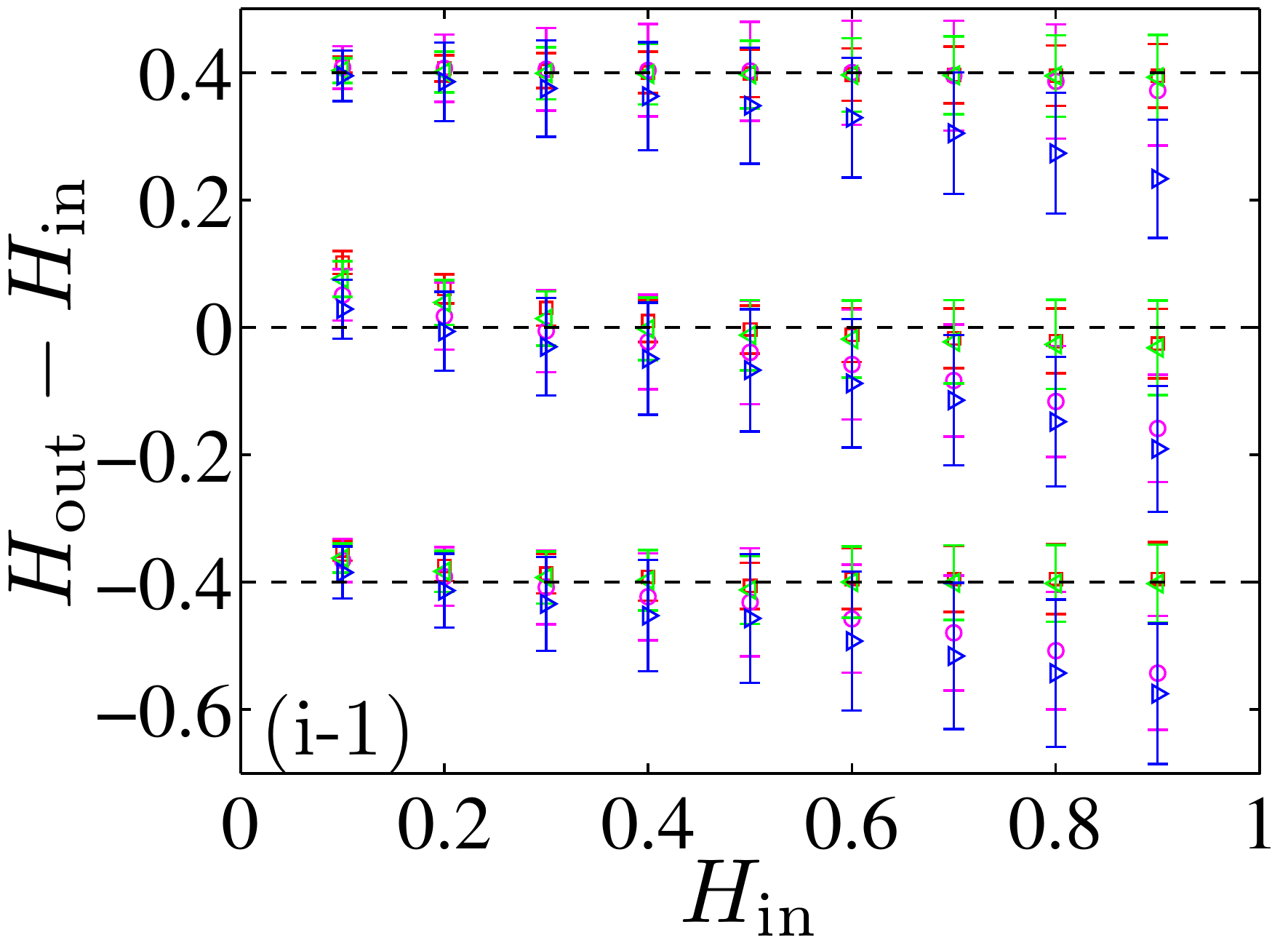}
\includegraphics[width=5.5cm,height=4cm]{Fig_H_20_999.pdf}
\includegraphics[width=5.5cm,height=4cm]{Fig_H_20_1992.pdf}
\includegraphics[width=5.5cm,height=4cm]{Fig_H_20_5000.pdf}
\caption{{\textbf{Comparing impacts of the scaling range on the Hurst index estimates.}} The plots labeled (a)-(o) are the same as in the paper where the length of time series is 20000, while the plots labeled with (a-1) to (o-1) are the results where the length of time series is 2000.}
\label{SFig:ScalingRange}
\end{figure*}

\end{document}